\documentclass[twoside,pra,a4paper,12pt,tightenlines,eqsecnum,nofootinbib,showpacs]{revtex4}
\usepackage{amsmath,amssymb}
\usepackage{float}
\newif\ifpdf
        \ifx\pdfoutput\undefined
        \pdffalse 
        \else
        \pdfoutput=1 
        \pdftrue
        \fi
\ifpdf
        \usepackage[pdftex]{graphicx}
        \DeclareGraphicsExtensions{.pdf, .jpg}
\else
        \usepackage[dvips]{graphicx}
                \DeclareGraphicsExtensions{.eps, .jpg}
\fi
\def\decale#1{\par\noindent\hskip 3em\llap{{\it 
#1}\enspace}\ignorespaces}
\catcode`\@=11
\def\eqalign#1{\null\,\vcenter{\openup\jot\m@th
\ialign{\strut\hfil$\displaystyle{##}$&$\displaystyle{{}##}$\hfil
     \crcr#1\crcr}}\,}
\catcode`\@=12
\catcode`\@=11
\def\Eqalign#1{\null\,\vcenter{\openup\jot\m@th
\ialign{\strut\hfil$\displaystyle{##}$&$\displaystyle{{}##}$\hfil
&&\qquad\strut\hfil$\displaystyle{##}$&$\displaystyle{{}##}$\hfil
     \crcr#1\crcr}}\,}
\catcode`\@=12

\def\vec#1{\boldsymbol{#1}}

\def\H{{\rm H}}
\def\D{{\rm D}}
\def\Ps{{\rm Ps}}

\renewcommand{\ap}{{\bar{\mathrm p}}}
\let\pa=\ap
\newcommand{\ad}{{\bar{\mathrm d}}}
\newcommand{\aH}{{\overline{\mathrm H}}}
\let\Ha=\aH
\newcommand{\br}{{\boldsymbol{r}}}
\newcommand{\bs}{{\boldsymbol{s}}}
\newcommand{\etal}{{\emph{et al.}}}
\newcommand{\had}{\hat{H}_{\mathrm{ad}}}
\newcommand{\hpad}{\hat{H}'_{\mathrm{ad}}}
\newcommand{\hint}{\hat{H}_{\mathrm{int}}}
\newcommand{\hpint}{\hat{H}'_{\mathrm{int}}}
\newcommand{\half}{{\scriptstyle{1\over 2}}}
\newcommand{\mL}{{\mathcal{L}}}
\newcommand{\rme}{{\mathrm{e}}}    \let\e=\rme
\newcommand{\p}{{\mathrm{p}}}
\newcommand{\Pt}{{\mathrm{t}}} \let\t=\Pt
\newcommand{\rmh}{{\mathrm{h}}}

\newcommand{\ro}{{\mathrm{o}}}
\newcommand{\Pd}{{\mathrm{d}}}
\let\d=\Pd
\newcommand{\Pn}{{\mathrm{Pn}}}
\newcommand{\Hy}{{\mathrm{H}}}

\newcommand{\Htwo}{{\mathrm{H}}_2}

\newcommand{\He}{{\mathrm{He}}}

\newcommand{\HeaH}{\mathrm{He}\overline{\mathrm{H}}}
\newcommand{\HyaH}{\mathrm{H}\overline{\mathrm{H}}}
\newcommand{\PsHy}{\mathrm{Ps}\mathrm{H}}
\newcommand{\Heap}{\mathrm{He}\bar{\mathrm{p}}}

\begin{document}
\ifpdf
\DeclareGraphicsExtensions{.pdf, .jpg}
\else
\DeclareGraphicsExtensions{.eps, .jpg}
\fi
\title{STABILITY OF FEW-CHARGE SYSTEMS\\[4pt]
IN QUANTUM MECHANICS}
\author{E.A.G.~Armour}
\affiliation{School of Mathematical Sciences, University of Nottingham, Nottingham NG7 
2RD, UK}
\author{J.-M.~Richard}
\affiliation{Laboratoire de Physique Subatomique et Cosmologie, \\
Universit\'e Joseph Fourier--CNRS-IN2P3, \\
53, avenue des Martyrs, 38026 Grenoble Cedex, France}
\author{K.~Varga}
\affiliation{Condensed Matter Science Division,
Oak Ridge, TN 37831, USA,\\
and \\
Institute of Nuclear Research of the Hungarian Academy 
of Sciences, ATOMKI, Debrecen PO Box 51, Hungary}
\date{Last update \today, by JMR}
\begin{abstract}
We consider non-relativistic systems in quantum mechanics  
interacting through the Coulomb potential, and discuss the existence of bound
states which are stable against spontaneous dissociation into smaller atoms or ions.  We 
review the studies that have been made of specific mass configurations 
and also the properties of the domain of stability in the space of 
masses or inverse masses.  These rigorous results are supplemented by 
numerical investigations using accurate variational methods.  A 
section is devoted to systems of three arbitrary charges and 
another to molecules in a world with two space-dimensions.
\end{abstract}
\maketitle
 \markboth{\sl Stability of few-charge systems}
        {\sl Contents}
\clearpage\tableofcontents
 \markboth{\sl Stability of few-charge systems}{\sl Introduction}
\clearpage\label{se:Intro}
\section{Introduction}
In classical mechanics, determining the motion of three bodies which 
attract each other according to Newton's universal law of gravitation 
is the most celebrated of all dynamical problems (see, for example, 
\cite{Whi52}, \cite{Roy78}).  Over the period since 1750, it has 
attracted the attention of some of the greatest mathematicians and, in 
particular, Euler, Lagrange, Jacobi, Poincar\'e and Levi-Civita.  It 
cannot be solved exactly.  However, recently the advent of high-speed 
computers has opened up a whole new approach to the problem by making 
possible step-by-step numerical integration of the differential 
equations of motion from the initial time to any desired later time. 
The quantum three-body problem also has a rather
well-known history, in particular for systems governed by Coulomb forces. 
Some contributions will be reviewed in this article.

Suppose we have a quantum system made up of three particles, two 
having the same charge and the third having a charge of the same 
magnitude but of opposite sign, interacting only through Coulombic 
forces.  According to non-relativistic quantum mechanics, such a 
system may, or may not, form bound states, depending on the relative 
masses of the particles.  For instance $\H_{2}^+ (\p,\p,\rme^-)$, 
$\Ps^-(\rme^+,\rme^-,\rme^-)$
or $\H^-(\p,\rme^-,\rme^-)$ are stable, while 
$(\p,\rme^+,\rme^-)$ or $(\p,\ap,\rme^-)$ are not.
The critical values of the relative masses 
at which bound states appear have not been established.  However, we 
shall describe in this article some of the interesting and often 
ingenious methods which have been devised to obtain bounds on the 
number of bound states of a given system and also on the critical mass 
ratios at which such states appear.
Our aim is also to describe the properties of the stability domain as a 
function of the masses or inverse masses of the particles involved.

We then extend our study in several ways. First we consider three-body systems
with arbitrary charges.
There are sets of charges for which stability is obvious, i.e., is achieved whatever values are assumed for the masses.
On the other hand, there are more interesting situations, where stability requires very large values of certain mass ratios.

Another field deals with systems containing more than three unit charges.
In the case of four charges, 
whose best known prototype is the hydrogen molecule ($\mathrm{p,p,e^{-},e^{-}})$,
we discuss for which masses there exists at least one bound state 
stable against spontaneous dissociation. 

The sections which follow will be devoted to systems with five, 
six or more unit charges. Here, the boson or fermion character 
has to be specified when dealing with more than two identical particles.

We will also repeat some of the studies with only $d=2$ space 
dimensions. In the one-body case, (or for two self- interacting
particles), the situation is dramatically different for $d\le 2$ and
$d>2$ since in the former case, any attractive potential, however
weak, will produce a bound state. 
One expects also qualitative changes for more complex systems. Some
conclusions, and a list of unsolved  problems will be included in the
last section.

Some useful results on Schr{\"o}dinger operators,
an introduction to methods for solving the few-body problems, in
particular to the Born--Oppenheimer approximation and 
variational calculations, and  a recollection of elementary 
calculations used historically to establish the stability of 3- and
4-body systems, are given in Appendixes.

This review contains both a summary of rigorous results and a
survey  of accurate numerical methods. The knowledge of precise binding energies of few-charge systems
is extremely useful in comparing the efficiency and accuracy 
of various quantum mechanical methods. It also helps
development of new methods and the improvement of those already in
   existence, as well as the understanding of electron--electron, 
electron--positron or electron--lepton  correlations and dynamics.

The investigation of the stability of few-charge systems
is not only a very good testing ground for different models
and methods but it also leads to important applications.
The study of stability provides very useful information 
for various physical systems, e.g.,  positronic atoms 
\cite{Mit02,Mit02a,Mit00,
Mit99,Ryz97}, antimatter compounds \cite{Sch98}, 
systems of atoms and antiatoms \cite{Reb03,Reb02}, 
or charged excitons (system of electrons and holes) 
in semiconductors \cite{Urb03}. New experimental 
investigations \cite{Urb03} of the fine structure 
of charged excitons in semiconductor quantum dots have 
shown the existence of bound states of three electrons and a 
positive charge or four electrons plus two-positive
charges. The stability studies presented in this
paper clearly show that such systems are not stable 
if only the pairwise Coulomb (or Coulomb like) interaction 
is included and their existence is therefore due to some other binding 
mechanism, e.g., confinement. There is also an interesting analogy
between multiquark systems in hadron physics, and small exotic
molecules: the property of flavour independence states that the
same potential applies to quarks of different masses, in the same
way that the same Coulomb interaction is felt by electrons, muons
or antiprotons. For such systems, similar patterns are observed for
configurations leading to maximal stability against spontaneous
dissociation~\cite{Ric92a}.

In the preparation of this review article, it has been necessary to read hundreds
of papers.  Unfortunately, it is impossible to discuss or even list all
the contributions that have been made to the subject of few-charge systems
in quantum mechanics.  For example, there are more than twenty papers dealing
with the binding energy and static properties of the positronium ion, with
considerable overlap between them.  In each section, the key pioneering papers
are mentioned as well as a wide range of more recent contributions which
should make it possible for the reader to explore the parts of the literature
that are not referred to.  We apologise in advance to any author of a
significant contribution that has inadvertently been omitted.

\markboth{\sl Stability of few-charge systems} {\sl Stability criteria}
\clearpage\section{Stability criteria}
\label{se:crit}
As we are considering self-interacting, translation-invariant Hamiltonians in this review, 
we shall assume that the centre-of-mass motion is removed
(see Appendix \ref{se:SO}), leading to the ``interaction Hamiltonian''  $\hint$.
In the case of three unit charges $(m_1^+,m_2^-,m_3^-)$, one can assume $m_2\le m_3$. 
Then, in simplified units,
 \begin{equation}\label{crit:eq:Ethr}
E_{\mathrm{th}} = -\half\mu_{13} = -\half { m_1 m_3 \over m_1+m_3 }~,
 \end{equation}
the ground-state energy of the hydrogen atom-like system made up of 
particles 1 and 3, can be shown to be the lowest continuum threshold 
of $\hint$. For systems with more than three charges, there is always such a threshold at which the system separates into two or more smaller
neutral or charged systems made up of one or more particles.

 Any eigenfunction with energy below this threshold will 
be square-integrable and thus correspond to a bound state \cite{Hun66}.

This behaviour will be called \emph{genuine stability} or simply, \emph{stability}. Of course, it is sufficient to establish that some upper bound, as given for instance by the Rayleigh--Ritz 
variational method (see Appendix \ref{se:SO}), lies below $E_\mathrm{th}$.
However, if it cannot be established using the Rayleigh--Ritz method that an
upper bound exists that is below $E_\mathrm{th}$, this does not necessarily mean that
the state under consideration is not bound.  It may be that the nature of the
variational trial function makes it impossible to obtain a sufficiently good
upper bound. One possible way of showing that a state is not bound, when the
Rayleigh--Ritz method is inconclusive, is to obtain a lower bound to the energy
of the state and show that this is above $E_\mathrm{th}$ 
(see Appendix \ref{se:SO}, Section \ref{SO:sub:comp}).

Except in a few cases, we shall not consider the possibility of bound states in the 
 continuum.  The existence of such states is conceivable but they are 
 rather pathological and are expected to be unstable under 
 perturbations \cite{Sim70}. 

Forms of stability weaker than genuine stability can also occur.

\decale{1.} Imagine two heavy nuclei or particles surrounded by 
electrons.  The particles behave almost classically in the 
Born--Oppenheimer potential $V(r)$.  If $V(r)$ has a local minimum at 
internuclear distance $r=r_0$, classical particles  with sufficiently 
small kinetic energy would remain in the vicinity of this 
minimum, even if its value $V(r_0)$ lies above the dissociation 
energy.  We shall call this behaviour  \emph{classical stability}.

\decale{2.} Imagine a 3-body state $(a^{+},b^-,c^-)$ with angular 
momentum and parity $J^P=1^+$, this not including intrinsic spins and 
parities.  Then the decay 
\begin{equation}
\label{crit:eq:Iabc}
(a^+,b^-,c^-)\to (a^+,b^-)_{\rm 1s}+c^-~,
\end{equation}
cannot take place if the atom in the final state is in its 1s 
ground-state with angular momentum and parity $j^p=0^+$. The decay 
requires higher states, for instance 2p. This means that the threshold is 
higher, as long as radiative corrections are neglected. Otherwise, 
the reaction
\begin{equation}
\label{crit:eq:abcgam}
(a^+,b^-,c^-)\to (a^+,b^-)_{\rm 1s}+c^-+\gamma~,
\end{equation}
would have to be considered. We shall call this behaviour  \emph{metastability}.

\markboth{\sl Stability of few-charge systems} {\sl Particular three-unit charge systems}
\clearpage\section{Particular three-unit-charge systems}
\label{se:3u1}
In this section, we present the methods applied to examine the stability of 
three unit-charge systems, and discuss in some details configurations which have 
been extensively studied. The next
sections will be devoted to global studies of the stability region in the parameter space.
\subsection{Notation}
\label{3u1:sub:not}
We consider systems with one charge, $q_1$, of mass $m_1$, 
and two charges opposite to $q_1$, say $q_2=q_3=-
q_1$, with masses $m_2$ and $m_3$, respectively.
By charge conjugation invariance,  it does not matter whether $q_1>0$ and 
$q_2=q_3<0$ or vice-versa. By scaling, one can take $q_1=-q_2=-q_3=1$.

The Hamiltonian thus reads
\begin{equation}
\label{3u1:eq:H }
\hat{H}={\vec{p}_1^2\over 2 m_1}+{\vec{p}_2^2\over 2 m_2}+{\vec{p}_3^2\over 2
m_3}+ {1\over r_{23}}-{1\over r_{12}}-{1\over r_{23}}~,
\end{equation}
where $r_{ij}=\vert\vec{r}_j-\vec{r}_i\vert$ is the distance between particles $i$ and $j$.

If we separate off the centre-of-mass motion, we obtain the interaction
Hamiltonian
\begin{equation}
\label{ 3u1:eq:hintz}
\hint=\hat{H}-{(\vec{p}_1+\vec{p}_2+\vec{p}_3)^2\over 2(m_1+m_2+m_3)}~.
\end{equation}
 Then if we take the origin of $\hint$ to be at particle 3,
 \begin{equation}\label{3u1:eq:hint}
\hint = -{1 \over 2\mu_1} \nabla^2_{\br_1} -
{1 \over 2\mu_2} \nabla^2_{\br_2} - {1\over m_3} \nabla_{\br_1}
\cdot \nabla_{\br_2} - {1 \over r_1} + {1 \over r_2}
-{1 \over r_{12}} \;,
 \end{equation}
where $\br_1$ and $\br_2$ are the position vectors of particles 1
and 2, respectively, with respect to the third particle as origin,
 $m_i$ is the mass of the $i$th particle, and
 \begin{equation}\label{3u1:eq:def-mui}
\mu_1 ={ m_1 m_3 \over m_1 + m_3 }~, \qquad
\mu_2 ={ m_2 m_3 \over m_2 + m_3 }~,
 \end{equation}
where $\mu_1$, for example, is the reduced mass of particles 1 and 3.

$\hint$ represents the motion of two particles (more properly, 
pseu\-do-par\-ti\-cles) of mass $\mu_1$ and $\mu_2$, respectively.  It 
differs from the usual form for the Hamiltonian for two particles as 
it contains a coupling term between the momenta of the particles.  
This is known as the Hughes--Eckart or mass-polarisation term 
\cite{Hug30,Bet77,Bra83}.

If $m_3$ is very much larger than $m_1$ and $m_2$, the effects 
introduced by locating the origin at particle 3 are very small.  For 
this reason, they are normally neglected in the analogous many-body 
problem of an $n$-electron atom.
 
The mass-polarisation term can be eliminated by using Jacobi 
coordinates, first introduced by Jacobi and Lagrange, in the context 
of the three-body problem in Newtonian mechanics \cite{Roy78}.  If we 
transform to new internal coordinates, (see Appendix \ref{se:SO}).
 \begin{equation}\label{3u1:eq:Jacobi}
\vec{x}=\br_1~,\qquad
\vec{y} = \br_2 - {m_1 \over m_1 + m_3}\br_1~, 
 \end{equation}
which corresponds to taking the origin of the coordinates of particle
2 to be at the centre of mass of particles 1 and 3, $\hint$ is of
the form \cite{Far75,Mes61}.
 \begin{equation}\label{3u1:eq:Hint2}
\hint = {1 \over 2\mu_x} \nabla^2_{\vec{x}}
- {1 \over 2\mu_y} \nabla^2_{\vec{y}} 
+ {1 \over | \vec{y}+ \eta' \vec{x} | } - {1 \over x}
- {1 \over | \vec{y} - \eta \vec{x} | }~,
 \end{equation}
where
 \begin{equation}\label{3u1:eq:redm}
\mu_x=\mu_2~,\quad
\mu_y= { m_2 (m_1+m_3) \over (m_1+m_2+m_3) }~,\quad 
\eta={ m_3 \over m_1+m_3 }~,\quad
\eta'= { m_1\over m_1+m_3 }~\cdot
 \end{equation}
Note that though $\vec{x} = \br_1$, $\nabla^2_{\vec{x}}$ differs from 
$\nabla^2_{\br_1}$ as $\vec{y}$, and not $\br_2$, is held constant 
during the partial differentiation.  As seen in Appendix \ref{se:SO}, using Jacobi 
coordinates simplifies the kinetic energy part of $\hint$ at the 
expense of making the potential energy part more complicated.
\subsection{H$^+_2$}\label{3u1:sub:H2+}
The hydrogen molecular ion, H$^+_2$, belongs to a special 
category of three-body systems that contain two identical 
particles and a third particle with charge of equal but opposite 
sign.  
 Using the comparison theorem, see Appendix \ref{se:SO}, Hill \cite{Hil77} 
was able to prove that systems of this type always have at least 
one bound state.
 He was able to show further that these systems have only one 
bound state if $m_i/m_d < 0.210101636$, where $m_i$ is the mass 
of the identical particles and $m_d$ is the mass of the remaining 
particle.

 In the case of H$^+_2$, where the two identical particles are 
protons and the remaining particle an electron, $m_i/m_d = 1836.2$.  
 Thus this system is far outside the region in which it can be 
shown that only one bound state exists.  
 As might be expected, it can be shown that this system has many 
bound states.
 
 The large difference in mass between the protons and the 
electron means that the Born--Oppenheimer approximation (Appendix \ref{se:BO}) 
gives accurate results when applied to this system.  
 In this approximation, the nuclei are fixed and the energy of 
the electron is calculated as a function of the internuclear
distance, $R$.  
 The energy of the system is then obtained by calculating the
energy of the nuclei moving in a potential made up of the 
electronic energy and the internuclear potential.

 The Hamiltonian for the electron moving in the field of the 
fixed nuclei is
 \begin{equation}\label{3u1:eq:H2+}
\hat{H} = -{1\over 2m_{\rme}} \nabla^2 - {1\over r_A}
- {1\over r_B}~,
 \end{equation}
where $m_{\rme}$ is the mass of the electron and $r_A$ and 
$r_B$ are the distance of the electron from nucleus $A$ and $B$, 
respectively.  
 The units are such that $\hbar=1$ and $e=1$, where $e$ is the 
charge on the proton.
 In terms of prolate spheroidal (confocal elliptical) coordinates
$\lambda$, $\mu$, $\phi$ (see, for example, \cite{Fla57}), the 
potential, $V$, in this Hamiltonian is of the form
 \begin{equation}\label{3u1:eq:V}
V = -{4\over R} {\lambda \over \lambda^2-\mu^2}~,
 \end{equation}
where
 \begin{equation}\label{3u1:eq:prol}
\lambda = {r_A+r_B \over R} \;, \quad
\mu = {r_A-r_B \over R}~,
 \end{equation}
and $\phi$ is the azimuthal angle with respect to Cartesian 
coordinates
with origin at the midpoint of the internuclear axis and $z$-axis
directed along this axis.

As is well known \cite{Bur27,Eis48}, the 
Schr\"odinger equation with this Hamiltonian is separable in 
terms of prolate spheroidal coordinates.
 Details of papers describing how the resulting ordinary 
differential equations can be solved to give physically acceptable 
solutions are listed on page 2 of Ref.~\cite{Fla57}.  
 Details of square-integrable wave functions and energies are given 
 by Bates \etal~\cite{Bat53}.  
 More accurate results for the ground-state energy have been obtained by 
Wind \cite{Win65} and Peek \cite{Pee65}.
 For a review of calculations of this type, see Bates and Reid \cite{Bat68}.

 Calculations of bound-state energies for H$_2^+$ and the very closely 
related system HD$^+$, that include corrections to the 
Born--Oppenheimer approximation, have been carried out by 
Beckel \etal~\cite{Bec70}, 
Wolniewicz and Poll \cite{Wol80}, 
Wolniewicz and Orlikowski \cite{Wol91},
Moss \cite{Mos93},
Taylor \etal~\cite{Tay99}, 
Korobov \cite{Kor00},  
Hilico \etal~\cite{Hil01}
and Frolov~\cite{Fro03}.
 The results of Wolniewicz and Poll's calculations for HD$^+$
are compared with experiment by Carrington and Kennedy \cite{Car85}, 
who also include a detailed discussion of the underlying theory 
and an extensive list of references to earlier work on HD$^+$.  
 Theory and experiment are in agreement that there are a large 
number of bound states for these systems.  
 The bound states are classified using a vibrational quantum 
number, $v$, and a rotational quantum number, $J$.  
 See, for example, Bransden  and Joachain \cite{Bra83}.  
 $v$ may have a value as high as $v=19$ for H$_2^+$ and
$v=21$ for HD$^+$.  In general, each value of $v$ is associated with a number of 
values of $J$, starting with $J=0$.

Recently, Carbonell \etal\ \cite{Car02} found a new state of H$_2^+$, with $J=0$, but with an antisymmetric coordinate-space wave-function for the two protons. The binding energy is extremely small, of
the order of $10^{-9}\,$a.u.\@ This state has been discovered by studying how the
$a^+\H$ scattering length evolves as a function of the mass of the projectile
$a^+$. The case of H$_2^+$ corresponds to $a$ being a proton. The scattering
equations at zero energy are solved in the configuration-space Faddeev formalism.
The energy of the new state  of H$_2^+$ was confirmed by a very accurate variational
calculation.

($\p,\rme^-,\mu^+$) and ($\mu^+,\rme^-,\mu^+$) are systems similar to
H$^+_2$ but with one or both protons replaced by a positively 
charged muon.  The mass of the muon is $206.8\,m_{\rme}$.  
 This is much less than the mass of the proton, but sufficiently 
greater than $m_{\rme}$ that the Born--Oppenheimer approximation gives
accurate energy values when applied to these systems.  
 In this approximation, the internal energy of ($\p,\rme^-,\mu^+$) or 
($\mu^+,\rme^-,\mu^+$) is obtained by calculating the energy levels 
of a pseudo-particle with mass equal to the reduced mass of 
($\p,\mu^+$) or ($\mu^+,\mu^+$) moving in the same electronic and 
Coulombic internuclear potential as in the case of H$_2^+$
and HD$^+$   (see Appendix~\ref{se:BO}).   
 The reduced masses of ($\p,\mu^+$) and ($\mu^+,\mu^+$) are $185.9\,m_{\rme}$ 
and $103.4\,m_{\rme}$, respectively.  
 These values are much less than $918.1\,m_{\rme}$, the reduced 
mass of (p,p).  It follows from this that these systems do have bound states 
but substantially fewer than in the case of H$_2^+$.  
 See, for example, Bransden and Joachain \cite{Bra83}.
 For a non-adiabatic calculation of ($\mu^+,\rme^-,\mu^+$), see, e.g., \cite{Fro99a}, and for the asymmetric  ($\p,\rme^-,\mu^+$), ($\mathrm{d},\rme^-,\mu^+$) and ($\mathrm{t},\rme^-,\mu^+$) systems,~\cite{Fro03a}.
\subsection{H$^-$}\label{3u1:sub:Hminus}
 H$^-$ contains two electrons and a proton and is thus another 
example of a three-body system that contains two identical 
particles of unit charge and a third particle with equal but 
opposite charge.   In this case, $m_i/m_d = 0.00054$.  
 It follows from the result of Hill \cite{Hil77} referred to earlier 
that only one bound state of this system exists. This is the singlet ground state.

 It has been known from early on in the development of quantum 
mechanics that the singlet ground state is bound.  See Appendix \ref{se:elem}.
 Variational upper bounds of sufficient accuracy to demonstrate 
this were obtained by Bethe \cite{Bet29}, Hylleraas \cite{Hyl30} and 
Chandrasekhar \cite{Cha44}.  Very accurate upper bounds for the energy of this state have 
been obtained by Pekeris \cite{Pek62}, Frolov and Yeremin \cite{Fro87}, 
Cox \etal~\cite{Cox94} and Drake \etal\cite{Dra02}.

 There exist a number of autodetaching or autoionising states 
of H$^-$ below the continuum thresholds corresponding to excited 
states of the H atom. See Massey \cite{Mas76}, Bhatia and Temkin \cite{Bha69} 
and Bhatia \cite{Bha74}.  
 In particular, there is a triplet P$^{\ro}$ state below the $n=2$ 
continuum threshold. This is not a bound state as it is embedded in the $n=1$ continuum; 
it has a small but non-zero width indicating that autoionisation 
can occur through interaction with states in this continuum.

 There is also a triplet P$^{\rme}$ state with energy 0.0095~eV below 
this threshold.   See Bhatia \cite{Bha70} and Drake \cite{Dra70}.  
 In this case autoionisation is forbidden if spin-orbit coupling 
is neglected as in this review article.
 This is because the $n=1$ continuum does not contain any states 
of the required angular momentum and parity for this process
to occur. This triplet P$^{\rme}$ state is often described as a bound state.  
 See, for example, Drake \cite{Dra70} and Ho \cite{Ho94}.  
 However, it should more properly be called a metastable state 
as it does not satisfy the criterion for stability specified in 
Sec.~\ref{se:crit} of this review, namely that its energy must be below 
the energy of the lowest continuum threshold of the system.  
 This is also the criterion used by Hill \cite{Hil77}.  
 There is some confusion over this point in the review article on 
the stability of three-body atomic and molecular ions by Armour 
and Byers Brown \cite{Arm93}.

 Mu$^-$ is a similar system containing a positively charged 
muon and two electrons.  The mass of the muon is 206.8 $m_{\rme}$.  
 Thus in this case $m_i/m_d = 0.0048$ and the system, like H$^-$, 
can only have one bound state.  Mu$^-$ has been observed by, for example, 
Kuang \etal~\cite{Kua87}.  Its ground-state energy has been calculated by, for example, 
Bhatia and Drachman \cite{Bha87}, Petelenz and Smith \cite{Pet87b} and
Frolov and Yeremin \cite{Fro89}.

For a review on two-electron atoms and ions, including $\mathrm{H}^-$ and its role in astrophysics, see
\cite{Bet77,Hyl63,Hyl64}.
\subsection{Ps$^-$}\label{3u1:sub:Psminus}
 Ps$^-$, i.e., ($\rme^-,\rme^+,\rme^-$), is a system that contains two identical
particles, in this case electrons, and a third particle, a positron 
of the same mass and equal but opposite charge.   It is thus the special case of the systems considered earlier 
for which $m_i/m_d = 1$.  
 It follows from the result of Hill \cite{Hil77} that this system must 
have at least one bound state.  
 The existence of a bound state of this system was predicted
by Wheeler \cite{Whe46} and shown to be the case by Hylleraas \cite{Hyl47}.  
 It was first observed experimentally by Mills \cite{Mil81}.
 
 The positron is the antiparticle corresponding to the electron.  
 As first shown by Dirac \cite{Dir30}, it can annihilate with an electron to form gamma rays. 
 The rate of this annihilation process for Ps$^-$ was first
measured by Mills \cite{Mil83}.
 
 Details of calculations carried out on Ps$^-$ are given by 
Ho \cite{Ho94} and Cox \etal~\cite{Cox96}.  See, also, Patil~\cite{Pat98}.
 Very accurate calculations have been carried out for the ground 
state of Ps$^-$ by Bhatia and Drachman \cite{Bha83}, Ho \cite{Ho83,Ho93}, 
Frolov \cite{Fro86,Fro87,Fro99}, Petelenz and Smith \cite{Pet87b},
Frolov and Yeremin  \cite{Fro89},  Cox \etal~\cite{Cox96}, Krivec \etal~\cite{Kri00} and Drake \etal~\cite{Dra02}.
 
 All the evidence indicates that Ps$^-$ has only one bound state.  
 It is of interest to note that all attempts to obtain an energy 
for a triplet P$^{\rme}$ metastable state below the $n=2$ threshold 
for Ps$^-$ have been unsuccessful.  
 See Mills \cite{Mil81a} and Ho \cite{Ho94}.

 Bhatia and Drachman \cite{Bha83} have examined the behaviour of the 
energy of the triplet P$^{\rme}$ state as $m_d/m_i$ is varied.  
 They show that this energy is below the energy of the $n=2$ 
continuum if $m_d/m_i > 16.1$ or $m_d/m_i < 0.4047$.  
 This agrees with the findings described earlier for H$^-$ and 
shows that the energy of this state of Mu$^-$ must be below the 
energy of the $n=2$ continuum. Note that $m_d/m_i = 1$ for Ps$^-$.
\subsection{$(\p,\rme^-,\rme^+)$}\label{3u1:sub:pemep}
 $(\p,\rme^-,\rme^+)$, an H atom and a positron, is an interesting case for
which the variational method fails.  
 Inokuti \etal~\cite{Ino60} were able to show by this method that if 
the mass of the proton is taken to be infinite, a ``positron'' with 
mass $\ge 7.8\,m_{\rme}$ would form a bound state with an H atom.  
 Frost \etal~\cite{Fro64} reduced this upper bound to $2.625\,m_{\rme}$ 
by using a more flexible trial function.  It was reduced still further to $2.20\,m_{\rme}$ by 
Rotenberg and Stein \cite{Rot69}.  
 They made their trial function even more flexible by including 
basis functions suitable for representing a weakly bound positron 
in a potential, $V(r)$, with the appropriate asymptotic form for 
an H atom and a positron \cite{Bra83,Tem59}, 
 \begin{equation}\label{3u1:eq:VLR}
V(r)\,\underset{ r\rightarrow\infty }{\sim}\, - {\alpha \over 2r^4} \;,
 \end{equation}
where $\alpha$ is the dipole polarisability of the H atom and $r$ 
is the distance of the positron from the proton.

The method described at the end of the section in Appendix~\ref{se:SO}
on the comparison theorem was used by Spruch \cite{Spr69} to devise a
method for showing that no bound state of the e$^+$H system exists
if the mass of the proton is taken to be infinite.

If we take particle 1 to be the positron and particle 2 to be
the electron, $\hint$ can be expressed in the form
 \begin{equation}\label{3u1:eq:pe-e+}
\hint = -{1\over 2m_1} \nabla^2_{\br_1} + {1\over r_1} 
+ \hat{H}_{\rme}~,
 \end{equation}
where
 \begin{equation}\label{3u1:eq:He}
\hat{H}_{\rme} = -{1\over 2m_2} \nabla^2_{\br_2} - {1\over r_2}
- {1\over r_{12}}~,
 \end{equation}
is the Hamiltonian for an electron in the field of two equal, fixed
positive charges, one at the origin and the other at $\br_1$.
 As discussed in the Section above on H$_2^+$, $\hat{H}_{\rme}$
is just the Hamiltonian which determines the electronic potential
energy of the H$_2^+$ ion in the Born--Oppenheimer approximation.
 The associated Schr\"odinger equation is separable in prolate
spheroidal coordinates.
 The eigenvalues of $\hat{H}_{\rme}$ are functions only of 
$r_1$ and the separability of the Schr\"odinger equation makes
it possible to calculate them to high accuracy \cite{Bat53,Win65}.

 Let us consider the adiabatic Hamiltonian
 \begin{equation}\label{3u1:eq:Had}
\had = \hat{I}(\br_2) \hat{H}_{\mathrm{p}} \;,
 \end{equation}
where
 \begin{equation}\label{3u1:eq:Hp}
\hat{H}_{\mathrm{p}} = -{1\over 2m_1} \nabla^2_{\br_1}
+ {1\over r_1} + E_0(r_1) \;,
 \end{equation}
$\hat{I}(\br_2)$ is the unit operator for allowed square-integrable
functions of $\br_2$ and $E_0(r_1)$ is the ground-state eigenvalue
of $\hat{H}_{\rme}$.
 It is referred to as adiabatic as the potential term, $E_0(r_1)$,
is calculated by fixing $\br_1$, then calculating the ground-state
energy of the electron in the resulting two-centre attractive
Coulombic potential.

 It is easy to show that $\hat{H}_1 = \had$ and
$\hat{H}_2 = \hint$, where $\had$ and $\hint$ are as given above, satisfy
the required conditions for the comparison theorem and have the
same continuum threshold.
 Hence if no bound state of $\had$ exists no bound state of 
$\hint$ exists.
 It is usual to adjust the potential in an essentially one
particle Hamiltonian such as $\had$ so that it tends to zero
as $r_1$ tends to infinity.
 Thus the problem of showing that $\had$ and hence $\hint$ have
no bound states reduces to the problem of showing that the
potential, $V(r)$, cannot support a bound state, where
 \begin{equation}
V(r) = {1\over r} + E_0(r) - E_{\mathrm{H}}^{(0)} \;,
\label{vr}
 \end{equation}
and
 \begin{equation}
E_{\mathrm{H}}^{(0)} = \lim_{r\rightarrow\infty} E_0(r) = 
 -\half m_2~,
\end{equation}
is the ground-state energy of $\H$.
 
 $V(r)$ is a central potential, i.e., it is spherically symmetric.
 Though it depends on the single radial variable, $r$, it is a
potential in three dimensions. This is very important.
 Any attractive well potential can bind a particle in one 
dimension (see, for example, Landau and Lifshitz \cite{Lan77}) but in
three dimensions it has to exceed a critical strength to produce
binding (see, for example, Dyson on p.~1225 of \cite{Dys56}, Wu and Ohmura
\cite{Wu62}, and Sec.~\ref{SO:sub:2D} of Appendix).

 Ways of calculating the number of bound states which $V(r)$
can support have been extensively studied \cite{Bar52,Sch61}.
 As $V(r) \rightarrow 0$ more rapidly than $r^{-2}$ as $r
\rightarrow \infty$ and behaves like $r^{-1}$ as $r \rightarrow
0$, a necessary but \emph{not} sufficient condition for
$V(r)$ to be able to support $N^{(l)}_{\mathrm{b}}$ bound
states corresponding to angular momentum $l$ for a particle
of mass $m$ is that
 \begin{equation}
N_{\mathrm{b}}^{(l)} \le { 2m \over (2l+1)\hbar^2 }
\int_0^\infty r[-V_-(r)] {\mathrm{d}}r~,
 \end{equation}
where
\begin{equation}
\Eqalign{%
&V_-(r) = V(r)\ &\hbox{if}\ \ &V(r) \le 0~, \cr
&V_-(r) = 0\ &\hbox{if}\ \ &V(r) > 0~.\cr
}
\end{equation}
 This is usually referred to as the Bargmann--Schwinger result \cite{Spr69}.
 Also important information about the number of bound states
can be obtained by an analysis of the phase shift of particles
of mass $m$ scattered by $V(r)$ (see, for example, Ref.~\cite{Bur77}).

 The exact number of bound states of the system for a given $l$
value can be obtained by determining the number of zeros (other than
at $r=0$) in the regular solution to the radial Schr\"odinger
equation with potential $V(r)$ and energy $E=0$ \cite{Bar52}.
 This is usually done by step-by-step numerical integration on
a computer.

Unfortunately, Gertler \etal~\cite{Ger68} found that when $m$ is the
mass of the positron, $V(r)$ as given in Eq.~(\ref{vr})
could support one bound state. Thus their attempt to prove that e$^+$H has no bound states
failed.

 Fortunately, Aronson \etal~\cite{Aro71} were able to get round this
difficulty.
 It is reasonable to assume that the ground state of $(\rme^+,\H)$
will be an s state.
 For such a state, in the infinite proton mass approximation
$\hint$ is of the form,
 \begin{equation}\label{3u1:eq:hint3}
\hint = t(r_1) + t(r_2) + \left( {1 \over 2m_1 r_1^2} +
{1 \over 2m_2 r_2^2} \right) \mL^2 + {1 \over r_1}
- {1 \over r_2} - {1 \over r_{12}}~,
 \end{equation}
where
 \begin{equation}
t(r_i) =  {1 \over 2m_i} {1 \over r_i^2} {\partial \over
\partial r_i} \left( r_i^2 {\partial \over \partial r_i} \right)~,\quad
\mL^2 = -{1 \over \sin\theta} {\partial \over \partial
\theta} \left( \sin\theta {\partial \over \partial \theta} \right)~,
 \end{equation}
and $\theta$ is the angle between the directions of the vectors
$\br_1$ and $\br_2$.

In this case if we make the adiabatic approximation of fixing
$r_1$, \emph{not} $\br_1$ as previously, we obtain the
adiabatic Hamiltonian
 \begin{equation}
\hpad = \hat{I}(r_2,\theta) \hat{H}'_{\mathrm{p}}~,
 \end{equation}
where
 \begin{equation}
\hat{H}'_{\mathrm{p}} = t(r_1) + {1 \over r_1} + E'_0(r_1) \;,
 \end{equation}
$\hat{I}(r_2,\theta)$ is a unit operator similar to $\hat{I}(\br_2)$
and $E'_0(r_1)$ is the ground-state eigenvalue of
 \begin{equation}
\hat{H}'_{\rme} = t(r_2) + \left( {1 \over 2m_1 r_1^2} +
{1 \over 2m_2 r_2^2} \right) \mL^2 - {1 \over r_2} 
- {1 \over r_{12}}~.
 \end{equation}
 As
 \begin{equation}
\left\langle \Psi | \mL^2 | \Psi \right\rangle \ge 0~,
 \end{equation}
for any allowed square-integrable function, $\Psi(\br_1,\br_2)$,
it follows that for functions of s symmetry
 \begin{equation}
\left\langle \Psi |\had| \Psi \right\rangle \le
\left\langle \Psi |\hpad| \Psi \right\rangle \;.
 \end{equation}
 However, it can also be shown \cite{Ger68,Aro71} that for functions of S symmetry
 \begin{equation}
\left\langle \Psi |\hpad| \Psi \right\rangle \le
\left\langle \Psi |\hint| \Psi \right\rangle \;.
 \end{equation}
 In addition, $\hpad$ and $\hint$ have the same continuum threshold,
$-\half m_2$.
 Thus if $\hpad$ has no bound states, this is also true for $\hint$.

 The potential $V'(r)$ associated with $\hpad$ is of the form
 \begin{equation}
V'(r) = {1 \over r} + E'_0(r) - E_{\mathrm{H}}^{(0)}~.
 \end{equation}
 Aronson \etal~\cite{Aro71} showed that it is extremely unlikely that
$V'(r)$ can support a bound state and hence it is highly
probable that no bound state of $(\rme^+,\H)$ exists.
 This result is supported by information from scattering
calculations \cite{Bur77,Hum86}.
 Aronson \etal~showed further that it is highly probable that
no bound state of e$^+$H exists for $m_1 < 1.46\,m_{\rme}$
and $m_2 = m_{\rme}$.

 However, Aronson \etal~\cite{Aro71} were unable to establish their
conclusion rigorously. Armour~\cite{Arm78} made their method of proof rigorous.
 He first of all calculated a very accurate wavefunction for
the system described by $\hat{H}'_{\rme}$ using the
variational method and basis functions in terms of prolate
spheroidal coordinates.
 He was able to calculate a good lower bound to $V'(r)$ using
this wavefunction and the method of Temple  and Kato 
(see, for instance, Ref.~\cite{Thi79})
and show that this lowest bound, and hence $V'(r)$, could not
support a bound state.
 In a later paper, Armour and Schrader \cite{Arm82a} showed that if 
$m_1 = 1.51\,m_{\rme}$ and $m_2 = m_{\rme}$, no bound state of
the $\rme^+\H (\rme^+,\rme^-,\p_\infty)$ system exists. To date this is the best lower bound 
on the critical positron mass required for binding.

So far we have assumed that the mass, $m_3$, of the proton in
the $(\rme^+,\H)$ system is infinite. Suppose we choose units in this case so that particle 2 has
unit mass. Then
 \begin{equation}
\hint = -{1 \over 2\bar{m}_1} \nabla^2_{\br_1}
- {1\over 2} \nabla^2_{\br_2} + {1\over r_1} - {1\over r_2}
- {1\over r_{12}} \;,
 \end{equation}
where $\bar{m}_1 = m_1 / m_2$. 
 We can see from this that it is the ratio $m_1 /r m_2$ which
determines whether or not a bound state exists.
 It follows that Armour's result \cite{Arm78}  for the case $m_1=m_2=
m_{\rme}$ shows that this Hamiltonian has no bound state for
$m_1=m_2$, whatever the value of $m_2$.
 
 It follows from this that if the system is to have a bound state
when $m_3$ is finite and thus
 \begin{equation}
\hint = -{1 \over 2\mu_1} \nabla^2_{\br_1}
- {1\over 2\mu_2} \nabla^2_{\br_2} - {1\over m_3} \nabla_{\br_1}
\cdot \nabla_{\br_2} + {1\over r_1} - {1\over r_2}- {1\over r_{12}}~,
 \end{equation}
 this must be due to the presence of the mass-polarisation term.
 As pointed out earlier, $(\e^-,\e^+,\e^-)$, and hence  $(\e^+,\e^-,\e^+)$,
is known to have a bound state \cite{Whe46,Hyl47}.
 Its existence must be due to the large mass-polarisation term
in this case.

 This term can be taken into account by an interesting method due
to Armour \cite{Arm82}.
 We are free to choose the origin of $\hint$ to be at any of the
three particles which make up the system.
 Thus if we choose the origin to be at the electron rather than
the proton,
 \begin{equation}
\hint = -{1 \over 2v_1} \nabla^2_{\bs_1}
- {1\over 2\mu_3} \nabla^2_{\bs_3} - {1\over m_2} \nabla_{\bs_1}
\cdot \nabla_{\bs_3} - {1\over s_1} - {1\over s_3}
+ {1\over s_{13}}~,
 \end{equation}
where
 \begin{equation}
\bs_3 = -\br_2 ~,\quad \bs_1 = \br_1 - \br_2~,\quad
v_1  =  {m_1 m_2 \over m_1 + m_2}~,\quad
\mu_3 = {m_3 m_2 \over m_3 + m_2} = \mu_2~.
 \end{equation}
 Note that $\mu_3=\mu_2$ on account of the symmetry of the reduced
masses of particles 2 and 3 under the operation of interchange of
these particles.
 
 Now we can write this form of $\hint$ as an operator, $\hat{A}$,
in terms of $\br_1$ and $\br_2$ i.e., with origin at particle 3,
 \begin{equation}
\hat{A} = -{1 \over 2v_1} \nabla^2_{\br_1} -
{1\over 2\mu_2} \nabla^2_{\br_2} - {1\over m_2} \nabla_{\br_1}
\cdot \nabla_{\br_2} - {1\over r_1} - {1\over r_2}
+ {1\over r_{12}} \;.
 \end{equation}
 Clearly, $\hat{A}$ has no physical significance.
 However, Armour was able to make use of it in a novel way to
take into account the mass-polarisation term within the framework
of the infinite proton mass approximation.
 
 Suppose $\hint$ does have a bound state, $\Phi(\br_1,\br_2)$, i.e.,
 \begin{equation}
\hint \Phi(\br_1,\br_2) = E \Phi(\br_1,\br_2) \;,
 \end{equation}
where
 \begin{equation}
E < -\half\mu_2 = E_{\mathrm{thr}} \;.
 \end{equation}
 We shall assume that $E$ is the lowest eigenvalue of $\hint$
and $\Phi$ is normalised so that
 \begin{equation}
\langle \Phi | \Phi \rangle = 1 \;.
 \end{equation}
 Because $\hat{A}$ corresponds to $\hint$ when the origin is taken
to be at particle 2 rather than particle 3, it follows that
$\hat{A}$ must have the same eigenvalue spectrum as $\hint$.
 In particular, $\hat{A}$ will also have lowest eigenvalue $E$.
 It follows that
 \begin{equation}
\left\langle \Phi \left| \hat{A}-\hint \right| \Phi \right\rangle \ge 
0 \;.
\label{a-hint}
 \end{equation}
 
 It can be shown to follow from Eq.~(\ref{a-hint}) by
straightforward manipulation (Armour 1982) that
 \begin{equation}
\left\langle \Phi \left| \hpint \right| \Phi \right\rangle \le E
< -\half\mu_2 \;,
\label{php}
 \end{equation}
where
 \begin{equation}
\hpint = -{1 \over 2m_1} \nabla^2_{\br_1} 
- {1 \over 2\mu_2} \nabla^2_{\br_2} + {Q \over r_1}
- {1\over r_2} - {Q \over r_{12}}~,
 \end{equation}
and
 \begin{equation}
Q = {m_3 + m_2 \over m_3 - m_2} \;.
 \end{equation}
 The crucial point is that $\hint$ represents the internal motion,
in the infinite proton mass approximation, of a system made up
of a ``positron'' of the usual mass but with charge $Q$ and an
electron of mass $\mu_2$ and the usual charge.

 Now $-\half\mu_2$ is also the continuum threshold for $\hpint$
and $\langle \Phi |\hpint| \Phi \rangle$ is an upper bound to
the lowest eigenvalue of $\hpint$.
 Thus it follows from the initial assumption and Eq.~(\ref{php})
that a necessary condition for $\hint$ to have a bound state is
that $\hpint$ has a bound state.
 Thus if it can be shown that $\hpint$ has no bound state, it
follows that $\hint$ has no bound state.

 As $\hpint$ does not involve the mass-polarisation term, the method
described earlier can be applied to it.
 In the case of $(\e^+,\H)$,
\begin{equation}
Q = 1.0011 \;\; {\mathrm{and}} \;\; \mu_2 = 0.9995\,m_{\rme} \;.
\end{equation}
Thus it is not surprising that Armour \cite{Arm82} was able to show that
no bound state of $(\e^+,\H)$ exists, even if the finite mass of the
proton is taken into account.

 $(\mu^+,\e^-,\e^+)$ is a very similar system to $(\p,\e^-,\e^+)$.
 However, the mass of the muon is $206.8\,m_{\rm e}$, which is 
approximately a tenth the mass of the proton.  
 Thus the effect of the mass-polarisation term will be greater for 
this system.  Armour \cite{Arm83} applied his method to this system for which
 \begin{equation}
Q=1.01 \;\; {\mathrm{and}} \;\; \mu_2 = 0.995\,m_{\rme} \;.
 \end{equation}
 He was able to show that no bound state of this system exists.

 He also applied it to $(\p,\mu^-,\e^+)$, which differs from $(\p,\e^-,\e^+)$
in that the electron is replaced by the $\mu^-$, which is much 
more massive.
 The $(\p,\mu^-)$ atom is much more compact than the H atom and has 
a very much smaller dipole polarisability. The reason for this is that the polarisability  this scales as 
$m_{\mathrm{r}}^{-3}$, where $m_{\mathrm{r}}$ is the reduced 
mass of the atom.  
 See, for example, Bransden and Joachain \cite{Bra83}.  
 As pointed out earlier, this polarisability determines the 
magnitude of the attractive potential between the positron and
the atom containing the other two particles in the asymptotic 
region.  
 Thus in this region, this potential is very much smaller for 
$(\p,\mu^-,\e^+)$ than for $(\p,\e^-,\e^+)$.

For $(\p,\mu^-,e^+)$,
 \begin{equation}
Q=1.254 \;\; {\mathrm{and}} \;\; \mu_2= 186\,m_{\rme} \;.
 \end{equation}
 Armour \cite{Arm83} was also able to show that no bound state of this 
system exists.
\subsection{$(\ap,\p,\e^-)$}\label{3u1:sub:ppbe}
$(\ap,\p,\e^-)$ is a system containing an antiproton, a 
proton and an electron.
 As we are only considering Coulombic forces, as pointed out 
in Sec.~\ref{3u1:sub:not}, the Hamiltonian for the systems we are considering 
is invariant under the operation of charge conjugation, i.e., 
change of sign of the charge of all particles in the system, 
this system is equivalent to $(\p,\ap,\e^+)$.

 In the absence of the electron, the proton and the antiproton
can form a strongly bound and very compact protonium
atom with ground-state energy $\mathrm{-\half\mu_{p\bar{p}}}$, 
where $\mathrm{\mu_{p\bar{p}}}$ is the reduced mass of the protonium 
atom.  
 The expectation value of the internuclear separation in this 
state is
 \begin{equation}
\mathrm{{3a_0\over 2\mu_{p\bar{p}}}= 1.6 \times 10^{-3} a_0} \;,
 \end{equation}
where~a$_0$ is the Bohr radius.

 The ground-state energy of protonium is the energy of the 
lowest continuum threshold of $(\ap,\p,\e^-)$.  
 It can be seen that in this state the expectation value of 
the dipole moment resulting from the equal and opposite unit 
charges is very small indeed.  
 Furthermore, the probability that the dipole moment has a value 
greater than 0.01 a.u. is less than 10$^{-6}$.  
 A consequence of this is that the dipole polarisability 
of the $(\p,\ap)$ atom, that determines the magnitude of 
the attractive potential between the electron and ground-state 
protonium asymptotically, is very small indeed, much smaller, even, 
than in the case of  $(\p,\mu^-)$.

It is clear, therefore, that
 no bound state of  ($\ap,\p,\e^-$) exists.  
 As we shall see in Sec.~\ref{3u2:sub:trian}, the position of 
$(\ap,\p,\e^-)$
on the triangular plot of the stability properties of three-body 
systems is so far into the unstable region as to be almost at 
one of the vertices associated with this region.  
 This is to be expected in view of the very small dipole moment 
and polarisability of the ($\p,\ap$) atom in its 
ground state.

 It is of interest to consider what would happen if the nuclei 
were fixed.
 Such a system has been extensively studied, in particular by Fermi and Teller 
\cite{Fer47}, Wightman \cite{Wig50}, Wallis \etal~\cite{Wal60}, Mittleman and 
Myerscough 
\cite{Mit66}, Turner and Fox \cite{Tur66}, Levy-Leblond \cite{Lev67}, 
Byers Brown and 
Roberts \cite{Bye67}, Coulson and Walmsley \cite{Cou67}, Crawford
\cite{Cra67}, and Turner \cite{Tur77}.
 
 It is well known that no bound state of this system exists
if the dipole moment of the system is less than 0.639 a.u.
 As the nuclei have unit charge, this corresponds to an 
internuclear distance, $R$, of $0.639\,\mathrm{a}_0$.

 The Schr\"odinger equation for this system is separable in
prolate spheroidal coordinates~\cite{Eis48}. It has been shown \cite{Cra67}
 that if $R > 0.639\,\mathrm{a}_0$, the system 
has an infinity of bound states.
 As is to be expected, the infinity is countable as it can be set in a one to one 
relationship with the natural numbers.  
In these states the electron or positron has zero angular 
momentum about the internuclear axis and no nodes in the part of
the wave function that is dependent on the hyperbolic coordinate.  
 As $R$ is increased, further infinities of bound states appear 
corresponding to higher values of one or both of these quantum 
numbers, the first appearing when $R > 3.792\,\mathrm{a}_0$.
\subsection{Muonic molecular ions}\label{3u1:sub:muonic}
 Muonic molecular ions are systems such as ($\p,\mu^-,\p$), 
($\p,\mu^-,\d$), ($\d,\mu^-,\d$), etc.  
 These systems are analogous to molecular ions such as H$^+_2$, 
but they contain a negatively charged muon rather than an electron.  
 There has been considerable interest in the properties of
these systems as ($\d,\mu^-,\d$) and, in particular, 
($\d,\mu^-,\Pt$), play a crucial role in the process known 
as muon catalysed fusion. 
 See, for example, Bhatia and Drachman \cite{Bha89}, Ponomarev \cite{Pon90} 
and Froelich \cite{Fro92}.

 In a muonic molecular ion, ($a,\mu^-,b$), where $a$ and $b$ are protons,
deuterons or tritons, the dimensions of the molecular ion scale as
$m_{\rme}/m_\mu$ relative to ($a,\rme^-,b$).  
 Thus the nuclei $a$ and $b$ are brought very much closer 
together than in ($a,\rme^-,b$).  
 Provided that ($a,\mu^-,b$) is in a state of zero angular momentum 
and $a$ and $b$ are not both protons, fusion occurs at a rate 
between 10$^5$ and $10^{12}\;\mathrm{s}^{-1}$, depending on the nuclei 
involved.  
 The proton-proton fusion process is very much slower as it 
involves the weak interaction.

 A muon has mean lifetime of $2.2 \times 10^{-6}$ s before 
undergoing beta-decay.
 See, for example, Semat and Albright \cite{Sem73}.  
 Clearly, if a muon is to catalyse a large number of fusions, 
it is important that the whole process takes place in a much 
shorter time than the lifetime of the muon.  
 There seemed little prospect of this until extensive theoretical 
work by Vesman \cite{Ves67} and Gershtein and Ponomarev \cite{Ger77} showed 
that  $(\d,\mu^-,d)$ and $(\d,\mu^-,\t)$ can be formed by a resonant process.

 This resonant process can take place because of a remarkable 
coincidence.
 Both $(\d,\mu^-,\d)$ and $(\d,\mu^-,\t)$ have weakly bound $(J,v)=(1,1)$ 
excited states, i.e. states with angular momentum $J=1$ and in 
the first excited vibrational state, $v=1$.  
 In the case of $(\d,\mu^-,\d)$, this state has binding energy 1.95~eV, 
whereas in the case of $(\d,\mu^-,\t)$ it is 0.66~eV.
 These binding energies are sufficiently small for energy 
conservation requirements to be satisfied by a process in which 
a low-energy $(\d,\mu^-)$ or $(\t,\mu^-)$ atom, in its ground-state, 
collides with one of the nuclei of a $(\D A)$ molecule $(A=\D,\,H$ or T), 
in its electronic and vibrational ground state, and attaches itself 
to the deuteron to form a muonic molecular complex, 
$[(\d,\mu^-,\d),\mathrm{a},\rme^-,\rme^-]$ or $[(\d,\mu^-,t),\mathrm{a},\rme^-,\rme^-]$, 
with the $(\d,\mu^-,d)$ or $(\d,\mu^-,\t)$ in the (1,1) state.

 The complex containing $(\d,\mu^-,\t)$ rapidly undergoes Auger decay to
a state of $(\d,\mu^-,\t)$ which has zero angular momentum.  
 Fusion then occurs at a rate of 10$^{11}$ -- 10$^{12}$ s$^{-1}$.  
 This is the fastest fusion rate of any muonic molecular ion.  
 So far, fusion rates of about 150 fusions per muon have been 
observed in this process.
 The breakeven value for energy production is about 480 fusions 
per muon.

 The key role played by $(\d,\mu^-,\t)$ in this process has focussed 
attention on the properties of muonic molecular ions.  
 It is known that ions that contain a proton have two bound states, 
whereas $(\d,\mu^-,\d)$ and $(\d,\mu^-,\t)$ have five bound states and 
$(\t,\mu^-,\t)$ has six bound states.  
 For details, see Ponomarev \cite{Pon90}.

 As  the mass of the muon is $m_\mu= 206.8\,m_{\rme}$, the 
Born--Oppenheimer approximation does not give accurate results 
for these systems.  
 Accurate results can be obtained starting from this approximation, 
using the adiabatic representation method \cite{Vin82,Pon90}.
 However, even more accurate results have been obtained using the
Rayleigh--Ritz variational method, which was first applied to 
muonic molecular ions by Ko{\l}os \etal~\cite{Kol60}.  
 This method has been extensively applied to the calculation of 
the energies of the bound states of ($\mathrm{d},\mu^-,\mathrm{t}$) by, for example, 
Bhatia and Drachman \cite{Bha84}, Korobov \etal~\cite{Kor87}, 
Frolov \cite{Fro87},
Hara \etal~\cite{Har87}, Hu \cite{Hu87}, Petelenz and Smith \cite{Pet87a}, 
Szalewicz \etal~\cite{Sza87}, Kamimura \cite{Kam88}, Alexander and Monkhorst 
\cite{Ale88} and Zhen \cite{Zhe90}.
 For further details, see Froelich \cite{Fro92}.

\markboth{\sl Stability of few-charge systems}{\sl A general approach to 3-body systems}
\clearpage%
\section{A general  approach to three unit-charge systems}
\label{se:3u2}
In this section, we study the stability of three-unit charge systems 
as a function of the masses of the constituent particles. This problem has
been addressed by many authors, in textbooks
\cite[p.~286]{Thi79}, in review papers \cite{Arm93}, or in articles, see for
instance
\cite{Pos85,Fro92a,Fro92b,Mar92,Din94,Reb95,Kai00,Reb02,Reb02a}, and references therein.

In the literature, results have been obtained on how the binding energy evolves one or two masses are varied, one of the precursors being \cite{Ore48}. There are also attempts to parametrise the energy as a function of the constituent masses $m_i$ and use this parametrisation for a guess at the stability border, see for instance \cite{Pop87,Bha87}. Some papers contain thorough numerical investigations, where one or two mass ratios are varied. It is noted that this is a slightly different art to estimate accurately the energy of a well-bound system  and to determine at which value of a mass ratio binding disappears. For instance, in Ref.~\cite{Pos85}, the stability is studied  as a function of the mass ratios $m_1/m_2$ and $m_3/m_1$ and an interesting comparison is made with existing results, but the matching between ``molecular'' states similar to $\H_2^+$ and atomic states similar to $\H^-$ gives an unphysical spike in the drawing of the stability border. In Ref.~\cite{Din94}, an astute changes of variables makes it possible to use harmonic-oscillator type of wave functions, but while the binding energy of symmetric states with $m_2=m_3$ are accurately computed, the stability domain of states with $m_2\not = m_3$ extends too far, with, e.g., $(M^+,M^-,m^\pm)$ leaving stability for $M/m>2.45$, while Mitroy \cite{Mit00} found it unstable. Clearly, more cross-checks of the published results are needed.
\subsection{Triangular plot}
\label{3u2:sub:trian}
Consideration  of the results already described shows that stability is a property 
of systems with either $m_{1}$ and $m_{2}$ both large or close to each 
other.  In other words, stability requires $\vert 
m_{2}^{-1}-m_{3}^{-1}\vert$ to be small compared to $m_{1}^{-1}$.  
This suggests that it would be advantageous to consider stability as a function of 
 the inverse masses $m_{i}^{-1}$ instead of the 
masses $m_{i}$ themselves. This is confirmed by observing that the inverse masses
 enter the Hamiltonian linearly 
and consequently the binding energies have simpler monotonic and 
convexity properties in terms of these variables.

One can combine inverse masses and scaling and represent the 
stability domain with the normalised coordinates
\begin{equation}\label{3u2:eq:alpha}
\alpha_{i}={ m_{i}^{-1} \over 
m_{1}^{-1}+m_{2}^{-1}+ m_{3}^{-1} }~,\qquad
\alpha_{1}+\alpha_{2}+\alpha_{3}=1~.
\end{equation}
\begin{figure}[H]
\centering{
 \includegraphics[width=.4\textwidth]{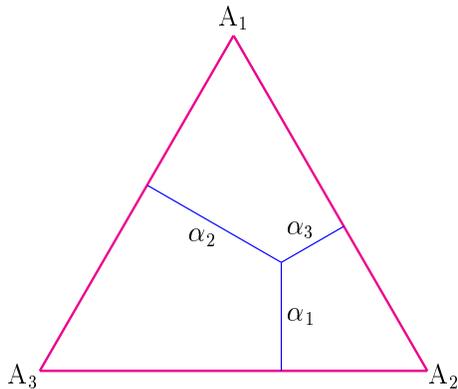}  }
\caption{\label{3u2:fig:trian1} 
The domain of possible inverse masses $\alpha_i$, such that 
$\sum\alpha_i=1$, is an equilateral triangle.}
\end{figure} 
 As seen in Fig.~\ref{3u2:fig:trian1}, each system can be represented as a point 
 inside an equilateral triangle ${\rm A}_{1} {\rm A}_{2} {\rm A}_{3}$, 
 the inverse mass $\alpha_{i}$ being the distance to the side opposite 
 to ${\rm A}_{i}$.  This is equivalent to barycentric coordinates.
 
\begin{figure}[H]
\centering{\includegraphics[width=.4\textwidth]{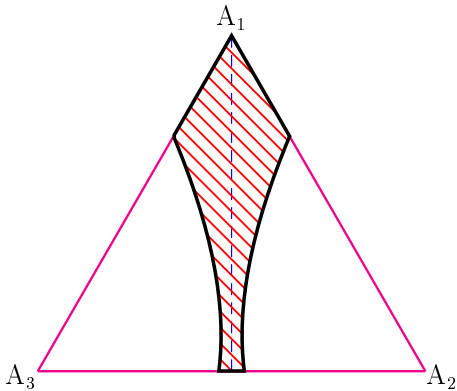}  }
\caption{\label{3u2:fig:shape} 
Schematic shape of the stability domain in the triangle of normalised 
inverse masses for three unit-charges $(+1,-1,-1)$.}
\end{figure}

 In this representation, the shape of the stability domain  is shown 
 in Fig.~\ref{3u2:fig:shape}.
 \subsection{General properties of the stability domain}
 \label{3u2:sub:gen-prop}
The patterns of the stability and instability regions shown in 
Fig.~\ref{3u2:fig:shape} result from three main properties, besides the 
obvious left--right symmetry corresponding  to 
$\alpha_{2}\leftrightarrow\alpha_{3}$ exchange. They are:
\decale{1.}  {\sl The domain of stability includes the symmetry axis 
where $\alpha_{2}=\alpha_{3}$.}

To our knowledge this result was first pointed out by Hill 
\cite{Hil77}, who used the variational wave-function
\begin{equation}\label{3u2:eq:HillWF}
\Psi=\exp(-a r_{12}-b r_{13})+(a\leftrightarrow b)~, 
\end{equation}
already used by Hylleraas \cite{Hyl47} for demonstrating the stability of $\H^-$ 
and $\Ps^-$.  If  the Rayleigh--Ritz principle is combined with the virial 
theorem (see Sec.~\ref{SO:sub:Virial}), then 
only the ratio $b/a$ has to be adjusted. The applicability of the virial theorem
to variational solutions was noticed by 
Hylleraas \cite{Hyl29} and  Fock \cite{Foc30}.
\decale{2.} {\sl The instability domain including ${\rm A}_{3}$ is star-shaped 
with respect to ${\rm A}_{3}$.}

\begin{figure}[H]
\centering{\includegraphics[width=.4\textwidth]{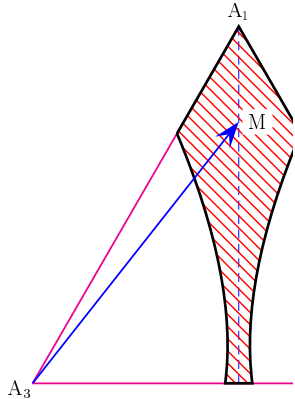}   }
\caption{\label{3u2:fig:star} 
Star-shape behaviour: the line A$_{3}$M shown  crosses the 
stability border only once. This is because the binding energy with 
respect to the threshold increases monotonically on progression from the border
   to M. }
\end{figure}

The same holds of course for ${\rm A}_{2}$. We have seen that ${\rm 
A}_{3}$ does not correspond to a stable configuration, for a 
point-like protonium atom does not bind an electron. Imagine a 
straight line  from ${\rm A}_{3}$ toward the inner part of the 
triangle, as pictured in Fig.~\ref{3u2:fig:star}. Moving along this line means keeping the mass ratio $\alpha_{1}/\alpha_{2}$ constant while $\alpha_{3}$ decreases. A 
suitable rescaling provides us with a system of inverse masses
\begin{equation}\label{3u2:eq:resc_alpha}
{\alpha_{1}\over \alpha_{1}+\alpha_{2}}~,
{\alpha_{2}\over \alpha_{1}+\alpha_{2}}~,
{\alpha_{3}\over 1-\alpha_{3}}~. 
\end{equation}
In this rescaled system, the masses $m_{1}$ and $m_{2}$ are 
both constant, and thus the threshold energy $E^{(2)}$ is fixed, while 
$m_{3}$ increases, thus strengthening the binding of any 3-body bound 
state.  This means that once  the stability domain is entered, the 
system becomes more and more strongly bound, as long as the $(1,\,2)$ atom is 
the lowest threshold.  
\decale{3.} {\sl Each instability domain is convex.}

Imagine two points $M'=\{ \alpha'_{i} \}$ and $M''=\{ \alpha''_{i} \}$ both 
corresponding to instability with respect to the same lowest 
threshold, say the $(1,\,2)$  atom. The same rescaling as above 
\begin{equation}\label{3u2:eq:resc_alpha_2}
 M=\{ \alpha_{i}\}\to 
 \tilde{M}=\{\beta_{i}\}, \quad \beta_{i}={\alpha_{i}\over \alpha_{1}+\alpha_{2}}~,
\end{equation}
transforms the half triangle where $\sum\alpha_{i}=1$ and 
$\alpha_{2}<\alpha_{3}$ into a triangle in which $\beta_{1}+\beta_{2}=1$ 
is fixed. If one moves on the straight line segment between $\tilde{M}'$ 
and $\tilde{M}''$, the images of $M'$ and $M''$, say
\begin{equation}\label{3u2:eq:sec}
 \tilde{M}=x \tilde{M}' + (1-x) \tilde{M}''~,
\end{equation}
one deals with an Hamiltonian
\begin{equation}\label{3u2:eq:H(x)}
H(x)=H(0)+x\left[(\beta'1-\beta''_{1})\vec{p}_{1}^{2}+
(\beta'_{2}-\beta''_{2})\vec{p}_{2}^{2}\right],
\end{equation}
where $x$ enters linearly.  Thus its ground state $E(x)$ is a concave 
function of $x$ (see Appendix).  If $E(0)\ge E^{(2)}$ and $E(1)\ge 
E^{(2)}$, then  $E(x)\ge E^{(2)}$ if $0\le x\le 1$.  This means this instability 
domain is convex in the $\{ \beta_{i} \}$ plane.  Now the convex 
projection $\tilde{M}\to M$ which moves the system back to the 
$\alpha_{i}$ plane transforms any convex domain into another convex 
domain.

\begin{figure}[H]
\centering{\includegraphics[width=.4\textwidth]{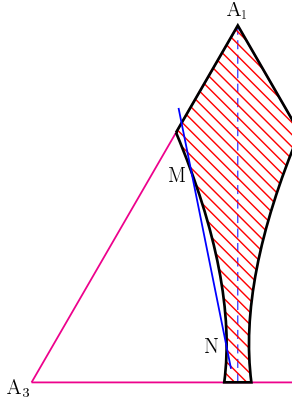}}
\caption{\label{3u2:fig:conv} 
Convexity: if the points M and N both lie on the same stability border, any 
intermediate point on the straight segment MN belongs to an 
instability domain.}
\end{figure}

The convexity property is illustrated  in Fig.~\ref{3u2:fig:conv}.
\subsection{Shape and width of the stability band}
\label{3u2:sub:shape}
\begin{figure}[H]
\centering{ \includegraphics[width=.4\textwidth]{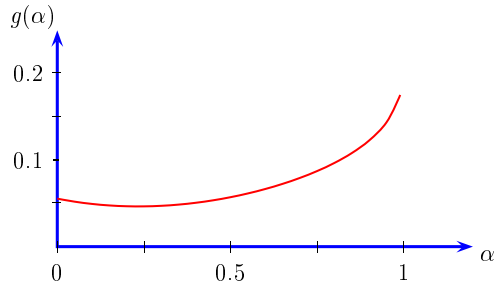}}
\caption{The function $g(\alpha)$ shows the relative excess of binding, 
relative to the threshold, for $(M^+,m^-,m-)$, as a function of 
$\alpha=M^{-1}/(M^{-1}+2m^{-1})$.}
\label{3u2:fig:galpha}
\end{figure}

H$_2^+$ is markedly more stable than H$^-$. 
A measure of this is the dimensionless ratio
\begin{equation}
\label{3u2:eq:def-g}
g={E_0-E_\mathrm{th}\over E_\mathrm{th}}~,
\end{equation}
that compares the three-body energy $E_0$ to that of the lowest threshold $E_{\mathrm{th}}$.
Values of $g$ are available in the literature (see the many papers cited in
the previous and the present sections). 
They are set out  in Table~\ref{3u2:tab:g}, for symmetric configurations.
\begin{table}[H]
  \centering
  \caption{ Relative excess energy for some symmetric configurations $(M^\pm,m^\mp,m^\mp)$}
\label{3u2:tab:g}
\begin{tabular}{ccc}
\hline\hline
State & $M/m$ & $g$ \\
\hline
 $\H_2^+$ & 1836.15 &  0.19495 \\
$(\mu,\d,\d)$ & 17.666 & 0.122\\
$(\mu,\p,\p)$ &8.8802  & 0.100  \\
Ps$^-$ & 1 & 0.047982 \\
$(\mu,\e,\e)$ & 0.483793 & 0.05519 \\
 $\H^-$ & 0.0005446 & 0.0553  \\
 ${}_\infty\H^-$ & 0 & 0.0555  \\
\hline\hline
\end{tabular}
\end{table}  
The corresponding curve is plotted in Fig.~\ref{3u2:fig:galpha}.
The numerical calculations used to calculate $g$ should be very accurate as they have been
 checked by several authors. 
 It is interesting that $g(\alpha)$ is not minimal for $\alpha=0$, 
 which would correspond to ${}_\infty\mathrm{H}^-$. 
 The minimum can be estimated to be located at about $\alpha=0.23$, 
 intermediate between H$^-$ ($\alpha\simeq0$) and Ps$^-$ ($\alpha=1/3$). 
 This explains why the stability band is narrower in this region.

\begin{figure}[H]
\centerline{\includegraphics[width=.4\textwidth]{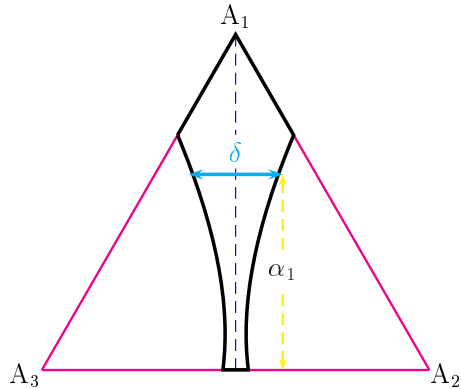}}
\caption{Definition of the width of the stability band.}
\label{3u2:fig:delta}
\end{figure}

The link between the relative excess of binding, $g$, and the width of the stability 
band, $\delta$, can be made more precise. The decomposition described in 
Appendix, (Sec.~\ref{SO:sub:sym-break}) reads
\begin{equation}
\label{3u2:eq:width1}
H(\alpha_1,\alpha_2,\alpha_3)=H(\alpha_1,\alpha_{23},\alpha_{23})+
\lambda (\vec{p}_2^2-\vec{p}_3^2)~,
\end{equation}
with $\alpha_{23}=(\alpha_2+\alpha_3)/2=(1-\alpha_1)/2$ and $\lambda=(\alpha_2-\alpha_3)/4$.
As a function of $\lambda$, for given $\alpha_1$ (and hence $\alpha_{23}$), the ground-state energy $E$ is concave, and maximum at $\lambda=0$. It thus fulfils
\begin{equation}
\label{3u2:eq:width2}
E(\alpha_1,\alpha_2,\alpha_3)\le E(\alpha_1,\alpha_{23},\alpha_{23})=[1+g(\alpha_1)]\,
E_\mathrm{th}(\alpha_1,\alpha_{23},\alpha_{23})~.
\end{equation}
But the threshold energies are  exactly known. Thus the above inequality reads
\begin{equation}
\label{3u2:eq:width3}
E(\alpha_1,\alpha_2,\alpha_3)\le E_\mathrm{th}(\alpha_1,\alpha_2,\alpha_3)
[1+g(\alpha_1)]{1+\alpha_1-(\alpha_3-\alpha_2)
\over 1+\alpha_1}~.
\end{equation}
Hence stability is guaranteed if the last factor is less than 1, i.e.,
\begin{equation}\label{3u2:eq:width4}
\delta={2\over\sqrt3}\,(\alpha_3-\alpha_2)\le {2\over \sqrt3} 
{g(\alpha_1)\over 1+g(\alpha_1)} (1+\alpha_1)~.
\end{equation}
Here, $\delta$ is the width of the band at height $\alpha_1$, 
as illustrated in Fig.~\ref{3u2:fig:delta}.
\subsection{Applications of the convexity properties}
\label{3u2:sub:appli}
To illustrate the potential of the stability plot, let us take the example 
of the $(1^-,m^+,1^+)$ configurations. For $m=1$, we have the positronium 
ion (or its conjugate), which is known to be stable. For $m$ large, 
we have $(\e^+,\p,\e^-)$ which is unstable, as reviewed in Sec.~\ref{se:3u1}.  
It was stated there that the best  rigorous limit for instability for this
system is $m\le1.51$.

\begin{figure}[H]
\centering{\includegraphics[width=.4\textwidth]{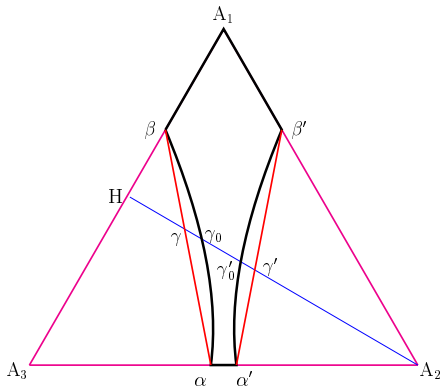}}
\caption{\label{3u2:fig:appli} 
A safe limit on $\alpha$ and $\beta$ ensure that $\gamma$, 
at the intersection of $\alpha\beta$ and A$_2$H ($m_1=m_3$), and corresponding to a configuration $(1^+,m^+,1^-)$, belongs to an instability domain.}
\end{figure}

Consider now an indirect approach. ${}_\infty\mathrm{H}^-$ is weakly bound. 
If one changes the mass of one of the electrons by about 10\%,  stability is
lost. 
This corresponds to the point $\alpha$ in Fig.~\ref{3u2:fig:appli}. 
A very conservative but rigorous estimate by Glaser \etal\ \cite{Gla80} is that
the system
 is unstable for $m_1=\infty$, $m_3=1$ and $m_2\ge1.57$.

Also, the system $(1^+,\infty^-,M^-)$ is stable for $M\to\infty$ 
(conjugate of ${}_\infty\H_2^+)$, and unstable for $M\to 1$ 
(this is again $(\e^+,\p_\infty,\e^-)$). The point $\beta$ in Fig.~\ref{3u2:fig:appli} 
corresponds to the critical mass $M$ at which instability occurs. It has been studied by 
several authors. See Sec.~\ref{se:3u1}. For $M=1.51$ or larger, the system is unstable. 
This value yields a conservative estimate for $\beta$ in Fig.~\ref{3u2:fig:appli}. 
Thus $\gamma$ at the intersection of $\alpha\beta$ with the A$_2$H axis ($m_1=m_3$) 
is in a region of instability. It follows from this that $(\e^-,m^+,\e^+)$ is
unstable
 at least for $m\ge 4.6\,m_\e$ \footnote{This corrects a misprint in
Ref.~\protect\cite{Mar92}}.

This is much better than the results obtained by direct study of  $(\e^-,m^+,\e^+)$.

A similar reasoning shows that $\gamma'$ in Fig.~\ref{3u2:fig:appli}
belongs to the other instability region. Hence  $(\e^-,m^+,\e^+)$ is unstable for at least $m\le 0.44\,m_\e$. 

These values compare well with the numerical study by Mitroy \cite{Mit00}, who
estimated that $(\e^-,m^+,\e^+)$ is stable in the region
\begin{equation}\label{3u2:eq:mitroy}
0.69778\lesssim m/m_\e\lesssim 1.6343~.
\end{equation}
If one takes for $\alpha$ and $\beta$ values corresponding to $m=1.1\,m_\e$ and $M=2.20\,m_\e$, then one gets for $\gamma$ and $\gamma'$ estimates corresponding to 
\begin{equation}\label{3u2:eq:gamma}
m/m_\e=0.64\ \ (\gamma')~,    \quad 2.02\ \ (\gamma)~.
\end{equation}
The curvature of the stability border can be determined by comparing the
values in (\ref{3u2:eq:gamma}) with those in (\ref{3u2:eq:mitroy}).
\subsection{Stability domain for excited states}
One needs to distinguish between two types of excited states.
\begin{enumerate}\itemsep-2pt
\item
States with the same quantum numbers as the ground state
\item
First state with a (conserved) quantum number different from the value in the ground state
\end{enumerate}
As explained in Sec.~\ref{se:crit},  the relevant threshold should be
identified. It might consist of an excited atom for states of the second category, 
if radiative processes are neglected.

For this category of states, the variational principle (\ref{SO:eq:var1}) is
almost unchanged.
We need only note  that the trial wave function used must belong 
to the subspace of the Hilbert space of the system with the relevant quantum
number, negative parity, for instance. Most general results on the stability domain are 
immediately applicable for the ground state with a specific quantum number, 
as, e.g., total angular momentum $\ell=2$, or negative parity. 
However, there is a crucial exception: stability is no longer 
guaranteed  along the symmetry axis.

\begin{figure}[H]
\centering{\includegraphics[width=.4\textwidth]{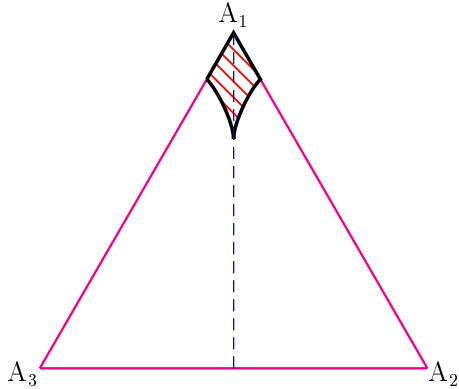}  }
\caption{\label{3u2:fig:exc}Schematic shape of the stability domain 
for the $2^+$ state of $(m_1^+,m_2^+,m_3^+)$ in the triangle of normalised inverse masses.}
\end{figure}

This corresponds to well-known properties: there are 
many excited states for H$_2^+$ without analogue in the case of  Ps$^-$ or H$^-$. 
The stability domain is then a small island near the upper vertex A$_1$, 
as schematically pictured in Fig.~\ref{3u2:fig:exc}. A similar shape will be 
encountered in the next section, when we consider the ground state with a reduced
 central charge $q_1$.

For excited states with the same quantum number 
as the ground state (generalising the radial excitations of 2-body systems), 
the variational principle no longer holds  in its simplest form. 
See Appendix \ref{se:SO}. Hence,  the convex behaviour of each instability 
region does not necessarily apply in the case of excited states.
However, the star-shape property remains, as it is a consequence of the fact that
$\vec{p}_2^2$ and  $\vec{p}_3^2 \ge 0$.

%
\markboth{\sc Stability of few-charge systems} {\sl Three-body systems with arbitrary charges}
\clearpage\section{Three-arbitrary-charge systems}
\label{se:3a}
Once the domain of masses or inverse masses 
insuring stability of three unit charges $(+1,-1,-1)$ has been established, 
it is interesting to go on and let the charges vary.  This is to some extent an academic exercise, 
to better analyse the binding mechanisms. 
However, non-integer effective charges are at work in media, 
due to screening or anti-screening effects.  Changing the charges may also be a very useful test of the accuracy of different calculational  methods and can help to understand the correlation effects.

The stability of three-particle systems with arbitrary masses and charges was the subject of Refs.~\cite{Reb90,Mar95,Kri00a}. We shall first review studies carried out on specific configurations.
\subsection{Specific configurations}\label{3a:sub:spe}
Detailed studies are available in the literature on how $\H^-$ and $\H_2$ survive if the charge $Z$ of the proton is increased. They will be summarise before other mass and charge configuration are considered.
\subsubsection{$\H^-$-like states}
In the limit where $m_p$ is infinite,  the Hamiltonian can be rescaled as
\begin{equation}\label{3a:eq:largeZ}
H={\vec{p}_1^2\over 2}-{1\over r_1}+{\vec{p}_2^2\over 2}-{1\over r_2}+{1\over Z r_{12}}\;,
\end{equation}
and the last term  treated as a perturbation around the unperturbed $(1s)^2$ ground state. The series of perturbation converges towards the bound-state energy and wave function provided $1/Z\lesssim 1.09766$, according to the detailed studies by Baker et al.~\cite{Bak90}, and Ivanov \cite{Iva95}. This means that $(\infty^Z,m^-,m^-)$ is bound for $Z\gtrsim 0.911$.
\subsubsection{$\H_2^+$-like states}
Another well document system is $(M^Z,M^Z,m^-)$ with a charge $Z$ assigned to the nuclei. In the limit $M/m\to\infty$, binding remains up to $Z\simeq 1.23667$. See, for instance, \cite{Hog92,Reb95} and references therein.

Note that for larger $Z$, till $Z\simeq 1.439$, the Born--Oppenheimer potential between the two protons has a local minimum,but the values of this minimum lies above the threshold energy. Thus the ion has classical stability. In the following, we shall concentrate on genuinely stable states.
\subsection{General properties}\label{3a:sub:gen}
Let $\{m_1,q_1\}$, $\{m_2,-q_2\}$, and $\{m_3,-q_3\}$ be the masses and charges, 
with, say, all $q_i>0$. By scaling, the properties of the $(1,2,3)$ system depend 
on two mass ratios and two charge ratios. One can for instance 
choose $q_1=1$ and let $q_2$ and $q_3$ vary.

For  a given set of charges $\{q_i\}$, one can search the stability and 
instability domains inside a triangle of
normalised inverse masses $\alpha_i$, as defined in Eq.~(\ref{3u2:eq:alpha}). 
The same general properties hold, namely
\begin{itemize}\itemsep -3pt
\item every instability domain is convex
\item every instability domain is star-shaped with respect the 
vertices A$_2$ and A$_3$, as defined in
Fig.~\ref{3u2:fig:trian1}.
\end{itemize}
Likely to be lost are the properties {\sl i)} that the frontier is symmetric
with respect to the vertical axis (if $m_2\neq m_3$), and {\sl ii)} that there is 
always stability along this axis.

\begin{figure}[H]
\centering{
\includegraphics[width=.45\textwidth]{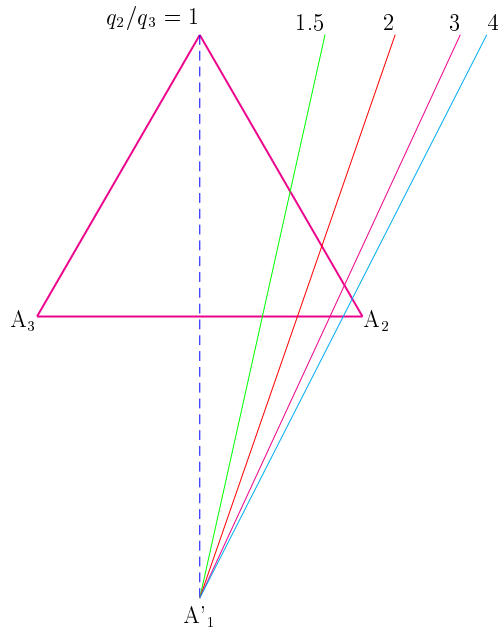}}
\caption{\label{3a:fig:sepa1}The threshold separation (T) is the line where 
the (1,2) and (1,3) atoms have the same energy. 
It is drawn here for $q_{2}/q_{3}=1$ (symmetry axis), 1.5, 2, 3 and 4. }\end{figure}

\begin{figure}[H]
\centering{
\includegraphics[width=.4\textwidth]{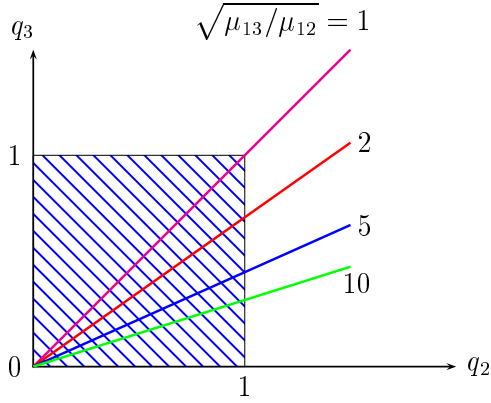}  }
\caption{\label{3a:fig:sepa2}The threshold separation (T) is the line where the (1,2) and (1,3) atoms have same energy. It is drawn here for a ratio of reduced masses $\mu_{13}/\mu_{12}=1$ (symmetry axis), 2, 5, and 10. }\end{figure}

 In fact, if $m_2\neq m_3$, the vertical axis plays no particular role, and one should instead concentrate on the line 
\begin{equation}
\label{3a:eq:sepa1}
\mathrm{(T)}:\qquad (q_1 q_2)^2(1-\alpha_2)=(q_1 q_3)(1-\alpha_3)~.
\end{equation}
This is a straight line issued from $\mathrm{A}'_1$, the symmetric of A$_1$ with respect to 
$\mathrm{A}_2\mathrm{A}_3$, corresponding to $-\alpha_1=\alpha_2=\alpha_3=1$.
Some examples are drawn in Fig.~\ref{3a:fig:sepa1}.
In the $\{q_2,\,q_3\}$ plane, for $q_1=1$ and some given masses, resulting into reduced masses $\mu_{12}$ and $\mu_{13}$, (T) is a line issued form the origin.
Examples are drawn in Fig.~\ref{3a:fig:sepa2}.

There is always a concentration of stability along this threshold separation (T).
This can understood as follows: near (T), one can seek  a wave function of the form
\begin{equation}
\label{3a:eq:sepa2}
\Psi=\{[1,2],3\}+\{[1,3],2\}~,
\end{equation}
where each term represents an atom linked to the third particle. 
Along (T), the two term have comparable energy, and can
interfere maximally.
The triangular diagram has different  possible shapes. Let us assume $q_2\le q_3$ without loss of generality.
\begin{itemize}\itemsep -4pt
\item
If $q_2\le q_3<1$, all mass configurations correspond to a stable ion.
\item
If $q_2< 1\le q_3$, the entire region on the left of (T), where (1,2) is the lowest threshold give stable configuration. There is instability at least near A$_3$.
\item
If $1\le q_1\le q_3$, there is instability at least near A$_2$ and near A$_3$.
\end{itemize}
The basic observation is that if, for instance, $q_2<1$, the (1,2) ion is negatively charged. 
Particle 3 feels a
potential that is asymptotically a Coulomb attraction,  
producing  bound states below the (1,2) energy. This state of affairs
holds in particular in the limit where $m_1=m_2=\infty$ (vertex A$_3$), and 
thus stability extends to the entire region from A$_3$ to (T), due to the star-shape property.

\begin{figure}[H]
\centering{
\includegraphics[width=.7\textwidth]{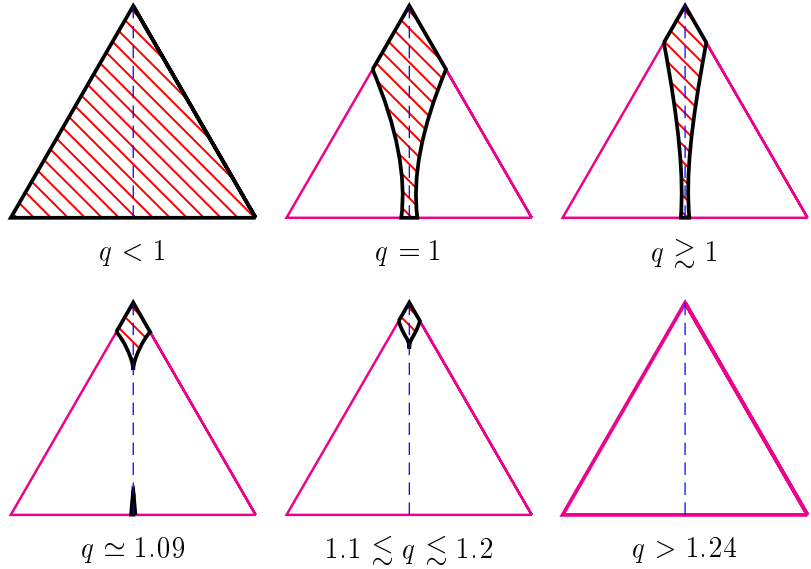} }
\caption{\label{3a:fig:tr1}Shape of the stability and instability regions for charge configurations $(1,-q,-q)$. }     \end{figure}

A plausible conjecture for the stability/instability frontier is 
shown in Fig.~\ref{3a:fig:tr1}, for charge
configurations $(1,-q,-q)$ with $2\leftrightarrow3$ symmetry. 
For $q<1$, there is stability everywhere. For $q=1$, we are in the case
studied in Sec.~\ref{se:3u2}. For $q$ slightly larger than 1, the stability 
band shrinks. It presumably breaks down somewhere
between Ps$^-$ and H$^-$, where the fraction of binding is smallest 
(see discussion in Sec.~\ref{3u2:sub:shape}). Then stability
around H$^-$ disappears, and stable ions are found only in the H$_2^+$ region. 
Stability there is lost where $q$ becomes larger than 1.24, as seen earlier in this section.

\begin{figure}[H]
\centering{\includegraphics[width=.7\textwidth]{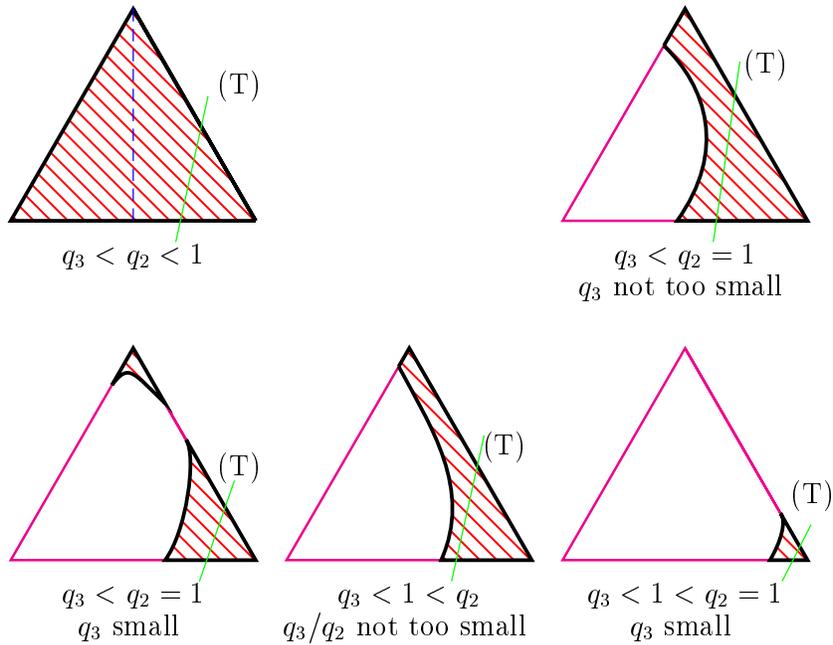} }
\caption{\label{3a:fig:tr2}Shape of the stability and instability regions for 
some asymmetric charge configurations
$(1,-q_2,-q_3)$. }      
\end{figure}

Some guess at the triangle plots are proposed in Fig.~\ref{3a:fig:tr2}, 
for asymmetric   configurations.
These stability domains would require thorough numerical investigation or more rigorous results. For instance, it has never been established that the domain sometimes splits into two parts, as suggested by one of these figures. What can be proved \cite{Mar95} is that the border shrinks if $q_1$ decreases (for the other $q_i$ fixed) and moves to the right as $q_2$ increases.

Another point of view, or, say, cross section of the parameter space consists of 
looking at stability for given masses $\{m_i\}$, as a function of the charges.
In this plot, the threshold separation (T) is a straight line starting at the origin.

\begin{figure}[!htpc]
\centering{\includegraphics[width=.35\textwidth]{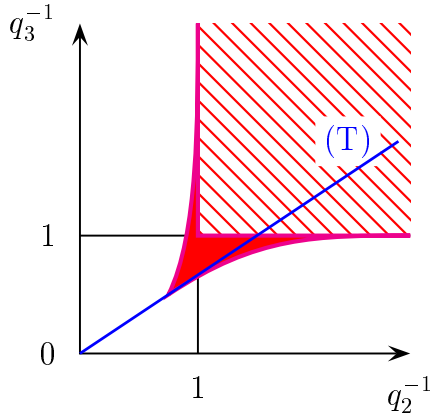} }
\caption{ Shape of the stability domain in the plane of inverse 
charges $(q_2^{-1},q_3^{-1})$, with normalisation $q_1=1$.  
The lines $q_2^{-1}=1$ and $q_2^{-1}=1$ are 
either asymptotes, or part of the border, starting from a value of $q_2^{-1}$ 
or $q_3^{-1}$ which might be less than 1, unlike the case shown in this 
figure.\label{3a:fig:conv-inv}}
\end{figure}

Suppose, for instance, that $q_2$ increases. In the region where $(1,2)$ is 
the lowest threshold, one can rescale all
charges by a factor $1/\sqrt{q_2}$, so that  the attraction $q_1 q_2$ 
(and hence the threshold energy), and the
repulsion strength $q_2q_3$ remain constant. The attractive term $q_1 q_3$ decreases, 
so stability deteriorates. 
If both $q_2$ and $q_3$ are increased by a factor $\lambda$, a similar 
rescaling leaves the attractive terms unchanged
and increases the repulsion by $\lambda$. Again, one moves away from stability.
It has been shown \cite{Kri00a} that each domain of instability 
is convex in the $\{1/q_2,\, 1/q_3\}$ plane (with, say, $q_1=1$
fixed.). This is illustrated in Fig.~\ref{3a:fig:conv-inv}.
\subsection{Critical charge for binding two electrons}\label{3a:sub:crit}
The symmetric case, where $m_2=m_3=m$, and $q_1=1$, $q_2=q_3=Z$, has been studied by Rebane~\cite{Reb95},  who addressed the following question:
 what is the maximal value, $Z_{\rm cr}$,  of $Z$ allowed for which the system is bound?
This is equivalent to finding the minimal charge $q_1$ required to bind two identical unit charges.

\begin{figure}[H]
\centering{\includegraphics[width=.4\textwidth]{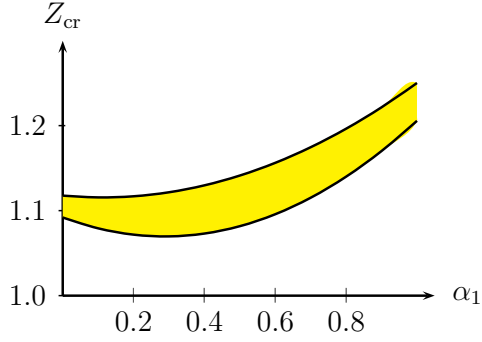}}
\caption{Estimate of the maximal value $Z_{\rm cr}$  of the charge $Z$ allowed to bind  i $(m_1^+,1^{-Z},1^{-Z})$, as a function of the normalised inverse mass $\alpha_1=1/(1+2m_1)$. The curves are rough fits to the upper and lower bounds given in Ref.~\protect\cite{Reb95}.\label{3a:fig:Reb}}
\end{figure}

It has been seen that $q\simeq 1.098$ for  $m_1/m\to\infty$, and $q\simeq 1.24$ for $m_1/m\to0$. 
The results for the general case are given in Fig.~\ref{3a:fig:Reb}.
There is a striking analogy between this curve and that of Fig.~\ref{3u2:fig:galpha} showing the relative  binding energy for $Z=1$. Again, the minimum is not exactly in the $\H^-$ limit, but for $\alpha_1\simeq 0.2$.
\subsection{Further limiting cases}\label{3a:sub:lim}
In addition to $\H^-$-like, $\H_2^+$-like and other symmetric cases, results have been obtained for the following configurations:
\subsubsection{Nearly symmetric states}
\begin{figure}[H]
\centering{\includegraphics[width=.36\textwidth]{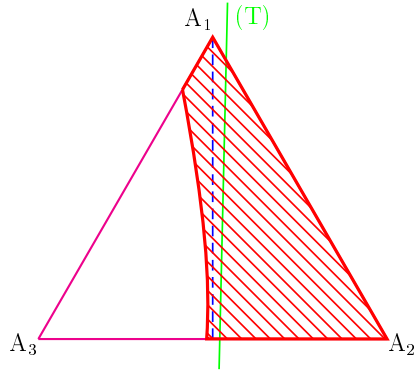}  }
\caption{%
\label{3a:fig:nearly-sym} %
For charges $q_1=1$, $q_2=1.02$, and $q_3=0.98$, the stability domain 
extends at least from A$_2$ to the median axis of
the triangle. }
\end{figure}

Equation (\ref{3u2:eq:width1}) was a decomposition into symmetric and antisymmetric parts for $q_2=q_3$ and $m_2$ slightly different from $m_3$. Similarly, one can consider the case $m_2=m_3$, i.e., the vertical axis of the triangle, and $q_2$ and $q_3$ slightly different. Introducing the strength factors $s_1=q_2q_3$, $s_2=q_3q_1$, and $s_3=q_1q_2$, and denoting $E[s_1,s_2,s_3]$ the ground state energy, one gets by the same reasoning
\begin{equation}
\label{3a:eq:nearly-sym1}
E[s_1,s_2,s_3]\le E\left[s_1,{s_2+s_3\over2},{s_2+s_3\over2}\right]~,
\end{equation}
with the later energy being $(1+g)$ times its threshold energy, as per Eq.~(\ref{3u2:eq:def-g}). If for instance
\begin{equation}
\label{3a:eq:nearly-sym2}
s_1=1~,\quad s_2=1-x~,\quad s_3=1+x~,
\end{equation}
stability is ensured if 
\begin{equation}
\label{3a:eq:nearly-sym3}
(1+x)^2\le 1+g~,
\end{equation}
which can be solved for the charges, rescaled to $q_1=1$, with the conclusion that stability remains at least up to the configuration
\begin{equation}
\label{3a:eq:nearly-sym4}
\alpha_i=\left\{\alpha_1,{1-\alpha_1\over 2}, {1-\alpha_1\over2}\right\},\quad
q_i=\left\{1,{1\over1-x},{1\over1+x}\right\}~.
\end{equation}
When $\alpha_1$ varies, the minimum of $g(\alpha_1)$ is between 4\% and 5\%.
The, if $x\le 0.02$, the vertical axis of the triangle belongs 
to the stability domain, as shown in Fig.~\ref{3a:fig:nearly-sym}.
\subsubsection{The asymmetric Born--Oppenheimer limit}
\begin{figure}
\centering{\includegraphics[width=.4\textwidth]{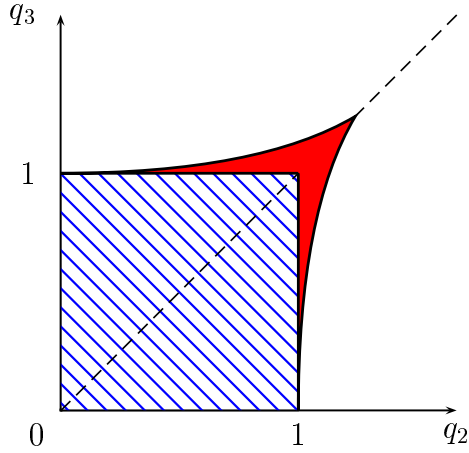}}
\caption{\label{3a:fig:BO} 
Schematic shape of the stability domain in the Born--Oppenheimer 
limit. The heavy particles have charges $q_{2}$ and $q_{3}$. 
The charge of the light particles is set to $q_{1}=-1$.}
 \end{figure}

Consider the stability domain in the $(q_2,\,q_3)$ plane in the limit where both
$m_2\to\infty$ and $m_3\to\infty$. The domain includes the unit s and 
square and extends till $q_2=q_3\simeq1.24$ on the symmetry axis.This is illustrated in Fig.~\ref{3a:fig:BO}.

It can be shown \cite{Mar95} that the boundary leaves 
the square vertically at the point ($q_2=1,\,q_3=0$) with a behaviour
\begin{equation}
\label{3a:eq:BO1 }
q_2-1\simeq 18 {q_3\over(-\ln q_3)^3}~.
\end{equation}
A similar pattern is observed near ($q_2=0, \,q_3=1$). 
\subsection{Numerical investigations}\label{3a:sub:num}
Modern computers and new computational algorithms  make it possible to perform intricate few-body 
calculations in a very short time, and thus to repeat the calculations with changes in the parameters.

\begin{figure}[!htpc]\centering{
\includegraphics[width=.4\textwidth]{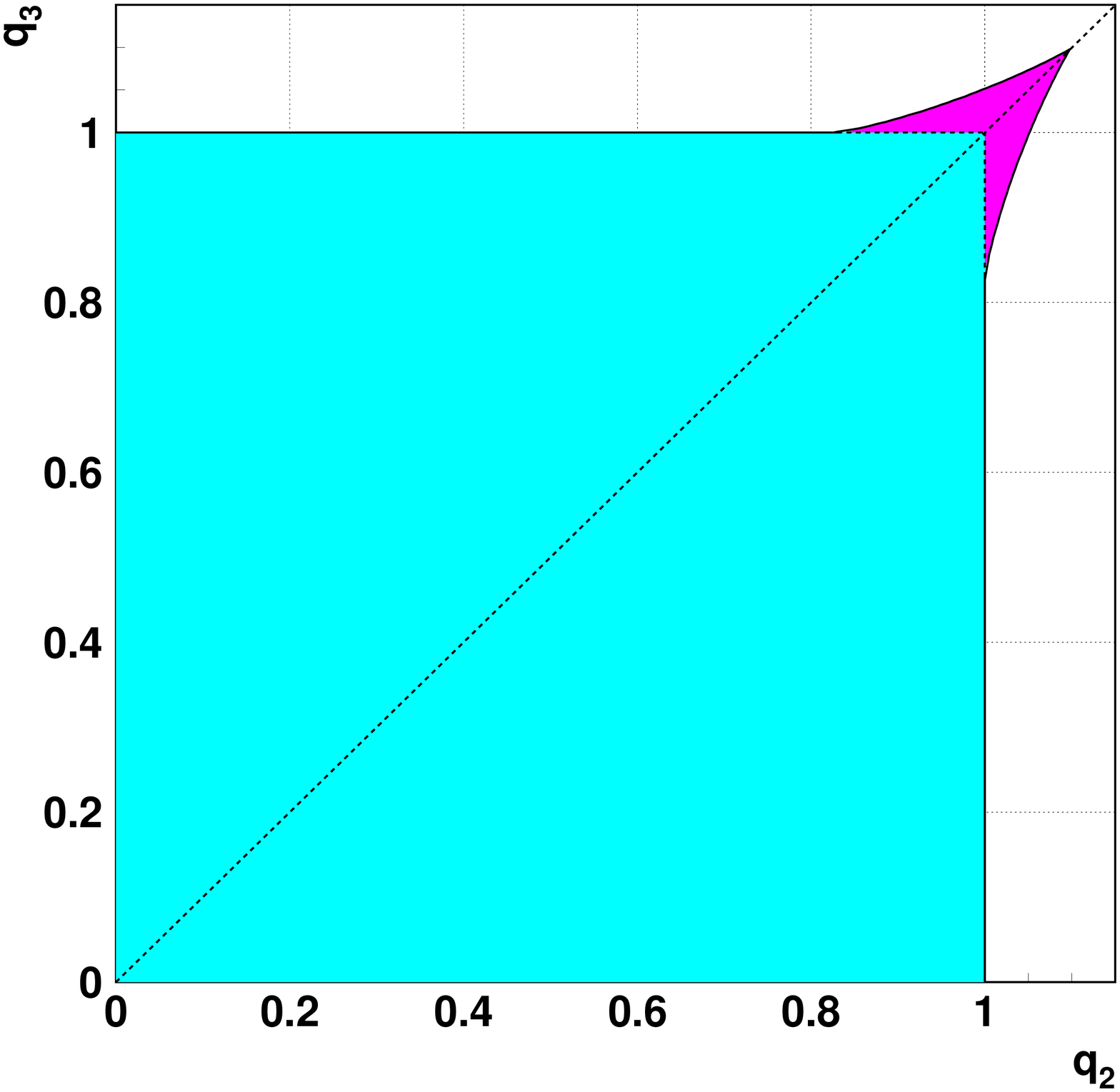}
\hspace*{.1\textwidth}
\includegraphics[width=.4\textwidth]{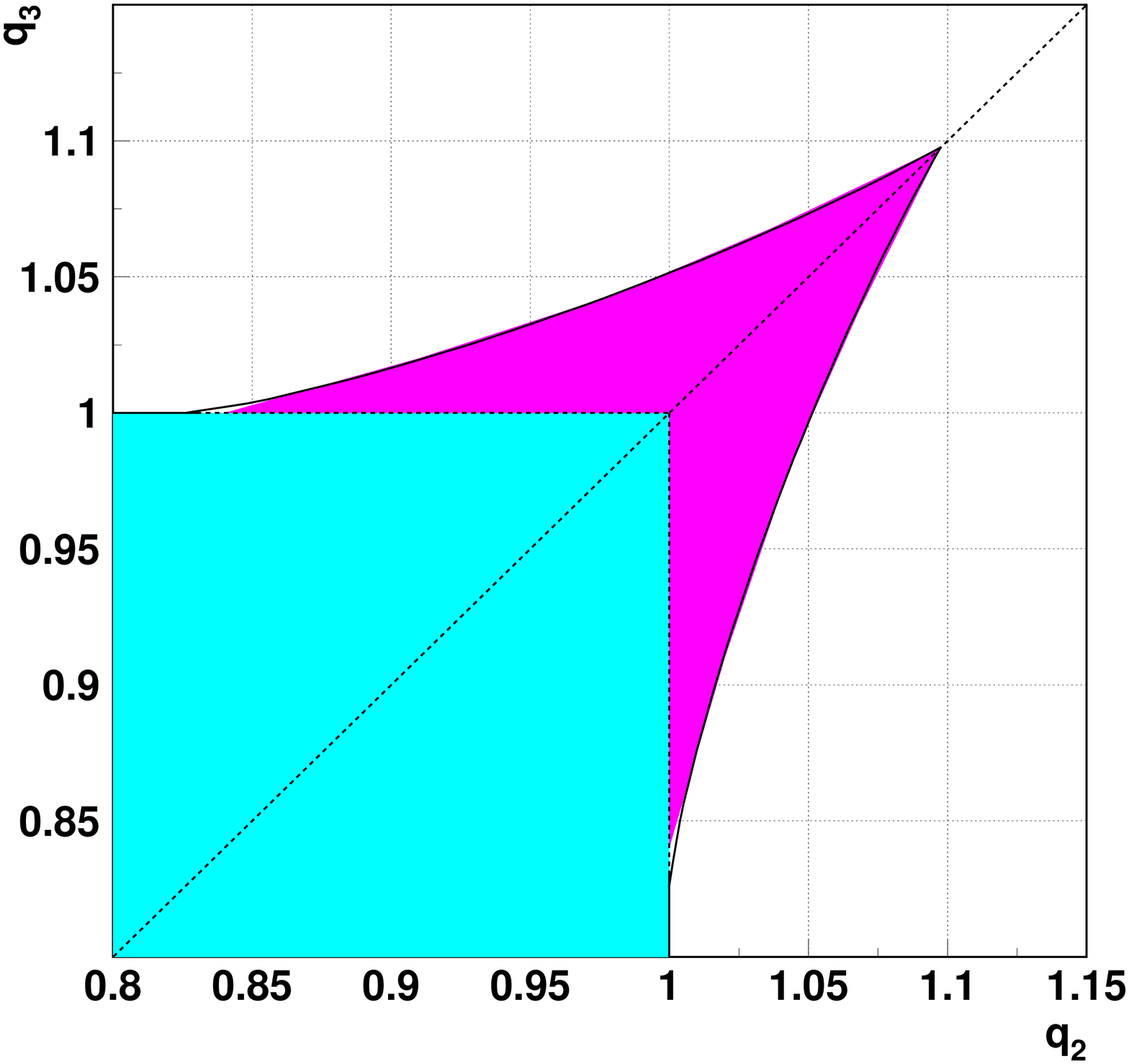} }
\caption{\label{3a:fig:Ali1}Variational estimate of the stability domain for $m_1=\infty$ and $m_2/m_3=1$, full view (left) and enlargement of the spike (right).}
\end{figure}

\begin{figure}[!htpc]\centering{
\includegraphics[width=.4\textwidth]{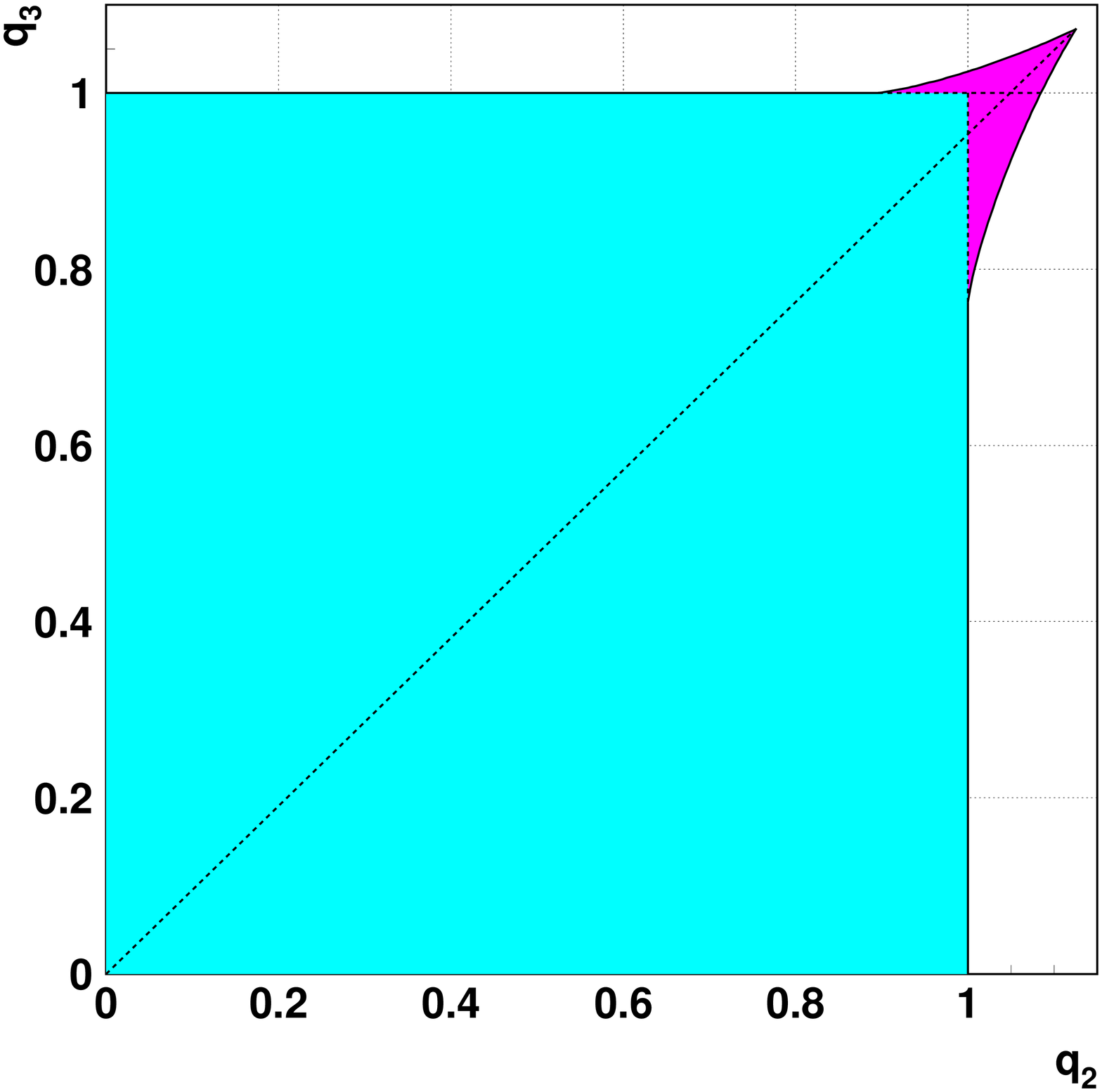}
\hspace*{.1\textwidth}
\includegraphics[width=.4\textwidth]{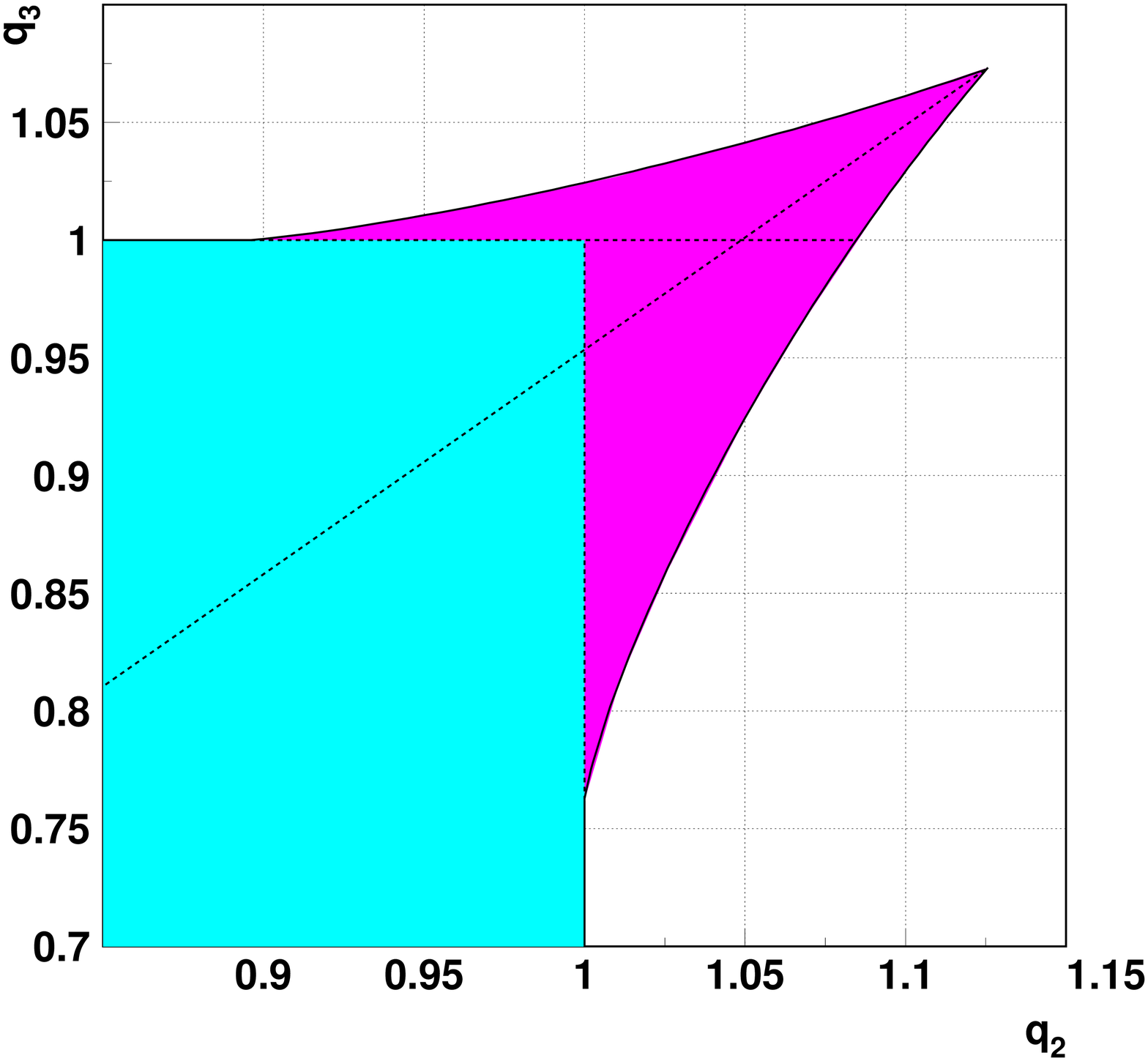} }
\caption{\label{3a:fig:Ali2}Same as Fig.~\protect\ref{3a:fig:Ali1}, for $m_2/m_3=1.1$.}
\end{figure}

\begin{figure}[!htpc]\centering{
\includegraphics[width=.4\textwidth]{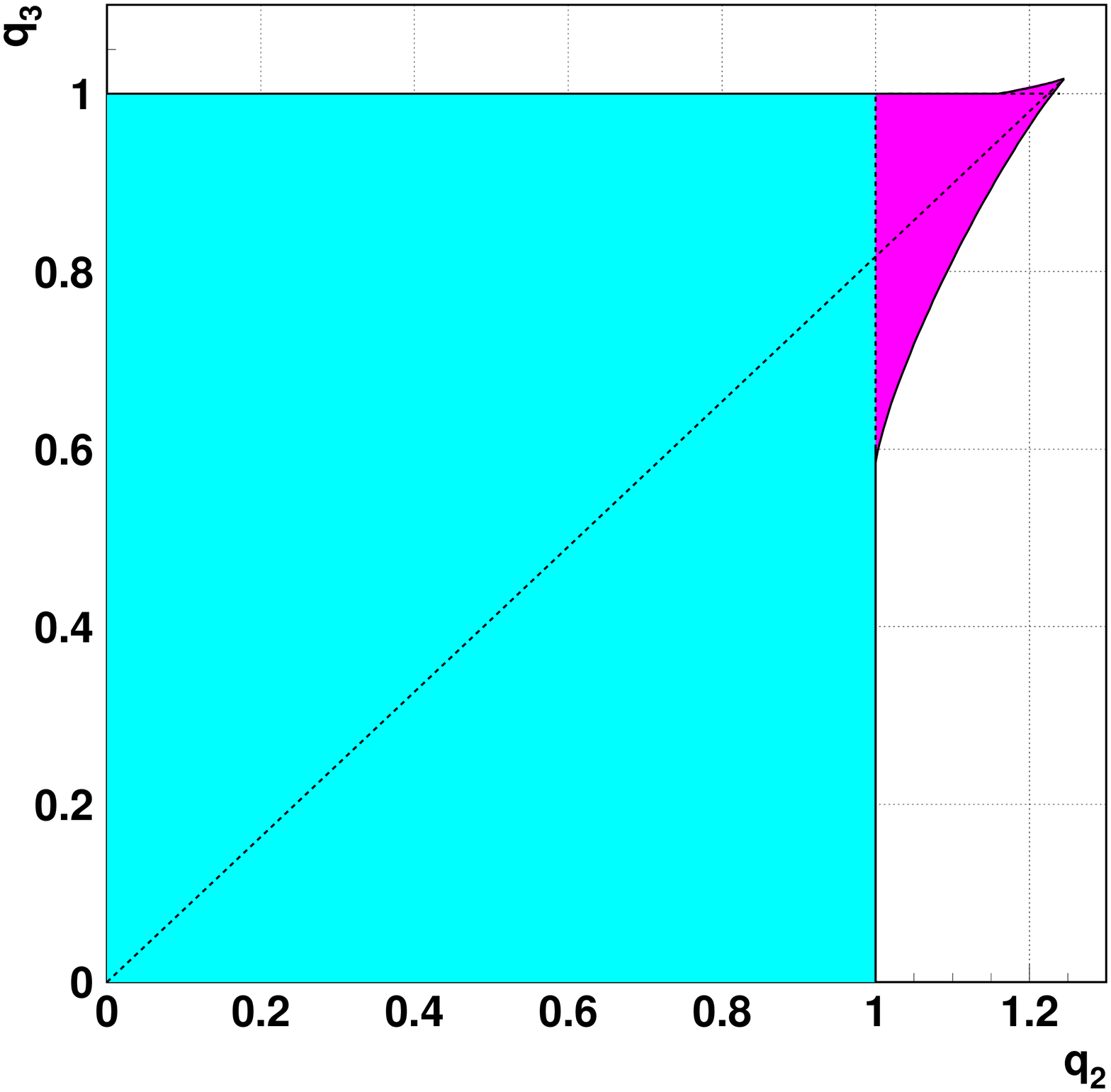}
\hspace*{.1\textwidth}
\includegraphics[width=.4\textwidth]{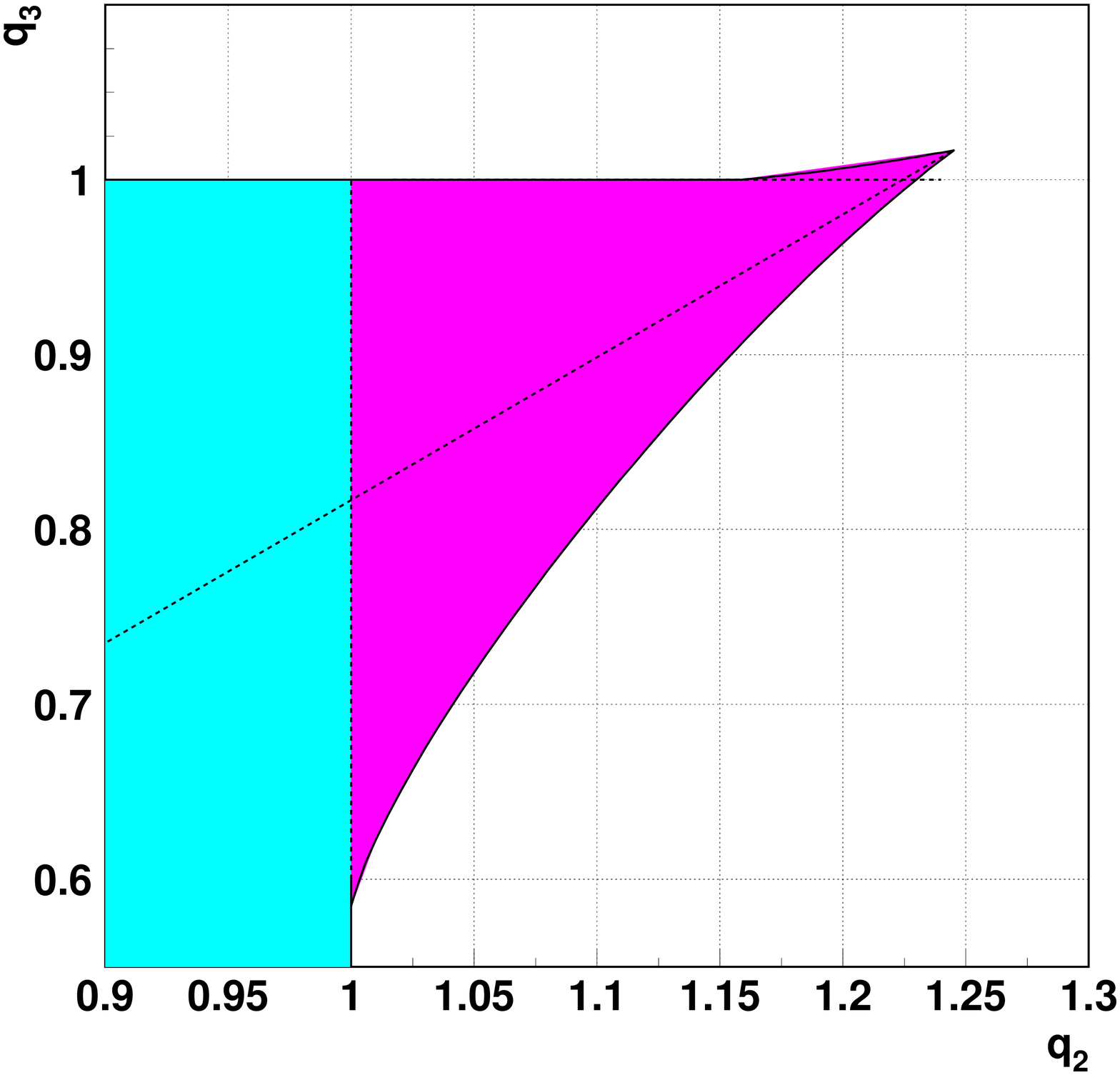} }
\caption{\label{3a:fig:Ali3}Same as Fig.~\protect\ref{3a:fig:Ali1}, for $m_2/m_3=1.5$.}
\end{figure}

\begin{figure}[!htpc]\centering{
\includegraphics[width=.4\textwidth]{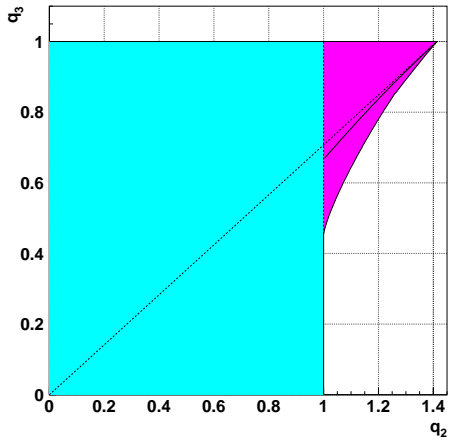}
\hspace*{.1\textwidth}
\includegraphics[width=.4\textwidth]{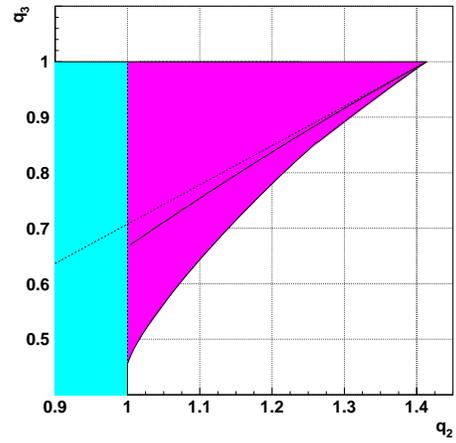} }
\caption{\label{3a:fig:Ali4}Same as Fig.~\protect\ref{3a:fig:Ali1}, for $m_2/m_3=2$.}
\end{figure}

For instance, Krikeb \cite{Kri98} has drawn the stability area (more precisely its minimal extension, as his calculation was variational) for different choices of constituent masses \cite{Kri98,Kri00a}. Some of his plots are reproduced in Figs.~\ref{3a:fig:Ali1}--\ref{3a:fig:Ali4}.
There is always a pronounced spike near the threshold separation (T).
\subsection{Open issues}\label{3a:sub:open}
\begin{figure}[H]\centering{
\includegraphics[width=.4\textwidth]{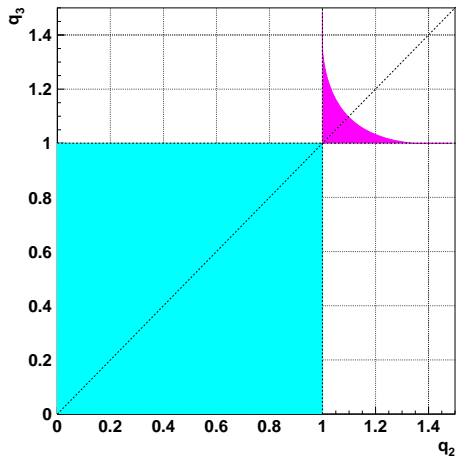}
\hspace*{.1\textwidth}
\includegraphics[width=.4\textwidth]{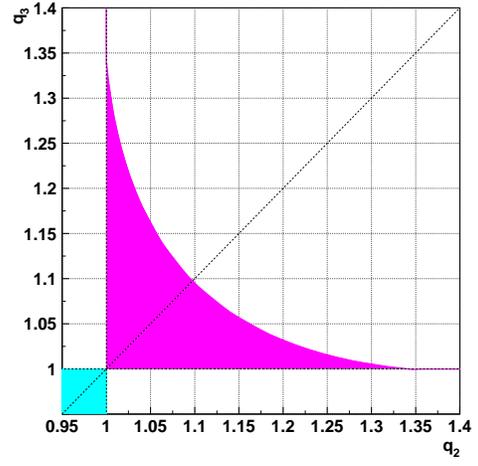} }
\caption{\label{3a:fig:Ali0} Variational estimate of the domain in the $(q_2,q_3)$ plane, for $q_1=1$, for which stability occurs at least for some specific values of the mass ratios.}
\end{figure}

Some questions remain, for instance:
\begin{itemize}\itemsep -2pt
\item In the $(q_2,q_3)$ plot, are there always points for which $q_3>1$?
\item What is  the overall shape of the areas of stability? This means a curve $f(q_2,q_3)=0$ 
such that beyond this curve, stability never occurs, whatever values are assumed for the constituent masses. A guess is proposed in Fig.~\ref{3a:fig:Ali0}.
\end{itemize}
%

\markboth{\sl Stability of few-charge systems} {\sl Four-charge systems}
\clearpage\section{Four-charge systems}\label{se:4u}
In this section, we consider systems 
$(m_{1}^{+},m_{2}^{+},m_{3}^-,m_{4}^-)$ with arbitrary masses $m_{i}$ and 
unit charges $q_{1}=q_{2}=+1$ and $q_{3}=q_{4}=-1$.  The question we address is:
for which values of the masses $m_{i}$ is the system stable against 
spontaneous dissociation?  The answer, of course, only depends on the mass ratios and not on the masses  themselves.
\subsection{Lowest threshold}\label{4u:sub:thr}
The first stage is to identify the lowest threshold.  Without 
loss of generality, one can assume $m_{1}\ge m_{2}$ and $m_{3}\ge 
m_{4}$, so that the lowest threshold consists of two neutral atoms 
is $(m_1^{+},m_{3}^-)+(m_2^{+},m_{4}^-)$.  This can be seen directly 
from the Bohr formula.  The results is in fact more general, as 
noticed by Bertlmann and Martin \cite{Ber80} and by Nussinov \cite{Nussinov:1999sx}: if a given 
two-body potential provides two constituent of masses $m$ and $m'$ with 
a ground-state energy $E_{2}(m,m')$, then
\begin{equation}\label{4u:eq:BertlmannM}
    E_{2}(m_{1},m_{3})+E_{2}(m_{2},m_{4})\le
    E_{2}(m_{1},m_{4})+E_{2}(m_{2},m_{3})~,
\end{equation}
when $m_{1}\ge m_{2}$ and $m_{3}\ge m_{4}$. This is easily shown using 
the property that the two-body energy $E_{2}(m,m')$ is a convex function 
of the inverse reduced mass $m^{-1}+m'^{-1}$.. 
For instance, in the limit of a strictly flavour-independent confining 
force between a quark and an antiquark, one anticipates a mass inequality
\begin{equation}\label{4u:eq:BertM}
({\rm Q}\overline{\rm Q}) + ({\rm q}\bar{\rm q})\le 2({\rm Q}\bar{\rm  q})~,
\end{equation}
which is observed in the actual spectrum of mesons \cite{PDG02}.

There is another type of threshold with a three-body ion and an 
isolated charge.  For instance, $(m_{1}^{+}m_{2}^{+}m_{3}^-)+m_{4}^-$ 
is certainly the lowest threshold if $m_{1}=m_{2}=m_{3}\gg 
m_{4}$, as typical Coulomb energies grow proportionally with masses. However, an 
ion with charge $\pm1$ can always bind a charge $\mp1$, as their 
interaction reduces to a Coulombic attraction at large separation (see Sec.~\ref{SO:sub:asymCoul} of Appendix  \ref{se:SO}). Thus, if the lowest threshold is an ion and a charge, the system is stable.

In short, to establish stability, it is sufficient to get below the 
lowest atom--atom threshold, i.e., to find a trial wave function $\Psi$ 
with a variational energy $E_{4}[\Psi]$ such that
\begin{equation}\label{4u:eq:SuffStab}
E_{4}[\Psi]< E_{2}(m_{1},m_{3})+E_{2}(m_{2},m_{4})~.
\end{equation}

Nevertheless, it is interesting to determine which is the 
lowest threshold. In Ref.~\cite{Ric93}, for instance, dealing with 
hydrogen-like configurations $(M^{+},M^{+},m^-,m^-)$, it is shown 
rigorously that the lowest threshold always consists of two atoms, 
namely, in an obvious notation,
\begin{equation}\label{4u:eq:ThresHydro}
    \eqalign{
 2 E_{2}(M,m)&<E_{3}(M,M,m)~,\cr
 2 E_{2}(M,m)&<E_{3}(m,m,M)~.\cr}
\end{equation}
In practical calculations, knowing that the lowest threshold of a 
system is made of cluster $C_{1}$ and cluster $C_{2}$ suggests a 
possible trial wave function associating the internal motion within 
$C_{1}$ and $C_{2}$ with the  motion of $C_{1}$ relative to 
$C_{2}$.

\subsection{Specific systems}\label{4u:sub:spe}
\subsubsection{$\H_2$}
The binding mechanism of the 
hydrogen molecule, $\H_2$,  is described in several textbooks.  A beautiful 
picture is provided by the method of Born, Oppenheimer, Heitler and 
London~\cite{Pau35}.  In the limit where the proton mass $M$ is 
very large compared with the electron mass $m$, the protons move almost 
classically in an effective potential $V(R)$.  Besides the direct 
electrostatic repulsion $-1/R$, this potential is generated by the energy 
of the two electrons in the static field of the two protons separated 
by a distance $R$.  At very large $R$, the system consists of two 
independent atoms, and one fixes the potential so that $V(\infty)=0$.  
At very small $R$, the repulsion dominates, and $V\to +\infty$.  In 
between, there is a well-pronounced pocket of attraction with $V<0$.  
In the extreme limit where $M\to\infty$, the minimum of $V$ is the 
binding energy of the molecule.  This minimum is deep enough to 
guarantee that for the actual molecule, binding survives numerical 
uncertainties, finite proton mass and quantum effects.

In fact, the potential $V$ between the two protons is deep enough not 
only to bind the molecule, but also to allows for many excited states.
See, e.g., \cite{Kar94}.

Several accurate, non-adiabatic, calculations of the $\H_2$ energy and properties have been published, see, e.g.,  \cite{Bub03}. Note that it was once believed that a problem existed, with a discrepancy between the experimental ionisation energy and the theoretical one \cite{Kol68}.  Nowadays, the  problem has disappeared.
 \subsubsection{$\Ps_2$}
The stability of $(\e^+,\e^+,\e^-,\e^-)$ with respect to dissociation into two positronium ions $(\e^+,\e^-)$ was speculated by Wheeler \cite{Whe46}. A first study by Ore in 1946 led him to conclude~\cite{Ore46} ``with reasonable assurance  that this structure is not stable against disintegration into two bi-electrons.'' He used, however, a sharply truncated harmonic-oscillator basis which is not appropriate for a weakly-bound molecule.

The following year, Hylleraas and Ore \cite{Hyl47a} introduced  a more suitable trial wave-function and the stability of the positronium molecule $(m^{+},m^{+},m^-,m^-)$ was 
established, and confirmed in 
several further studies using more refined wave functions. In natural 
units, the positronium atom $(m^{+},m^{-})$ has a binding $E_{2}=-1/4$, 
so that the dissociation threshold for the molecule is $E_\mathrm{th}=2E_{2}=-1/2$.
It is convenient to define the fraction $x$  of binding below 
threshold as
\begin{equation}
\label{4u:eq:Def-x}
E({\rm e}^+,{\rm e}^+,{\rm e}^-,{\rm e}^-)=(1+x)E_\mathrm{th}~.
\end{equation}
The result of Ref.~\cite{Hyl47a}, corresponding to $x\simeq 
0.0084$, has been improved by several authors. The latest (and best) 
values are $x=0.03186$ \cite{Koz93} and $x=0.03201$ \cite{Var98}. See, also, \cite{Bre97,Reb00}.

The proof of stability of $\Ps_2$ by Hylleraas and Ore 
\cite{Hyl47a} relies on  an elegant variational method.
They first got rid of the scale by noticing that if
$\Psi(\vec{\rm r}_i)$ is a trial wave function, with
norm and expectation values of the kinetic and potential
energies written as
\begin{equation}
\label{4u:eq:n-t-v}
n=\langle\Psi|\Psi\rangle~,\quad
t=\langle\Psi|T|\Psi\rangle~,\quad
v=\langle\Psi|V|\Psi\rangle~,
\end{equation}
then the best rescaling  of the type
$\phi=\Psi(\vec{\rm r}_i/\lambda)$ yields a minimum
\begin{equation}
\label{4u:eq:virial}
\widetilde{E}=-{v^2\over4tn}~,
 \end{equation}
which corresponds to $\langle\phi|T|\phi\rangle
=-\langle\phi|V|\phi\rangle/2$, i.e., the same sharing of the
kinetic and potential energies as for the exact
solution. This extension of the virial theorem to
variational approximations is well known \cite{Hyl29,Foc30}.

The frozen-scale wave function of Hylleraas and Ore only contains
a single parameter:
\begin{equation}
\label{4u:eq:Hylleraas-wf}
\Psi=\exp[-(r_{13}+r_{14}+r_{23}+r_{24}/2)]
\cosh[\beta(r_{13}-r_{14}-r_{23}+r_{24})/2]~.
\end{equation}
Explicit integration leads to (a misprint in Ref.\ \cite{Hyl47a}
is corrected below)
\begin{eqnarray}
\label{4u:eq:Hylleraas-ev}
&& n={33\over16}+{33-22\beta^2+5\beta^4\over16(1-\beta^2)^3}~,\qquad
t={21\over8}-{3\beta^2\over2}
  +{21-6\beta^2+\beta^4\over8(1-\beta^2)^3}~,\\
&& v={19\over6}+{21-18\beta^2+5\beta^4\over4(1-\beta^2)^3}
 -{1\over(1-\beta^2)^2}\left[1-{5\beta^2\over8}-{1\over4\beta^4}
   +{7\over8\beta^2}+{(1-\beta^2)^4\over4\beta^6}\ln
{1\over1-\beta^2}\right]~,\nonumber
\end{eqnarray}
to be inserted in (\ref{4u:eq:n-t-v}) and (\ref{4u:eq:virial}), leading to a
minimum $\widetilde{E}=-0.5042$ near $\beta^2=0.48$. More details of
the derivation of Eq.~(\ref{4u:eq:Hylleraas-ev}) are provided in Appendix
\ref{se:elem}.
It is rather easy to generalise the calculation of Ref.~\cite{Hyl47a}
to a trial wave function of the type
\begin{equation}
\label{4u:eq:Hylleraas-wf-gen}
\Psi=\sum_i
c_i\exp[-a_i(r_{13}+r_{14}+r_{23}+r_{24})]
\cosh[ b_i(r_{13}-r_{14}-r_{23}+r_{24})]~,
\end{equation}
but one does not gain much \cite{Ore47}.  As analysed for instance by 
Ho \cite{Ho93} and by the authors he quotes, some explicit $r_{12}$ 
and $r_{34}$ dependence is needed in the wave function to improve the 
accuracy.  Such a dependence is included in all accurate variational 
calculations \cite{Koz93,Var98}, as well as in Monte-Carlo 
variants \cite{Bre97}.

As $\Ps_{2}$ is weakly bound, one might guess that there are no bound 
excited states, and, indeed, for many years, no excited state was 
found or even  searched for.  However, a recent 
highly accurate calculation \cite{Var98} indicates the existence of a 
state $\Ps^{*}_{2}$ with angular momentum $L=1$ and negative parity. This
cannot decay into two positronium atoms in their ground 
state, as one of the atoms should be in a state with $L=1$. The threshold 
energy for spontaneous dissociation is thus
\begin{equation}
\label{4u:eq:Thr-Ps2-exc}
E_\mathrm{th}^{*}=-{1\over4}\left(1+{1\over4}\right)=-0.3125~,
\end{equation}
while $\Ps^{*}_{2}$ is found \cite{Var98} to have an energy $E^{*}=-0.3344$,
i.e.,
 bound by a fraction $x^{*}=0.07$, where $x^{*}$ is defined as in 
(\ref{4u:eq:Def-x}).

Electromagnetic transitions from $\Ps^{*}_{2}$ to $\Ps_{2}$ might be 
used as a signature of the formation of positronium molecules.  The 
main decay channels, $\Ps_{2}\to 4\gamma$ or $\Ps_{2}^{*}\to 4\gamma$, 
are not of great use, unfortunately, as the energy of the emitted 
$\gamma$-ray differs very little from those coming from annihilation of 
ordinary $\Ps$ atoms. This is discussed in Ref.~\cite{Usu98}.
 \subsubsection{$\Ps\H$}
The hydrogen ``hydride'' $\Ps\H$ is the system $(\p,\e^{+},\e^-,\e^-)$, or 
in our other notation, $(M^{+},m^{+},m^-,m^-)$ in the limit of a 
large $M/m$ mass ratio.  The chemistry of positrons is beyond the 
scope of this review.  See, for example, Ref.~\cite[p.~263]{Sch01}.  What is
   of interest to us is the fact that, if annihilation is neglected,
   there are many examples of stable configurations where a positron 
is attached to an atom or an ion \cite{Mit02}.

The stability of $\Ps\H$ is not completely obvious, if one starts 
building this configuration by adding the constituents one by one.
In a possible such construction scheme, one is faced with the instability 
of $(p,e^{+},e^-)$, as seen in Sec.~\ref{3u1:sub:pemep}; it is thus a pleasant 
surprise that the whole system is stabilised by adding a second 
electron.  On the other hand, when one starts from a $\Ps^-$ ion and a 
proton, one expects some attraction between them.  The question is 
whether this attraction brings the system down not only below 
the $(\Ps^-,\p)$ threshold but also the $(\H,\Ps)$ threshold.  A 
similar question arises when the system is viewed as a stable 
$\H^-$ ion combined to a positron.

The first proof of stability of $\Ps\H$ is due to Ore \cite{Ore51}.
 He used a trial wave function similar to (\ref{4u:eq:Hylleraas-wf}), namely,
 with a similar notation ($1$ is the proton, $2$ the positron)
 \begin{equation}
     \label{4u:eq:PsH-Ore-wf}
  \Psi=\exp\left[-k(r_{13}+\alpha r_{14}+\beta  r_{24})\right]+
             \{3\leftrightarrow 4\}~.
 \end{equation}
Again, the scale parameter $k$ can be fixed by the virial theorem, and 
the minimisation is restricted to the parameters $\alpha$ and $\beta$.  
Some details on the calculation of the relevant matrix elements are 
given in Appendix \ref{se:elem}.  For $\alpha=0.25$ and $\beta=0.5$, Ore 
obtained an energy $E=-0.75256$ corresponding to a fraction $x\simeq 
0.00334$ of extra binding, defined as in Eq.~(\ref{4u:eq:Def-x}). With a 
slightly more general wave function
 \begin{equation}
     \label{4u:eq:PsH-Ore-wf2}
  \Psi=\exp\left[-(r_{13}+\gamma r_{23}+\alpha r_{14}+\beta 
  r_{24} )\right]+\{3\leftrightarrow 4\}~,
 \end{equation}
one obtains a better value $x=0.0112$. With a polynomial added, 
Navin \etal\ obtained $x\simeq 0.039$ \cite{Nav74}.

Positronium hydride was identified experimentally in 1992 
\cite{Sch92}. Meanwhile, variational calculations have been 
pushed to higher accuracy. See, for instance, 
\cite{Ho78,Koz93,Bre98,Bre96,Str98,Ryz98,Usu98,Jia98,Yan99}, and references
therein. For instance, Zong-Chao Yan and Ho \cite{Yan99} obtained $x\simeq
0.052$.

The system $(\p,\e^{+},\e^-,\e^-)$  also has a resonant state with angular 
momentum $L=0$.  It is located at $E=-0.602$ with respect to complete 
dissociation, and thus is relatively narrow (width $\Gamma=0.0028$) 
since it is below the threshold for dissociation into a hydrogen atom and 
an excited positronium atom $(E_\mathrm{th}=-0.5625)$ \cite{Ho78}. There is 
also a metastable $L=1$ state at $E=-0.615$ \cite{Bre98}, which is thus 
below the threshold $E_\mathrm{th}=-0.5625$ for dissociation into a hydrogen atom and 
a  Ps atom with $L=1$. The situation is similar for the  metastable $L=2$ state at
$E=- 0.566$ \cite{Bre98}, which is below the threshold $E_\mathrm{th}=-0.5267$ 
corresponding to H and  Ps with $L=2$.
\subsection{Properties of the stability domain}
\label{4u:sub:domain}
\subsubsection{Tetrahedral representation}
The stability domain of the system,($m ^ + _ 1,\,m ^ + _2,\,m ^ - _ 3,\,m^-_4$),
made up of four particles, two having the same positive charge
and two having the opposite negative charge, interacting only through Coulombic
forces, can be represented in terms of points in the interior, or on the
surface of a regular tetrahedron \cite{Ric94,Ric02}. This is a generalisation
of the equilateral triangle diagram in the case of three bodies, two of which
have the same positive charge and the third has the opposite negative charge
\cite{Mar92,Arm93}.  Note that as the Coulombic potential between charges
is invariant under charge conjugation, stability properties are invariant
under this operation.

Let $\alpha_i$ =  1/$m_i$, were $m_i$ is the mass of particle $i$. As
only Coulombic forces are being considered, the stability properties are
dependent only on the ratios of the masses.  It is thus
possible to normalise the {$\alpha_i$} so that 
$\sum_{i=1}^4\alpha_i$ = 1.  The sum of the distances from each face to any point X,
inside or on the surface of the tetrahedron, is a constant.  If we choose the
length of the sides of the tetrahedron to be $\sqrt(3/2)$, this constant is $1$.
Take $\alpha _ j$ to equal the distance of the point X from face $j$.  Then
the set {X ($\alpha_1$,$\alpha_2$,$\alpha_3$,$\alpha_4$), with  $0 \le
\alpha _ j\le1$, $\sum_{i=1}^4\alpha _i = 1$} represents
all points within the tetrahedron and on its surface and hence all possible
mass ratios of the system ($m ^ + _ 1,\,m ^ + _2,\,m ^ - _ 3,\,m ^ - _ 4$).

\begin{figure}[H]
\centerline{\includegraphics[width=.72\textwidth]{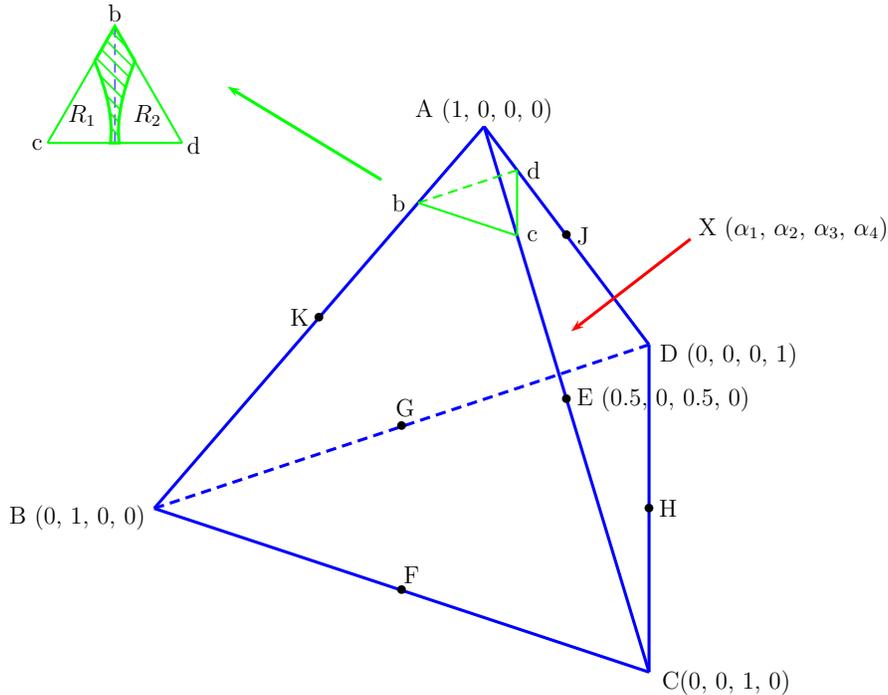}}
\caption{\label{4u:fig:FigTetra}%
The regular tetrahedron diagram representing the system
$(m_1^+,m_2^+,m_3^-,m_4^-)$. Faces BCD, ACD, ABD and ABC are labelled
1,2,3 and 4 respectively. Points E, F, G, H, J and K are the
midpoints of the sides of the tetrahedron. Regions $R_1$ and $R_2$ correspond to
the domain of instability on the triangular surface bcd. }
\end{figure}

The regular tetrahedral diagram is shown in Fig.~\ref{4u:fig:FigTetra}.  
The vertices correspond
to cases when three of the particles are infinitely massive.  The side
AB corresponds to the set of inverse masses $(1-x,x,0,0)$ with  $0\le x \le 1$, where two of
the particles, in this case 3 and 4, are infinitely massive, and similarly
for the other sides.  
It is interesting to consider the following special configurations:
\begin{itemize}
\item 
$\Hy$-$\Ha$ corresponds to $(\mathrm{p},\mathrm{e}^+,\pa, 
\mathrm{e}^-)$, i.e.
to (0.4997,0.0003,0.4997,0.0003) which is very close to $E (0.5,0,0.5,0)$,
the midpoint of AC.  Any permutation of particles 1 and 2 and/or 3 and 4 also
corresponds to $\Hy$-$\Ha$ and thus the midpoints F,G and J of BC, BD and
AD, respectively, also correspond very nearly to $\Hy$-$\Ha$.  All these points
are thus deep inside the domain of instability.
\item 
The plane KDC corresponds to systems with identical particles 1 and 2, 
with $m_1=m_2$, and the same charge. It can be seen that they are all stable. 
Similarly, the points of the plane ABH represent systems with identical particles 3
and 4, which are stable.
\item  
The midpoint H of CD lies very close to $(\p,\p,\e^-,\e^-)$, which is H$_2$ and
is known to be bound.  Similarly, the midpoint K of AB lies very close to
$(\mathrm{e}^+,\mathrm{e}^+,\pa,\pa)$, which is the antihydrogen molecule with
the same bound states as $\Hy_2$.
The segment KH corresponds to hydrogen-like configurations $(M^+,M^+,m^-,m^-
)$. 
In the middle,  the centre of the tetrahedron represents the positronium molecule.
\end{itemize}

There are the following obvious symmetries of the stability domain of 
($m_{1}^{+},m_{2}^{+},m_{3}^{-},m_{4}^{-}$): $1\leftrightarrow2$, 
$3\leftrightarrow4$ or $\{1,2\}\leftrightarrow\{3,4\}$ exchanges. As a 
consequence, the binding energy is stationary around the 
corresponding symmetry axes. For instance, it is of second order in 
$\delta m=m_{2}-m_{1}$, for given $m_{1}+m_{2}$.

Richard \cite{Ric94,Ric02} considers that the lines AC, BC, BD and AD represent
unstable systems whereas AB and CD represent stable systems.  He gives
schematic pictures of the stability domain on each of the faces of the
tetrahedron.  It takes the form of a sharp peak with base AB or CD, which
is symmetric about the median perpendicular to AB or CD on the face under
consideration.

\begin{figure}[H]
\centering
 \setlength{\unitlength}{0.3pt}
  \begin{picture}(100,600)(50,0)
\put(-80,0){\includegraphics*[width=137pt]{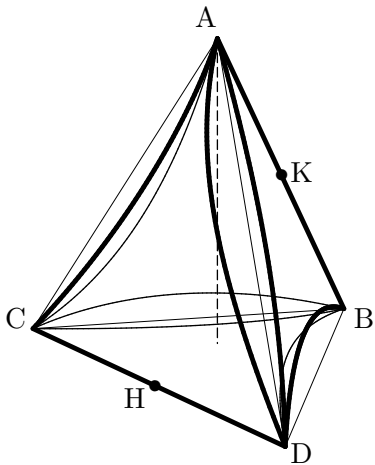}}
\put(-100,180){C}
\put(140,560){A}
\put(340,180){B}
\put(260,10){D}
\put(80,99){$\bullet$}
\put(50,80){H}
\put(240,365){$\bullet$}
\put(260,365){K}
\end{picture}
\caption{Guess at the shape of the stability domain for four unit charges, 
in the tetrahedron of normalised inverse masses. \label{4u:fig:shape}}
\end{figure}

Richard describes what is known about the form of the stability domain
within the tetrahedron.  The form on a triangular surface close to
a vertex and parallel to the opposite face is an interesting case.  This
surface is in the shape of an equilateral triangle and represents systems
in which one particle is very much lighter than the other three.  As Richard
points out, in this situation the existence or otherwise of four-body
bound states depends on whether the three very much heavier particles have
a bound state that can form a charged 'nucleus' that can bind the very much
lighter particle, which is of opposite charge.  It follows that the form of
the stability domain on the above surface is the same as for three
particles, two of which have the same positive charge and the other has
the opposite negative charge \cite{Ric02,Arm02}, with upper vertex on AB or CD,
the sides of the tetrahedron that are considered to correspond to bound
states.  See Fig.~\ref{4u:fig:FigTetra}.
To summarise, the shape of the stability domain is likely to be 
that schematically pictured in Fig.~\ref{4u:fig:shape}. Some guessed properties remain to be
demonstrated, and more numerical investigations would be needed for a more precise drawing.
\subsubsection{Hydrogen-like configurations}
We have seen that $(M^{+},M^{+},m^-,m^-)$ is stable for $M\gg m$ and for 
$M=m$, but this is shown with different approaches: the 
Heitler--London method for hydrogen and the Hylleraas--Ore variational 
method for $\Ps_{2}$. In fact one can show that the stability of 
$\Ps_{2}$ implies the stability of $(M^{+},M^{+},m^-,m^-)$ for any value 
of the mass ratio $M/m$.

The proof follows from the property that the ground state of an even 
Hamiltonian in one-dimensional quantum mechanics is lowered by any odd 
admixture.  If $h(\lambda)=h_\mathrm{even}+\lambda v_\mathrm{odd}$, 
the ground-state energy, $e(\lambda)$, of $h(\lambda)$ cannot  be greater
than the expectation value obtained with the even wave-function corresponding to 
$\lambda=0$, which is, of course, $e(0)$.  An illustration is 
$h(\lambda)=p^{2}+x^{2}+\lambda x$, whose ground state lies at  $e(\lambda)=1-
\lambda^{2}/ 4$.
Here, parity is replaced by charge conjugation \cite{Ada71,Ric93}.
The Hamiltonian for $(M^{+},M^{+},m^-,m^-)$ can be written as 
\begin{equation}\label{4u:eq:split-H}
    \eqalign{
H=&H_\mathrm{S}+H_\mathrm{A}~,\cr
H_\mathrm{S}=&\left({1\over4M}+{1\over 4m}\right)
\left(\vec{p}_1^2+\vec{p}_2^2+\vec{p}_3^2
+\vec{p}_4^2\right)+V~,\cr
H_\mathrm{A}=&\left({1\over 4M}-{1\over 4m}\right)
\left(\vec{p}_1^2+\vec{  p}_2^2-\vec{p}_3^2
-\vec{p}_4^2\right)~.\cr
            }
\end{equation}
The term $H_\mathrm{A}$ is odd under charge conjugation, i.e., simultaneous
$\vec{p}_1\leftrightarrow\vec{p}_3$ and $\vec{p}_2\leftrightarrow4\vec{p}_4$
exchanges. It improves the binding of the system. More precisely, 
the  ground state $E$ of $H$ is lower that the ground state of $H_{\rm S}$ 
alone, which is none other than the Hamiltonian of a rescaled 
positronium molecule with an inverse constituent mass 
$(m^{-1}+M^{-1})/2$, which provides $H_{\rm S}$ with the {\em same} 
threshold as $H$. Thus the stability of $\Ps_{2}$ implies that of 
$(M^{+},M^{+},m^-,m^-)$ for any $M/m$ ratio. The additional binding provided by
the term $H_\mathrm{A}$ is seen on the dimensionless fraction $x$ measuring
the fraction of binding below the lowest threshold.
In the best variational calculation, the fraction $x$, indeed, increases from 
$x\simeq 0.03$ for $\Ps_{2}$ to $x\simeq 0.17$ for $\H_{2}$.
\subsubsection{Minimal extension of the stability domain}
Once the binding energy $E=(1+x) E_\mathrm{th}$ of configurations with 
$m_{1}=m_{2}$ and $m_{3}=m_{4}$
is taken as firmly established by  detailed variational calculations, 
a minimal extension of the stability domain can be derived.

Consider first $(M^{+},m^{+},M^-,m^-)$ configurations as extensions of 
the $\Ps_{2}$ case. One can fix the scale by setting
$M^{-1}=1-y$ and $m^{-1}=1+y$. Then
\begin{equation}
\label{4u:eq:H-y}
H(M^+,m^+,M^-,m^-)={\vec{p}_1^2+\vec{p}_2^2+\vec{p}_3^2
+\vec{p}_4^2\over2}+V 
+{y\over2}\left(-\vec{ p}_1^2+\vec{ p}_2^2-\vec{ p}_3^2
+\vec{ p}_4^2\right)~.
\end{equation}
The reasoning is the same as previously:
as the last term is antisymmetric under simultaneous
$\vec{p}_1\leftrightarrow\vec{p}_2$
and $\vec{p}_3\leftrightarrow\vec{p}_4$  exchanges, it lowers the
ground-state energy.  Thus,
\begin{equation}
\label{4u:eq:upper-y}
E(y)\le (1+x) E_{\rm th}(0)~.
\end{equation}
Meanwhile the threshold becomes
\begin{equation}
\label{4u:eq:Threshold-y}
E_{\rm th}(y)=-{1\over4(1-y)}-{1\over 4(1+y)}=
{E_{\rm th}(0)\over 1-y^2}~.
\end{equation}
Thus stability remains at least as long as $y^2\le 1-(1+x)^{-1}$,
i.e., for
\begin{equation}
\label{4u:eq:M-over-m}
0.70\le{M\over m}\le 1.43~,
\end{equation}
if one accepts the value $x=0.032$ for $\Ps_{2}$ 
\cite{Var98}. For 
comparison, a recent numerical study \cite{Bre98} show that the 
actual range of stability should be very close to 
\begin{equation}
\label{4u:eq:M-over-m-2}
0.476\le{M\over m}\le 2.1~,
\end{equation}

Now, if one starts from $(M^+,m^+,M^-,m^-)$, and introduces
four different masses $m_i$ such that
\begin{equation}
\label{4u:eq:mass-distribution}
\Eqalign{
m_1&\ge m_2~,\qquad & m_1^{-1}+m_3^{-1}&=2M^{-1}~,\cr 
m_3&\ge m_4~, & m_2^{-1}+m_4^{-1}&=2m^{-1}~,\cr
        }
\end{equation}
one can rewrite the Hamiltonian as
\begin{equation}\label{4u:eq:anotherdec}
    \eqalign{
&H(m_1^+,m_2^+,m_3^-,m_4^-)=H(M^+,m^+,M^-,m^-)\cr
&\quad {}+{1\over4}\left[(m_1^{-1}-m_3^{-1})(\vec{ p}_1^2-
     \vec{ p}_3^2) +(m_2^{-1}-m_4^{-1})
        (\vec{p}_2^2-\vec{p}_4^2)\right]~.   }
\end{equation}
In Eq.~(\ref{4u:eq:anotherdec}), the first term is symmetric under simultaneous
$\vec{p}_1\leftrightarrow\vec{p}_3$ and $\vec{p}_2\leftrightarrow\vec{p}_4$
exchanges, and the second one is antisymmetric. By the same reasoning as
previously, the last term improves the stability of the system, without
changing the threshold energy.
 Equations~(\ref{4u:eq:M-over-m}) and 
(\ref{4u:eq:mass-distribution}) represent a minimal extension of the 
stability domain around $\Ps_{2}$,  based solely on the variational 
principle and the basic symmetries.

One can thus look in the tables \cite{PDG02} for charged hadrons or 
leptons which are long-lived and list many highly-exotic molecules 
which are stable.  One example is the strangeness $-3$ and charm $+1$ 
system $(\Omega^-,\Sigma^-,D^{+},p)$ system, as the constituent masses 
are (in GeV/$c^{2}$) 1.672, 1.197, 1.869 and 0.938, respectively, and 
satisfy Eqs.~(\ref{4u:eq:M-over-m}) and (\ref{4u:eq:mass-distribution}).

Going back to more realistic cases, one can see that Eqs.~(\ref{4u:eq:M-over-m}) 
and (\ref{4u:eq:mass-distribution}) fail by a narrow margin to establish the stability 
of $\Ps\H$ from the binding energy of $\Ps_{2}$.

Suppose now one assumes a binding energy 
$E=(1+x(s))E_\mathrm{th}$ for a hydrogen-like configuration $(M^{+}M^{+}
m^-m^-)$ with mass ratio  $s=M/m$. Introducing 
\begin{equation}
    \label{4u:eq:mi-domain}
    \Eqalign{
 m_{1}^{-1}&=M^{-1}-\delta~,\qquad &m_{3}^{-1}&=m^{-1}-\delta'~,\cr 
 m_{2}^{-1}&=M^{-1}+\delta~,\qquad &m_{4}^{-1}&=m^{-1}+\delta'~,\cr
            }
\end{equation}
 This leads to a decomposition of the Hamiltonian
\begin{equation}
H(m_1^+,m_2^+,m_3^-,m_4^-)=H(M^+,M^+,m^-,m^-)
-{\delta\over2}\,(\vec{ p}_1^2-\vec{ p}_2^2)
     -{\delta'\over2}\,(\vec{p}_3^2-\vec{p}_4^2)~,
\end{equation}
which implies $E(m_{1},m_{2},m_{3},m_{4})\le (1+x(s)) 
E_\mathrm{th}(M,M,m,m)$. Since the threshold for the $\{m_{i}\}$ 
configuration is 
\begin{equation}
    \eqalign{
    E_\mathrm{th}(\{m_{i}\})&=
    {(-1/2)\over M^{-1}+m^{-1}-\delta-\delta'}+
    {(-1/2)\over M^{-1}+m^{-1}+\delta+\delta'}\cr
    &=E_\mathrm{th}(M,M,m,m)/\left[
    1-\left({\delta+\delta'\over
     M^{-1}+m^{-1}}\right)^{2}\right],\cr
            }
\label{4u:eq:thr-mi}
\end{equation}
we have stability at least as long as
\begin{equation}
   \left({\delta+\delta'\over M^{-1}+m^{-1}}\right)^{2}\le 
   1-{1\over 1+x(s)}~.
    \label{4u:eq:stab-from-MMmm}
\end{equation}
Thus if it can be shown that a system $(M^+,M^+,m^-,m^-)$ 
with mass ratio $s=M/m=2$ is bound by a fraction $x(2)$ larger than 
$1/8$, then it follows that $\Ps\H$ is stable.
\subsubsection{Stability for equal-mass negatively-charged particles}
In the Born--Oppenheimer approximation, all molecules such as
$(\p,\d,\e^-,\e^-)$ or $(\p,\Pt,\e^-,\e^-)$ are stable. We have seen that all
hydrogen-like
configurations $(M^{+}, M^{+}, m^-, m^-)$ are stable, even for $M\simeq m$. The
positronium hydride $(\p,\e^{+},\e^-,\e^-)$ is stable. 
For $M\sim m\gg M'$, the 
configuration $(M^{+},{M'}^{+},m^-,m^-)$ is stable, as the lowest 
threshold consists of the stable ion $(M^{+},m^-,m^-)$ and an isolated 
${M'}^-$.

These observations make it reasonable to conjecture that all molecules 
$(m_1^+, m_2^+, m_3^-, m_4^-)$
with equal-mass electrons ($m_{3}=m_{4}$) are stable.  Of course, a 
similar result holds for $m_{1}=m_{2}$. This can be understood as 
follows: the system $(M^{+},{M'}^{+},m^-,m^-)$ has two degenerate 
thresholds, one with $M$ associated with the  first electron and $M'$ 
with the second and another in which $M$ and $M'$ are interchanged. There are
thus two ways of
 describing the system as two deformed atoms with a slow relative 
motion. These two components can interact strongly with each other,
leading to binding. In the 3-body case, the same phenomenon leads to enhanced stability of 
$(M^{+},m^-,m'^-)$  around the symmetry axis $m=m'$.

The conjecture of stability for $m_{3}=m_{4}$ has been checked in 
Ref.~\cite{Var97}.  The authors used two variational calculations.  
The first one is based on the exponential wave function 
(\ref{4u:eq:PsH-Ore-wf2}), the second is the stochastic variational method with
Gaussian wave functions.
As all matrix elements are calculated in a close analytic form, there are no
numerical uncertainties as to the values of the integrals involved, so the 
investigation can be considered as free from  ambiguity.
\subsubsection{Borromean binding}
The concept of Borromean binding is inspired by the Borromean rings, which are
interlaced in such a subtle topological way that if any one of them is removed,
the two other become unlocked. The name ``Borromean binding'' was introduced in
Nuclear Physics \cite{Ban93} to denote those isotopes with two external neutrons
which are stable against dissociation, while the partner with only one external
neutron is unstable. 
The best-known example is the Helium family. The nucleus
$\alpha={}^4\mathrm{He}$ is very stable and compact. The isotope
${}^5\mathrm{He}$ is unstable, but ${}^6\mathrm{He}$ is stable. 
If considered as a three-body system $(\alpha,n,n)$, it has the remarkable property that none 
of the two-body subsystems has a bound state. In other words, with short-range
potentials, 
a three-body system can be bound, with coupling strength that are too weak 
to bind the two-body subsystems. For an investigation of the domain of coupling 
constant where Borromean binding is possible, 
see e.g., \cite{Ric94a,Goy95,Mos00}

Borromean binding is related to the 
``Thomas collapse''  \cite{Tho35},  which occurs when the ratio of
3-body to 2-body  binding energies, $E_3/E_2$, is very large, and to the Efimov effect 
 \cite{Efi70}, which is associated with the existence of very weakly bound 3-body
 bound states together with a vanishing 2-body ground-state energy $E_2$.

Borromean binding is defined to occur in four-body binding as the property 
that all three-body subsystems are unbound. In other words, there is no 
way to build the system by adding the constituents one by one, forming a bound
state at every stage. Of course, for systems such as $(a^{+},b^{+},c^-,d^-)$, 
there  always exist stable two-body subsystems.

Now, Mitroy \cite{Mit00} established that $(m'^+,m^+,m^-)$
 ions remain stable for 
\begin{equation}\label{4u:eq:Mitroy}
0.698\le{m'\over m}\le 1.63~.\end{equation}
On the other hand, Bressanini \etal\ \cite{Bre97} found that 
the $(M^+,m^+,M^-,m^-)$ systems are stable with respect to dissociation 
into $(M^+,M^-) + (m^+,m^-)$ for
\begin{equation}\label{4u:eq:Bressanini}
{1\over 2.2}\le{M\over m}\le 2.2~.\end{equation}
This result is also obtained by Varga \cite{Var99a}.
Thus for $M/m\simeq 2$, the 4-body molecule is bound, but none 
of its 3-body subsystems is stable! This means, for instance, that the molecule 
$(\p,\d,\ap,\ad)$ made up of a
proton, a deuteron and their antiparticles, is Borromean. A minimal extension of
the domain of Borromean binding can be 
established if one knows the binding for $m_1=m_3$ and $m_2=m_4$ \cite{Ric02a}.

\begin{figure}[H]
 \begin{center}\includegraphics[width=.5\textwidth]{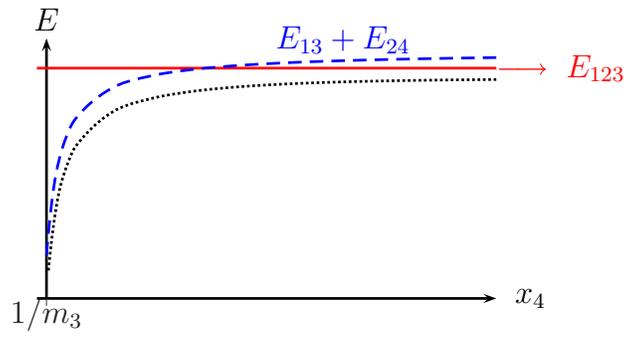}
 \end{center}
 \caption{\label{4u:fig:borro} Guess at the shape of  the four-body energy of $(m_1^+,m_2^+,m_3^-,m_4^-)$ (dotted curve) as a function of $x_4=1/m_4$, from $x_4=1/m_3$ to large $x_4$. The threshold is the lowest of the $(1,3)+(2,4)$ energy  (dashed curve) and the energy  of $(1,2,3)$ (solid line), which is assumed to be stable.}
 \end{figure}

The reciprocal conjecture, stating  that if a four-body systems has at least one stable three-body
 system, then the overall system is stable, remains, however,  to be demonstrated. An argument is the following. Assume $m_3\le m_4$ and suppose that $(1,2,3)$ is stable (if $(1,2,4)$ is stable the stability of $(1,2,3)$ is implied), i.e., with the current notation, $E_{123}<E_{13}$, and consider the system $(m_1^+,m_2^+,m_3^-,m_4^-)$ as a function of $x_4=1/m_4$. For $x_4=1/m_3$, the four-body system is stable, as seen in the previous subsection. For $x_4\to\infty$, the lowest threshold becomes the charged $(1,2,3)$ ion, which attract the charge $m_4^-$. As seen in Fig.~\ref{4u:fig:borro}, it is very unlikely that the curve of the four-body energy would not remain below the lowest threshold between $x_4=1/m3$ and $x_4=\infty$.
%

\markboth{\sl Stability of few-charge systems}{\sl Five-body systems}
\clearpage\section{Systems with five unit charges}\label{se:five}
Calculating the binding energy of a five-particle system
is obviously a very complicated task \cite{Usu99,Var99,Mez01}, especially 
when the system is loosely
bound. A five-particle system can have various bound subsystems 
including $3+2$, $4+1$ or $2+2+1$ decompositions. The binding energy is 
the energy of the system with respect to the lowest 
dissociation threshold. We shall consider the case of systems of
three particles with like charges and two particles with the opposite 
charge. Systems of unit charges where four particles have identical 
charges and the last  particle has an opposite  charge are not expected 
to be bound. 
\subsection {$(m^+,m^+,m^+,m^-,m^-)$}\label{mpmpmpmmmm}
The simplest system consists of five 
particles with equal mass and unit charge. 
In the  case of where the three positively-charged particles are identical fermions with spin 1/2, the antisymmetry requirement restricts the configuration space and no bound
state exists. In particular, the system of three electrons and two positrons, or its conjugate, is not bound.

The $(m^+,m^+,m^+,m^-,m^-)$
system is, however,  bound if the three positively-charged particles are bosons or if one of them is distinguishable, since the Pauli principle does not restrict the allowed states.
 An example is the (e$^+$,e$^+$,e$^-$,e$^-$,$x^\pm$) system,
where $x^\pm$ is a fictitious particle which has the same mass
as the electron but is distinguishable from both the electron
and the positron. No such system exists in the real world. What is of
practical interest is that binding often remains when some masses are changed, 
in particular when $m_{x}$ is increased. For example, a system made of a Ps$_2$ 
molecule and a proton, (e$^+$,e$^+$,e$^-$,e$^-$,p), is stable 
(see Subsecs.~\ref{epepeexp} and \ref{epPsH} below).

The energies of systems of $N$ equal-mass constituents (with mass equal to $m_\e$) are listed
in Table \ref{vkt1}, for both the 
boson and fermion cases. The Ps$^-$ ion and Ps$_{2}$ molecule are well known examples of
such systems.
\begin{table}[H]
\caption{\label{vkt1}Energies of $N$-particle systems of unit charges and equal masses. 
The total charge is 0 and 1 for $N$ even and odd, respectively. 
Atomic units are used. }
\vspace{0.4cm}
\begin{center}\begin{tabular}{ccc}   \hline\hline
$N$  & fermions        & bosons           \\ \hline  
2  & $-$0.250000   & $-$0.250000   \\
3  & $-$0.262005     & $-$0.262005   \\
4  & $-$0.516004   & $-$0.516004   \\
5  & unbound        & $-$0.556489   \\ \hline\hline
\end{tabular}\end{center}
\end{table}

In Table \ref{vkt1}, the energies of the bosonic and fermionic systems
are equal up to $N=4$. In the bosonic case, the particles
are considered to be spinless and the spatial part of the wave
function is symmetric in the coordinates of the identical
particles. In the lowest-energy state of the fermionic system (particles with spin 1/2), 
the spins of the pairs of identical particles are
coupled to zero. In this case the spin part of the wave function
is antisymmetric and the space part therefore has to be symmetric.
Consequently, both the bosonic and fermionic systems have symmetric
spatial parts and their ground state energies are equal.

\subsection{$(M^+,M^+,M^+,m^-,m^-)$}
\label{MpMpMpmmmm}
\begin{figure}[H]
\centerline{\includegraphics[width=.6\textwidth]{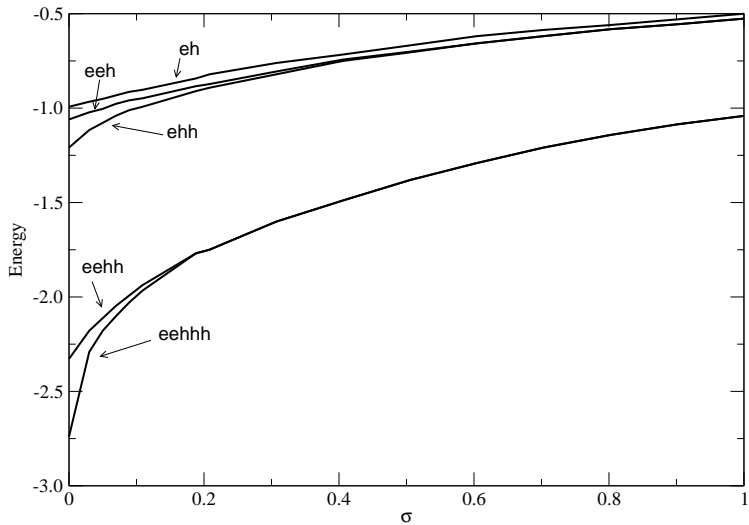}}
\caption{\label{vkf0}The total energies of three-hole and two-electron systems 
as a function of the electron to hole mass ratio $\sigma=m_\rme/m_h$.
The energies of the relevant three- and four-body thresholds are also
shown.}
\end{figure}

A well-known stable Coulombic five particle system
is the H$_3^+$ molecule where the three protons form an equilateral
triangle  sharing  the two electrons. The stability of this system is due to the
slower motion of the heavier particles. If the mass of the
protons were equal to that of the electrons, there would be no 
bound five particle system, as we know from  the previous section. The question
is
 at what heavy/light particle mass ratio is  the stability lost in the  case of
$(M^+,M^+,M^+,m^-,m^-)$. A system of three holes of mass $m_{h}$ and two 
electrons in 
semiconductors is a realistic example of this. The stability of such 
systems has been studied in \cite{Usu99}. The  energy as the 
function of the electron/hole mass ratio is shown in Fig.~\ref{vkf0}.
The system is bound provided that the
positively charged particles are at least five times heavier
than the two with negative charge $(0 < \sigma=m/M <0.2)$ ($m=m_\rme$, $M=m_h$). 
At higher value of the mass ratio, the system dissociates into a 
($M^+,M^+,m^-,m^-$) molecule 
and an isolated $M^+$. Another possible dissociation channel is
($M^+,M^+,m^-$) plus ($M^+,m^-$). However, the energy of this $3+2$ channel
 is always higher than that of the $4+1$.

This study shows that the H$_3^+$ molecule would remain stable
even if the protons were be much lighter. The system consisting of three
protons and two negatively charged muons, i.e.,
$(\p,\p,\p,\mu^-,\mu^-)$ can be mentioned as an exotic example
where $\sigma<0.2$ (see Table \ref{vkt2}). The mass ratio between the
muon and the proton is about $\sigma=0.11$ which is much larger
than that in the hydrogen atom ($\sigma=0.0005$). The energy of
the proton-muon atom is $-92.92$ a.u.
The molecule formed by two proton-muon
atoms is deeply bound just like the hydrogen molecule
[any ($M^+,M^+,m^-,m^-$) system is bound irrespective of the
$M/m$ ratio, as seen in Sec.~\ref{se:4u}]. The binding energy divided by
the reduced mass of the proton--muon atom is 0.07 in
$(\p,\p,\mu^-,\mu^-)$ and 0.02 in $(\p,\p,\p,\mu^-,\mu^-)$.
The corresponding ratios for H$_{2}$ and H$_3^+$
are 0.16 and 0.18, that is the $(\p,\p,\p,\mu^-,\mu^-)$
is much more loosely bound 
than the H$_3^+$ .
The energy of the $(\p,\p,\mu^-)$ ion is $-102.22$ a.u., 
corresponding to $-195.14$ for  $(\p,\p,\mu^-)+ (\p,\mu^-)$, to be compared to 
$ -1999.63$ for the $(\p,\p,\mu^-\mu^-)$ molecule,
an illustration of the fact that the energy of the $3+2$ dissociation
channel is higher than that of the $4+1$ one.
\begin{table}[H]
\caption{\label{vkt2}Energy $E$ of selected exotic five particle systems (in atomic units). 
The energy $E_{\mathrm{th}}$ of the lowest threshold (low.thr.) is also included. 
($m_\p=1836.1527\,m_\e$, $m_d=3670.4827\,m_\e$, 
$m_\t=5496.92158\,m_\e$, 
$m_{\mu}=206.76826\,m_\e$)}
\begin{center}\begin{tabular}{lllll}\hline\hline
 Subsec.       &     System & \ \ $E$   &    low.thr.    & \ \ $E_{\mathrm{th}}$ \\
\hline      
\protect\ref{MpMpMpmmmm} \ \ & $(\p,\p,\p,\mu^-,\mu^-)$ \ \ &
$-203.10453$\ \ & $(\p,\p,\mu^-,\mu^-)$\ \ & $-199.63069$ \ \   \\
\protect\ref{ppeexp} &$(\p,\p,\rme^-,\rme^-,\mu^+)$  &$-1.296583 $&
$(\p,\p,\rme^-,\rme^-)$&$-1.164023$    \\
\protect\ref{Mpxpxpee}&$(\p,\mu^+,\mu^+,\e^-,\e^-)$& $-1.271788$&
$(\p,\mu^+,\e^-,\e^-)$ & $-1.149679$   \\
\protect\ref{Mpxpxpee}&$(\p,\rme^+,\rme^+,\rme^-,\rme^-)$   & $-0.8099127$&
$(\p,\rme^+,\rme^-,\rme^-)$   &$-0.788865$     \\
\protect\ref{Mpxpxpee}&
$(\mathrm{d}^+,\rme^+,\rme^+,\rme^-,\rme^-)$\ \  & $-0.81007844$   &
 $(\mathrm{d},\rme^+,\rme^-,\rme^-)$    &$-0.7890280$  \\
\protect\ref{MpMpeexm}&$(\p,\p,\rme^-,\rme^-,\mu^-)$  & $-102.750286$  &
$(\p,\p,\mu^-,\rme^-)$ &$-102.723336$  \\
\protect\ref{MpMpeexm}&$(\mathrm{d}^+,\mathrm{t}^+,\mu^-,\rme^-,\rme^-)$  &$-111.889612$ &
$(\mathrm{d}^+\mathrm{t}^+,\mu^-,\e^-)$ \ \  & $-111.864106 $       \\
\protect\ref{MpMpMmmmmm}&   $(\p,\p,\ap,\rme^-,\rme^-)$  & $-481.605173$\ \  &
$(\p,\p,\ap,\rme^-)$   & $-481.580324$          \\
\hline\hline  
\end{tabular}\end{center}
\end{table}
\subsection{$(\p,\p,\rme^-,\rme^-,x^+)$}\label{ppeexp}
\begin{figure}[H]
\centerline{\includegraphics[width=0.5\textwidth]{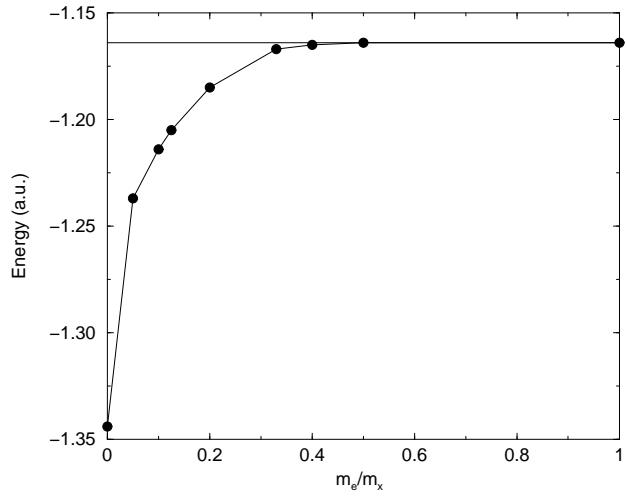}}
\caption{\label{vkf1}Energy of $(\p,\p,\rme^-,\rme^-,x^+)$ as a function of $m_\rme/m_x$.
The dots shows the mass ratios where the energies were calculated.
The horizontal line marks the H$_2$ threshold. Atomic units are used.}
\end{figure}

The difference between this case and the previously discussed
H$_3^+$-like systems is that one of the heavy particle
is different from the other two. 
Figure \ref{vkf1} shows how the total energy varies with $\sigma=m_\e/m_x)$. 
The total energy rapidly decreases toward the
energy of the H$_{2}$ threshold. The system becomes unbound
around $m_x/m_\rme=2.5$. This result shows that an H$_{2}$ molecule
can bind a positively charged particle provided that it is at
least 2.5 times heavier than an electron. So while the H$_{2}$
cannot bind a positron it forms a bound system with a positive
muon $\mu+$ (see Table \ref{vkt2}). The properties of this system are
very similar to those of H$_3^+$: The two protons and the muon form
an isosceles triangle where the two protons are somewhat
closer to each other than to the muon and the electrons
are slightly closer to the protons than to the muon.
If the mass of  $x^+$ approaches the mass of the electron
the distance between the protons and $x^+$ increases and 
the electrons become more and more localized 
around the protons. Eventually, for $m_x /m_\rme>52.5$, the
system dissociates into H$_{2}$ plus $x^+$.
\subsection{$(M^+,x^+,x^+,\rme^-,\rme^-)$}\label{Mpxpxpee}
By adding two positive charges to the H$^-$ ion,  one obtains a
$(M^+,x^+,x^+,\rme^-,\rme^-)$-like system. The stability of this system can
be predicted in the limit of very large  $m_{x}$ by noting that the H$^-$-like 
($M^+,\rme^-\rme^-$) negative 
ion behaves like a negative point charge and can bind 
two positive charges, forming a H$_2^+$-like system. In the limit where $m_{x}$
is very small, ($M^+,\rme^-\,\rme^-$) 
binds two identical light particles
in the same manner as a proton binds two electrons in H$^-$.

\begin{figure}[H]
\centerline{\includegraphics[width=.5\textwidth]{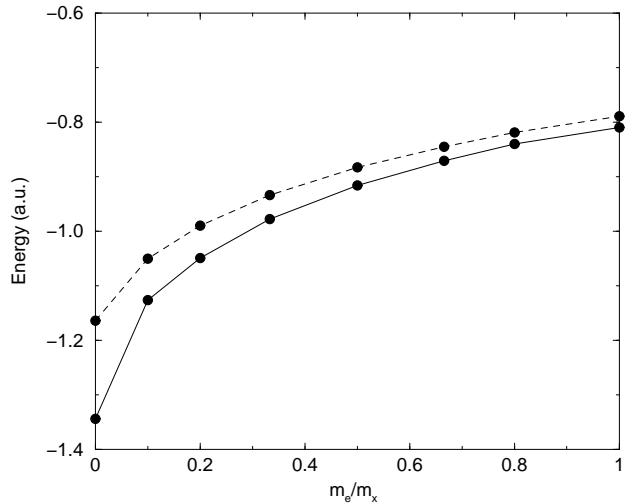}}
\caption{\label{vkf2}Energy of $(M^+,x^+,x^+,\rme^-,\rme^-)$ as a function of $m_\rme/m_x$ 
(solid line). The dashed line shows the energy of the 
$(M^+,x^+,\e^-,\e^-)$ threshold. Atomic units are used.}
\end{figure}

The  $(M^+,x^+,x^+,\rme^-,\rme^-)$ system can dissociate
into $4+1$ [$(M^+,x^+,\rme^-,\rme^-)+x^+$ and
$(x^+,x^+,\rme^-,\rme^-)+M^+$)] and $3+2$ [$(M^+,x^+,\rme^-)
+(x^+,\rme^-)$ and $(x^+,x^+,\rme^-)
+(M^+,\rme^-)$] subsystems.
Figure~\ref{vkf2} shows the binding energies as a function of
$m_\rme/m_x<1$, in the case where the mass of the heavy particle is
equal to the mass of the proton, $M=1836.1527\,m_\e$. For these configurations
with $m_\rme<m_x$, the lowest  threshold is is of the type $4+1$.
Examples of such bound systems are $(\p,\mu^+,\mu^+,\rme^-,\rme^-)$
 or  $(\p,\rme^+,\rme^+,\rme^-,\rme^-)$, see Table \ref{vkt2}. 
 This latter system will be
investigated in detail later. Table \ref{vkt2} shows that the energy of
$(\p,\mu^+,\mu^+,\rme^-,\rme^-)$, just like that of 
$(\p,\p,\mu^+,\rme^-,\rme^-)$ in
the previous example, is close to that of H$_3^+$. The proton and
the muon are likely to form an isosceles triangle but now the
like particles are further away from each other so the base of
the triangle is longer than the sides in this case. The most
important difference is that so long as the $m_\rme/m_x$ ratio lies
between 0 and 1, this system remains bound.
\subsection{$(M^+,M^+,\rme^-,\rme^-,x^-)$}
\label{MpMpeexm}
Another Coulombic five-body system which has attracted attention
is the H$_2^-$ ion. This ion is not bound, but the H$-$H$^-$ potential
energy curve has an attractive part beyond 3.5\,a.u. This 
leads to speculation about the possibility of resonant states of this 
system. The fact that the H$_2^-$ is not bound is a consequence of the
Pauli principle. Adding a negatively charged particle $x^-$  
which has the same mass as the electron (but is distinguishable 
from it) to the hydrogen molecule gives a bound system. 
Its binding energy is about 0.096\,a.u. The $x^-$ particle can attach itself to the 
H$_2$ molecule because the Pauli principle does not constrain its motion. 

\begin{figure}[H]
\centerline{\includegraphics[width=.5\textwidth]{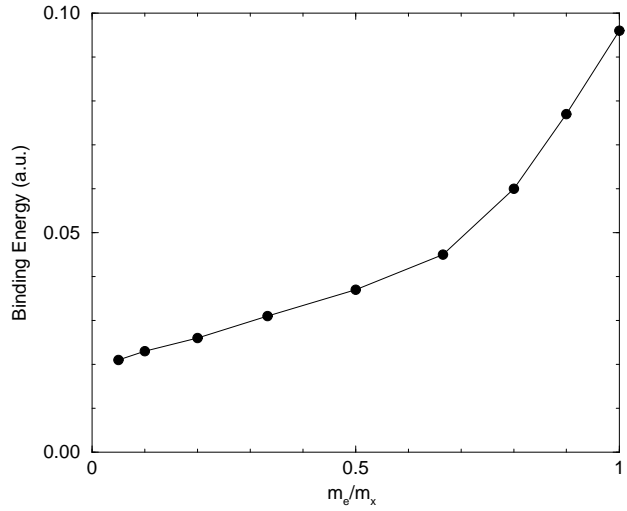}}
\caption{\label{vkf3}Binding energy of $(\p,\p,\e^-,\e^-,x^-)$ 
as a function of $m_\rme/m_x$. Atomic units are used.}
\end{figure}

Figure~\ref{vkf3} shows the dependence of the binding energy on the mass ratio
$m_x/m_\rme$. The threshold in this case is the energy of the 
$(M^+,M^+,\rme^-,x^-)$ four-body system. 
The calculation is not trivial because the energy and
structure very strongly depend on the $m_x/M$ mass ratio. For
$m_x/M\approx 0 $ we practically have a hydrogen molecule. In the case of
$m_x/M\approx 1 $, the $(M^+,M^+,x^-)$ system forms a Ps$^+$ ion-like system.
Due to the heavy masses, the size of this system will be very small compared to
that of Ps$^+$ and this small $(M^+,M^+,x^-)$ system will act as a positive
charge and bind the electron. The distances between the particles
in $(M^+,M^+,x^-)$ will be very small compared to the distance between the
centre of mass of $(M^+,M^+,x^-)$ and the the electron. This system 
can bind one more electron forming 
$(M^+,M^+,x^-,\rme^-,\rme^-)$, which is akin to H$^-$.
A possible choice of $x^-$ is  $\mu^-$. The ($\p,\p,\mu^-$) system is 
bound, and
as the present calculation shows the ($\p,\p,\rme^+,\mu^-$) and 
the ($\p,\p,\rme^-,\rme^-,\mu^-$) systems are also
bound. These systems remain bound even if the masses of the heavy particles
are slightly different, e.g., the $(M_1^+,M_2^+,\rme^-,\rme^-,x^-)$ system is
bound as a rough estimate for $1/3<M_1/M_2<1$. 
\subsection{$(M^+,M^-,m^+,m^-,x^+)$}
\label{MpMmmpmmxp}
The next system considered is $(M^+,M^-,m^+,m^-,m_{x^+})$. The
four-body system $(M^+,M^-,m^+,m^-)$ is akin to the hydrogen-antihydrogen
system and it is not found to be bound if the mass ratio $m/M$ is smaller
than about 0.45, see Sec.~\ref{se:4u}. If the mass ratio $m/M$ is small, the two heavy particle
with opposite charges form a small neutral particle and the ion formed
by the $m^+$, $m^-$and $x^+$ particle will not be able to form a bound 
five-body system with it.
\subsection{$(M^+,M^+,M^-,m^-,m^-)$}
\label{MpMpMmmmmm}
\begin{figure}[H]
\centerline{\includegraphics[width=.5\textwidth]{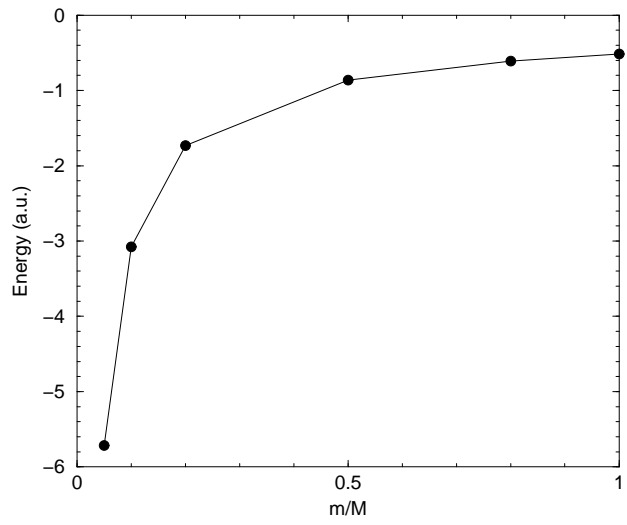}}
\caption{\label{vkf4}Energy of $(M^+,M^+,M^-,m^-)$ as a function of $m/M$ for $m<M$. 
Atomic units are used and it is assumed that $m=m_\rme$, otherwise the energy 
unit should be multiplied by $m/m_\rme$.}
\end{figure}

\begin{figure}[H]
\centerline{\includegraphics[width=.5\textwidth]{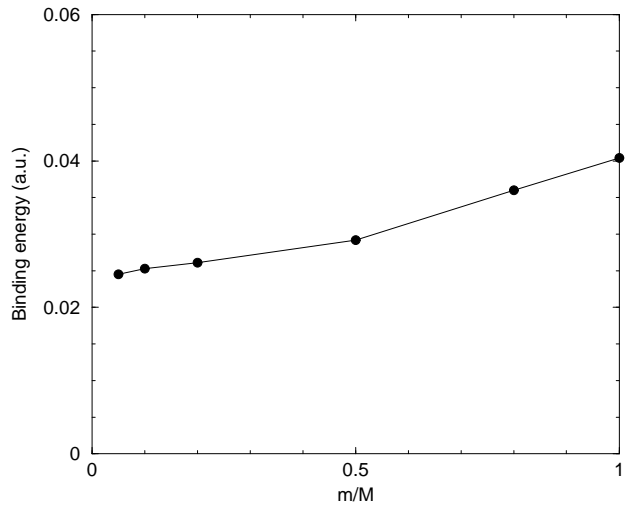}}
\caption{\label{vkf5}Binding energy of   $(M^+,M^+,M^-,m^-,m^-)$ as a function of $m/M$ 
for $m<M$. Atomic units are used and it is assumed that $m=m_\rme$ assumed.}
\end{figure}

\begin{figure}[H]
\centerline{\includegraphics[width=.5\textwidth]{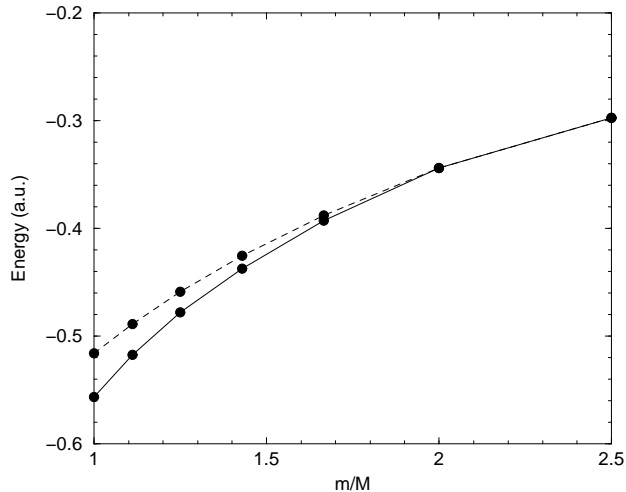}}
\caption{\label{vkf6}Energy of   $(M^+,M^+,M^-,m^-,m^-)$ as a function of $m/M$ for 
$m>M$ (solid line). The dashed line shows the energy of the $(M^+,M^+,m^-,m^-)$
threshold. Atomic units are used
and it is assumed that $m=m_\rme$.}
\end{figure}

This system can be characterized by a single mass ratio $\sigma=m/M$.
If $m<M$ then the dissociation threshold is the energy of the
$(M^+,M^+,M^-,m^-)$ system. The energy of $(M^+,M^+,M^-,m^-)$ as a 
function of $\sigma$ is shown in Fig.~\ref{vkf4}. The $(M^+,M^+,M^-,m^-,m^-)$ 
system is bound with respect to this threshold (see Fig.~\ref{vkf5}). 
$(\p,\p,\ap,e^-,e^-)$ is an example of this system
(see Table \ref{vkt2}).  This shows that a hydrogen molecule 
is capable of binding an antiproton to form a system similar to H$^-$.  
If $m>M$ then the relevant dissociation threshold is given by the energy of 
$(M^+,M^+,m^-,m^-)$. The $(M^+,M^+,M^-,m^-,m^-)$ system is bound in the interval
$1 < m/M < 2 $ (see Fig.~\ref{vkf6}).

There is a very interesting difference between these two cases. In the 
first case $\sigma$ is between 0 and 1. For small $\sigma$ values the 
three heavy  $M$  particles form a positive ion. Its size is small 
and it behaves like a single structureless positively-charged particle $c^+$.
This makes it able to  bind the two lighter charges forming
$(c^+,m^-,m^-)$. 
The resulting system is very similar to H$^-$. 

In the second case $1/
\sigma$
varies between 0 and 1. Here in the limiting case where $\sigma$ is infinite,
one has two heavy $m^-$ particles and a composite positive charge  ``$C^+$''
formed by $(M^+,M^+,M^-)$. This composite particle, however, cannot
be viewed as structureless in the presence of the heavier $m^-$
particles. Energetically it is more favorable to form a $(M^+,m^-)+(M^+,m^-)$
molecule than a $(C^+,m^-,m^-)$ system so the binding is lost at some $\sigma>1$.
\subsection{$(\rme^+,\rme^+,\rme^-,\rme^-,x^+)$}
\label{epepeexp}
The previous examples started from systems with two heavy positive 
and two light negative charges.  The other end of the mass 
spectrum where one has two light positive and two light negative 
charges has also been investigated.  In this case the two negative 
particles were electrons and the two positive particles were 
positrons.  The sign of the charge of $m_x$ is not important 
in this case.  For the calculations reported in this section the
extra charged particle $x^+$ is assumed to be distinguishable
from the electron and the positron.  

\begin{figure}[H]
\centerline{\includegraphics[width=.5\textwidth]{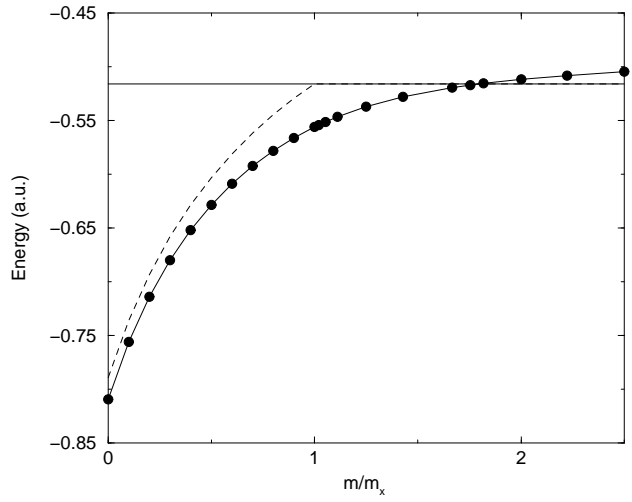}}
\caption{\label{vkf7}Energy of $(m^+,m^+,m^-,m^-,x^+)$ (solid line).
The horizontal line shows the energy of the Ps$_2$ atom, the
dashed line is the energy of the $(x^+,m^+,m^-,m^-)$
threshold.  Atomic units are used and it is assumed that $m=m_\rme$.}
\end{figure}

The five-body binding energy versus the $m_x/m_\rme$ ratio is shown 
in Fig.~\ref{vkf7}.  When the system has a mass ratio satisfying 
$m_x > m_\rme$, the lowest energy threshold is the energy of the
$(x^+,\rme^-,\rme^-,\rme^+) + \rme^+)$ dissociation channel.  This 
system is bound for all mass ratios such that $m_x > m_\rme$
and  it can be seen that that the binding energy of the five-particle system 
increases with increasing $m_x$. 

When the mass of the distinguishable particle is lighter
than that of the electron, i.e., $m_x < m_\rme$, the threshold 
energy is the energy of $\Ps_{2}+x^+$.  The binding
energy decreases steadily as $m_x$ is decreased.  The system
is no longer capable of forming a 5-particle bound state
when $m_x = 0.56 \, m_\e$.  The structure of the 
$(\rme^+,\rme^+,\rme^-,\rme^-,x^+)$ system increasingly 
resembles the structure of a system best described as
$x^+ +\Ps_2$. The dissociation limit is approached as $m_{x} \rightarrow 0.56\,m_\rme$. 
\subsection{The  $\rme^+\Ps\H$ system}\label{epPsH}
\begin{figure}[H]
\centerline{\includegraphics[width=3.5cm]{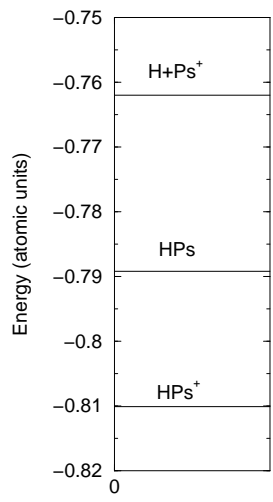}}
\caption{\label{vkf11}Energy levels of the $\H\Ps\rme^+$ 
and the $\H\Ps+\rme^+$ and $\H+\Ps^+$ dissociation channels.}
\end{figure}

The $\rme^+\Ps\H$ system, $(\p,\rme^-,\rme^-,\rme^+,\rme^+)$, corresponds
to the case where $m_x = m_\p$, 
and it is clear from Fig.~\ref{vkf7}
that this system is bound.  The system is stable against
dissociation into the $\H + \Ps^+$, $\p + \Ps_{2}$ or the
$\Ps\H + \rme^+$ channels.  The lowest threshold is the energy of the
$\Ps\H + \rme^+$ channel (0.789197\,a.u.) and  $\rme^+\Ps\H$ is bound by
0.021050\,a.u.\ with respect to this threshold (see Fig.~\ref{vkf11})
A model for this system is  a positron orbiting the $\Ps\H$
subsystem at a relatively large distance from the nucleus.
%

\markboth{\sl Stability of few-charge systems} {\sl Larger systems}
\clearpage\section{Larger systems with bosons or fermions}
\label{se:more}
Molecules are built up of multielectron atoms by various binding mechanisms. The
characteristic property of molecules is that multiple charged heavy atomic
nuclei share a delocalised electron cloud. The question considered in  this
section is the existence of bound states 
of a system made of  $m$ particles of unit positive charge and $n$ of unit
negative charge. 
Such states probably exist in the case of bosons with $\vert m-n\vert$ zero or
small. We have investigated various systems with $m$ electrons and  $n$
positrons, but we were unable to find bound states for $4<m+n<9$. 
Another possibility that we have investigated is a system comprising a proton,
$m$ electrons and $n$ positrons, similar to $\H\Ps$ and $\H\Ps\rme^+$. 

\begin{figure}[H]
\centerline{\includegraphics[width=8.cm]{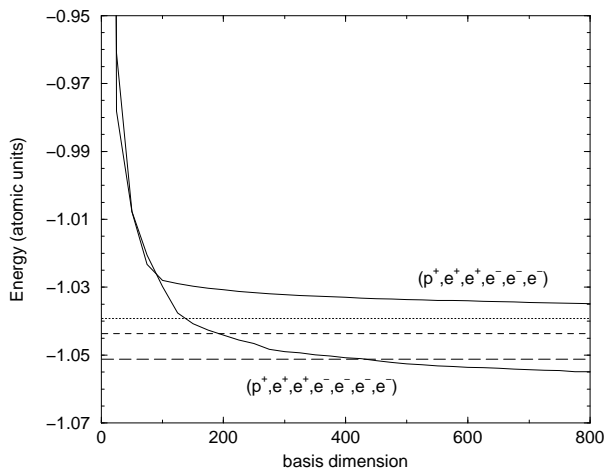}}
\caption{Convergence of the energy of the
(p$^+$,e$^+$,e$^+$,e$^-$,e$^-$,e$^-$) and
(p$^+$,e$^+$,e$^+$,e$^-$,e$^-$,e$^-$,e$^-$)
systems. The dotted line is the $\H\Ps+\Ps$, the dashed line is the
$\H^-+\Ps_2$,
the long dashed line is the $\H\Ps+\Ps^-$ threshold.}
\label{vkf10}
\end{figure}

The $\H\Ps\rme^+=(\H^-,\rme^+,\rme^+)$ is a positively charged system so one may
try to add one more electron to see if it remains stable. The
convergence of the energy is shown in Fig.~\ref{vkf10}. The energy of the system
slowly converges to the lowest $(\H\Ps+\Ps)$  threshold and the size of the
system continuously increases showing that this system is unlikely to be
bound. Surprisingly, however, by adding two electrons to
(H$^-$,e$^+$,e$^+$)
one gets a bound system, as shown in Fig.~\ref{vkf10}. This system ``$\H^-\Ps_2$''
contains a proton, two positrons and four electrons, and can also be
considered as a three-body system made up of a proton and two  Ps$^-$ ions
analogous
to the H$^-$ ion (with the electrons replaced by composite
Ps$^-$ ions).

In the case of neutral, self conjugate,  bosonic systems $(m^+)^n,(m^-)^n$, 
where the mass  can be chosen as $m=1$, a crude variational calculation
\cite{Fle95} 
gives energies $E(3)=-0.789$ and $E(4)=-1.046$.  An improved value $E(3)=-0.820$
was obtained in Ref.~\cite{Var95}. A large-$n$ behaviour $E(n)\propto n^{7/5}$
has been established \cite{Dys67}, and heuristic arguments \cite{Con88}
suggest that
$E(n)\simeq-0.148\,n^{7/5}$. The exact or approximate \cite{Fle95} values
correspond to $E(n)/n^{7/5} = -0.25$, $-0.196$, $-0.169$, $-0.150$ for
$n=1,\ldots,4$, respectively. It is not known whether $E(n)/n^{7/5}$ is expected
to be converge monotonically to the value suggested by heuristic arguments.
If so, it is remarkable that it is so
 close to the suggested limiting value for $n$ as low as 4.
\markboth{\sl Stability of few-charge systems} {\sl Systems containing antiprotons}
\clearpage\section{Systems containing antiprotons}\label{sec:anti}
This review mostly deals with how small atoms and molecules evolve when the
familiar negatively-charged electrons and positively-charged nuclei are replaced
by other elementary constituents. Investigations with positrons, and to a less
extent, pions, muons or  kaons have been carried out for many years,
 and are in the process of further development.  In recent years, however, the
most spectacular progress have been observed in the domain of antiprotons, thanks
to the commissioning of sources of intense and cooled antiprotons, in particular
at the European Laboratory for Particle Physics (CERN). In this section, we
shall review the current state of knowledge in this area and prospects for the
future.

\subsection{Antihydrogen, protonium and other antiprotonic atoms}
\label{anti:sub:proto}
\subsubsection{Antihydrogen atoms}
Antihydrogen, $\aH=(\e^+,\ap)$, is the charge conjugate of the
familiar hydrogen atoms. It has been produced for the first time at
CERN \cite{Wat96,Bau96}, by directing a high-energy antiproton beam onto the
electric field of a nucleus, where it encountered virtual $(\e^+,\e^-)$
pairs. The experiment was successfully repeated at Fermilab
\cite{Bla98}.

Antihydrogen production is 
of considerable current interest on
account of the successful preparation at CERN in 2002 of about 50 000 atoms
of $\aH$ atoms, cooled to within 15 degrees of absolute zero. See Amoretti
\etal~\cite{Amo02}.  This followed on from
several years preparatory work by the ATHENA project (ApparaTus for High
precision Experiments on Neutral Antimatter), Charlton \etal\ \cite{Cha94}
and Holzscheiter \etal\ \cite{Hol97,Hol99}, and the competing ATRAP (Antihydrogen
TRAP) project of Gabrielse \etal\ \cite{Gab01}.  The most recent
results obtained by ATRAP are described by Gabrielse \etal\ \cite{Gab02}.

Now that cold $\aH$ atoms have been prepared, it should be possible to trap
$\aH$ at a sufficiently low temperature that the laboratory frame is essentially is essentially its rest frame.
This would make possible tests of the predictions of two fundamental theories
of modern physics: quantum field theory and Einstein's general theory of
relativity.  In particular, tests could be made of the charge, parity and
time-reversal (CPT or TCP) symmetry of quantum field theory.  See, for example,
Hughes \cite{Hug93},  Schweber \cite{Sch64}, and for a recent 
discussion, Ref.~\cite{Mavromatos:2003qc}.

It follows from the CPT symmetry that a charged particle and its antiparticle
should have equal and opposite charges.  Also their masses and gyromagnetic
ratios should be equal and their lifetimes should be the same.  In addition,
this symmetry predicts that $\Hy$ and $\aH$ should have identical spectra
\cite{Gab01,Cha94,Hol97,Hol99}.  Experimentalists plan to test, as far as
possible, whether $\Hy$ and $\aH$ do have these properties.  In particular,
they intend to compare the frequency of the 1s-2s two-photon transition in
$\Hy$ and $\aH$.

The equivalence principle, according to which all bodies fall at the same rate
in a gravitational field, led Einstein to his general theory of relativity in
which gravitation manifests itself as a metric effect, curved spacetime, as
opposed to the flat spacetime of special relativity in the absence of
gravitation.  See, for example, Weinberg \cite{Wei72} and Hughes \cite{Hug93}.
The exact form of the curvature is determined by Einstein's field equations.

It is proposed to test the validity of the equivalence principle by carrying
out a null red shift experiment in which the frequency of the two-photon
1s-2s transition is observed for $\Hy$ and $\aH$ as both are moved through the
same gravitational field.  Any difference between the two sets of values would
indicate that the two atoms experienced a different gravitational red shift,
which can be shown to be a violation of the equivalence principle \cite{Hug93,
Cha94}.  However, it must be said that the possibility of such a violation is
much more remote when gravitation is viewed from the standpoint of general
relativity as resulting from the curvature of spacetime, rather than as a force
obeying an attractive inverse square law, as in the Newtonian
theory.
\subsubsection{Protonium}
Protonium, $(\p,\ap)$ often denoted by Pn,  is the heavy-mass analogue of the positronium
atom. The 1s energy and radius are of the order of 12\,keV and 50\,m.

When antiprotons are stopped  in a dilute hydrogen target, antiproton capture takes place
in a state of high principal quantum number $n$ of the order of $n=40$. The electron is expelled and
the antiproton quickly cascades down to lower orbits, with circular orbit $\ell=n-1$ preferentially populated. Strong interaction effects are negligible
as long as $n\ge2$. For the 2p state, strong interaction becomes important. States are shifted, but due to the pronounced spin dependence of the nucleon--antinucleon interaction, the singlet state, and the triplet states with $J=2$, 1 or 0 are moved differently.
The annihilation with also becomes dominant with respect to radiation to the 1s ground-state.
If the target is chosen to be more dense, protonium states often penetrate
inside the hydrogen atoms, where they experience a strong electric field which produces a Stark mixing of $(n,\ell=n-1)$ states with states of lower $\ell$, and eventually states with $\ell=0$ or $\ell=1$,
in which annihilation takes place. This is the Day--Sucher--Snow effect, which is important to understand the initial state of annihilation experiment. For a recent review on protonium, and references, see, e.g., Ref.~\cite{Kle02}.

Exotic atoms such as protonium give rise to an interesting quantum-mechanical
problem, in which the Coulomb potential
$-Z/r$ is supplemented by a short-range term $U(r)$, which is sometimes chosen
to  be complex, to mimic the effect of absorption or annihilation.
A naive application of perturbation theory suggests that the energy $E$ changes with respect to the pure Coulomb energy $E_{\mathrm{c}}$ (with wave function $\Psi_{\mathrm{c}}$) by
\begin{equation}\label{anti:eq:naive}
E-E_\mathrm{c}\simeq \int \vert \Psi^{(0)}\vert^2 U(r)\,\d^3 r~,
\end{equation}
which is unsatisfactory, as it gives too large a correction. The shift, indeed,
remains small, even if
$U(r)$ is an infinite hard core. The shift, in fact, is proportional not to the 
strong-interaction potential $U$, but to its scattering length $a$, and this shift is small as long as 
$|a|$ remains small compared to the Bohr radius $R$. More precisely  for the $n$-s state\cite{Kle02},
\begin{equation}\label{anti:eq:Trueman}
E-E_\mathrm{c}\simeq -E_\mathrm{c} {4\over n}\, {a\over R}~.
\end{equation}
\subsubsection{Ordinary antiprotonic atoms}
When an antiproton is captured in the field of a nucleus, it
usually cascades down very quickly toward low-lying orbits close to
the nucleus. The electrons surrounding the nucleus are either
expelled or remain at large distances. Hence, with the remarkable
exception of metastable antiprotonic Helium states, on which more
below, the few-charge problem decouples: within a distance of a few
fermis, or a few tens of fermis, we have a two-body atom consisting
of a nucleus of charge $Z$ and an antiproton; then, at the usual
atomic scale, we have an ion with an effective nucleus of charge
$(Z-1)$ and a few electrons. 

The energy levels of the antiproton are shifted and broadened by strong
interaction. This gives a measure of the antiproton--nucleus
interaction, to be compared with data from antiproton--nucleus
scattering experiment, and theoretical expectations based on
folding the antiproton--nucleon amplitude with the wave-function of
the nucleus. For a review and references, see, for instance,
Ref.~\cite{Bat97}.
\subsection{Three unit-charge systems with antiprotons}
If one combines and antiproton with protons, positrons, and electrons, one gets the following three-body systems.
\begin{itemize}
\item $(\ap,\e^+,\e^+)$ is the conjugate of H$^-$. It could in principle be formed and compared
to H$^-$ to test CPT symmetry,
\item $(\ap,\e^+,\e^-)$ is unstable,
\item $(\ap,\p,\p)$ is stable with Coulomb forces only. Its study could probe the long-range part
of the strong interaction between a proton and and antiproton.
\end{itemize}
\subsection{Antiprotonic helium}
In the limit where the positive charges are very heavy compared to the negative
charges, 
the four-charge configurations $(M^+,M^+,m^-,m'^-)$ reduce to Helium-like atoms,
say $(M^{++},m^-,m'^-)$. The spectroscopy of the He atom 
is well known and documented \cite{Bet77}, and thus will not be discussed
further. The existence of at least one bound state of $(M^{++},m^-,m'^-)$ is
rather intuitive: the $M^{++}$ nucleus can bind the heaviest of $m$ and $m'$, and
there is a long-range Coulomb force left to bind the lightest negative charge.

Let us take first the limit where $M$ is infinite. If $m=m'$, we have the
familiar Helium atom, with the two electrons in the 1s orbit. Indeed, unlike the
case of $\Hy$, for which the correlation 
between the electrons is crucial (see Secs.\ \ref{3u1:sub:Hminus} and
\ref{elem:He}). If, for instance, $m'\gg m$, the $\alpha$ particle and the
particle of mass $m'$ form 
a compact core of charge $+1$, which  forms a hydrogen-like atom with the light
particle $m$. The binding of $m$ and that of $m'$ occur at different scales, and
there is almost no three-body effect for this ground state.

Kondo, and Russel, inspired by anomalies in early data on kaonic atoms, have
shown that the dynamics becomes more intricate for some excited states. For
references, see, e.g.,
\cite{Yamazaki:2002he}. If the heavy negative charge $m'$ is in an excited
state, such that its mean separation from $M^{++}$ becomes comparable to that of
$m$, then the radiative and Auger transitions become suppressed. Experiments
with negative muons $\mu^-$, and later with antiprotons on Helium have revealed,
indeed,
delayed transitions to lower states.

The utility of these long-lived states is even better than might be expected.
 Some of these antiprotonic ``atomcules'' have 
smaller intrinsic width than conventional excited states of ordinary atoms, and thus can serve as a improved
prototypes to define and measure fundamental quantities (antiproton mass or magnetic moment, fine structure constant, etc.).

It is not our aim here to review all properties of these antiprotonic Helium states, 
nor to discuss the elaborate theoretical calculations by
Korobov \etal\  and  Y.Kino \etal\@ We refer the reader to the 
recent review by Yamazaki \etal\ 
\cite{Yamazaki:2002he},
 and to the Proceedings of the LEAP03 conference \cite{LEAP2003}.
 
 In the spirit of our discussions on the how the stability of three- or four-charge 
 systems depends on the involved masses, we note that the metastability of these
new atomcules
 has been studied with $^4\He$ and $^3\He$, and also with muons and
antiprotons.The use of antideuterons \cite{MartinKorobov}. The CERN antiproton
source could be tuned to produce, collect and cool
 antideuterons, at a rate of about $10^{-3}$ that of antiprotons, largely sufficient for capture experiments.
 
 An interesting question is also whether the dramatic metastability observed in
CERN experiments could survive a smearing of the Helium core.
 Suppose one replace $\He^{++}$ by two deuterons: are there metastable states 
 in the four-body molecule $(\Pd,\Pd,\bar\p,\rme^-)$?

\subsection{Four unit-charge systems with antiprotons}
\subsubsection{Symmetry tests}
The antihydrogen molecule $(\ap,\ap,\e^+,\e^+)$, the protonium molecule
$(\ap,\ap,\p,\p)$ and the positronium antihydride $(\ap,\e^-,\e^+,\e^+)$ are examples
of configurations that are stable if the interaction is purely Coulombic. They will only become interesting if they can be used in high-precision
experiments to test the CPT symmetry of quantum field theory \cite{Gab04} or measure
the proton-antiproton interaction at a range of distances at which it cannot
be obtained accurately from a spectroscopic study of protonium.
\subsubsection{$\H\overline{\H}$}

There has been interest in  calculations on the interaction of $\aH$ with
$\Hy$, since the 1970s, Junker and Bardsley \cite{Jun72}, Morgan and Hughes
\cite{Mor73}, Ko{\l}os \etal\ \cite{Kol75}, Campeanu and Beu \cite{Cam83},
Shylapnikov \etal\ \cite{Shl93}.  However, the work on the preparation
of $\aH$ described above has generated greater theoretical interest in
this system,  Armour \etal\  \cite{Arm98,Arm98a,Arm99,Arm01,Arm02,Arm02a}
Jonsell, Froelich \etal\ \cite{Jon99,Fro00,Jon00,Jon01,Jon04,Zyg04,Fro04a}, Sinha and Ghosh \cite
{Sin00}, Voronin and Carbonell \cite{Vor01}.

The $\Hy$-$\aH$ system is a four-body system similar to the $\Htwo$ molecule.
in that it contains two nuclei of the same mass as the proton and two light
particles with the same mass as the electron.  However, both the nuclei and
light particles have equal but opposite charges.  As might be expected, this
leads to very significant differences from $\Htwo$.

One of the most obvious is that if the nuclei coincide in $\Htwo$, the
resulting system is $\He$, whereas when this happens in $\Hy$-$\aH$, the
resulting nucleus has zero overall charge and thus clearly cannot bind the
light particles.

It is reasonable to assume that the critical internuclear
distance $R_c$, below which a fixed proton and antiproton cannot bind
the electron and the positron is greater than $0.639 a_0$.  This can be seen
as follows.  We know from the work on $(\p,\ap,\rme^-)$ described 
in Sec.~\ref{3u1:sub:ppbe}
 and the invariance of the energy levels of this system if the electron is replaced by
a positron that, if $R<R_c$, the electron and the positron would be
unbound if they did not attract each other.  In addition, the attraction
between them makes possible separation into ground state positronium, which is
$-0.25\,$a.u.\ lower in energy.

It might be thought that such a result would be easy to prove.  However, this
does not seem to be the case.  

Ko{\l}os \etal\ \cite{Kol75} showed that $R_c < 0.95\,\mathrm{a}_0$, using
the Rayleigh--Ritz variational method and a trial function containing
Hylleraas-type functions.  Armour \etal\  \cite{Arm98a} went on to show
$R < 0.8\,\mathrm{a}_0$, a fixed proton and an antiproton can bind an electron and 
a positron.   They used a method similar to Ko{\l}os \etal, but included
a basis function to represent weakly bound Ps, a long way from the nuclei.
More recently, Strasburger \cite{Str02} has shown that $R_c < 0.744\,\mathrm{a}_0$,
using the same variational method but with a basis set made up of explicitly
correlated Gaussian functions, similar in type to those used by Kamimura
\cite{Kam88} in his calculation of the energy levels of (d,$\mu^-,$t).

The lowest continuum threshold of $\Hy$-$\aH$ is $\Pn$ + $\Ps$, where both are
at rest in their ground states, at infinite separation.  Note that we can obtain
continuum states with any angular momentum, as close in energy as we please to
this threshold, by associating the angular momentum with the motion of the
centre of mass of either the $\Pn$ or the $\Ps$.  The threshold has a very low
energy, $-459.29\,$a.u.  This is due to the very low energy of the ground state
of Pn.  For $\Pn$ in this state, the expectation value of
$R$ is $0.0016\,\mathrm{a}_0$, which is far below the value of $R_c$.  In view of these
two properties of $\Pn$ + $\Ps$, it is very unlikely that a bound state
of $\Hy$-$\aH$ exists, though this has not so far been proved rigorously \cite{Ric94}.

More definite evidence for the absence of an $\Hy$-$\aH$ state has been obtained
by Bressanini \etal~\cite{Bre97} using the variational Monte Carlo method (VMC)
\cite{Bre02}.  They investigated the stability of systems made up of two pairs
of particles ($M^+,\,M^-$) and ($m^+,\,m^+$), where each pair has the same mass
and charge $\pm e$, where $e$ is the charge on the proton.  According to the
VMC method, if $M$ is chosen so that $M\ge m$, then no bound state
of the system exists if $M/m \ge 1.7$.  Using the diffusion quantum Monte Carlo
(DMC) method, they obtain $M/m\ge 2.2$.  These results are likely to be
accurate but are not rigorous.  However, in view of the fact that $M/m = 1836$
for $\Hy$-$\Ha$, it seems incredible that there should be a bound state in this
case.  The high value of the mass ratio strongly suggests that $\Hy$-$\Ha$
must lie far inside the domain of instability of systems of the above type.

\subsection{Antihydrogen--Helium}
We have seen that in the case of $\HyaH$ if the nuclei are fixed then
there is a critical value $R_c$, of the internuclear distance, $R$,
of the order of a Bohr radius, below which the nuclei are unable to bind the
electron and the positron.  The situation is different in the case of
$\HeaH$ \cite{Arm01}.  This is because the He nucleus has charge +2.  Thus
the nuclear potential is Coulombic asymptotically, rather than dipolar as in
the case of $\HyaH$. It is attractive for the electron and repulsive for the
positron.

At $R=0$, $\HeaH$ corresponds to positronium hydride ($\PsHy$).  This is
known to have a bound state with binding energy, 1.06\,eV \cite{Ho86}.
There are continuum thresholds for $\HeaH$ corresponding to antiprotonic
helium + positronium, $\He^+\ap$+Ps, and $\Heap$ + a positron ($\rme^+$), as well
other thresholds involving more extensive decomposition of $\HeaH$.

At $R=0$, $\He^+\ap$ corresponds to $\Hy  + \Ps$ and $\Heap  +{} $a positron to
$\H^- + ${} a positron.  They have energies of $-20.41\,$eV and $-14.45\,$eV,
respectively, with respect to completely separated systems.  Thus the
the continuum threshold involving antiprotonic helium is lower at $R=0$.
However, as $R\to\infty$,
\begin{equation}
\He^+\ap + \Ps \to \He^+  +  \ap  + \Ps~,
\end{equation}
whereas
\begin{equation}
\Heap  +  \e^+  \to  \He  +  \ap  + \e^+~.
\end{equation}
In this limit, the continuum threshold associated with the antiprotonic helium
plus positronium has an energy of $-61.2\,$eV, whereas the threshold associated
with $\Heap$ has an energy of $-79.0\,$eV.

It follows that  these two thresholds must cross at some point between $R=0$
and $R= \infty$.  As the Schr\"odinger equation for the Born-Oppenheimer potential
of $\He^+\ap$ helium is separable in prolate spheroidal coordinates,
the threshold energy for $\He^+\ap$ + Ps can be calculated very accurately
\cite{Shi92}.  The Born-Oppenheimer potential for $\He\ap$ can be calculated
accurately using the Rayleigh-Ritz variational method.  The results obtained
for both thresholds indicate that crossing occurs in the vicinity of 
$R=~1.25\,\mathrm{a}_0$ \cite{Arm04}.
A very accurate Born-Oppenheimer potential has been obtained by
Strasburger and Chojnacki \cite{Str02a}, using  the Rayleigh-Ritz variational
method and a basis set made up of 768 explicitly correlated Gaussian
functions. Comparison of the value of this potential with the energies of the
two continuum thresholds considered above, shows that the helium nucleus
($\alpha$) and the antiproton are able to bind the electrons and the positron
at all $R$ values.  This is in contrast with the situation in  the case of
$\HyaH$ in which the electron and the positron become unbound if $R$ is less
than the critical value, $R_c$.

In fact, of course, the nuclei are not fixed.  We have seen that in the
case $\HyaH$ the ground state energy of protonium is very low and the
expectation value of $R$ in this state is very small, so small that is
incredible that the dipole formed by the proton and the antiproton would be
able to bind the electron and the positron.

The ground state of $(\alpha,\ap)$ has an energy of $-2938.2\,$a.u.\ and the
expectation value of $R$ in this state is $0.0005\,\mathrm{a}_0$, both of which are
considerably lower than the corresponding values for $\HyaH$.  However, as
noted earlier, there is a key difference between the two systems: whereas Pn
forms a dipole, $\alpha$ + $\ap$ has a net charge of $+1$.  Thus, it is very
close to being a single particle with the same charge as the proton but a
much larger mass.

As mentioned earlier, $\Hy$ + Ps is known to form a bound state, positronium
hydride ($\Ps\H$).  If the inclusion of the three leptons only perturbs the
ground state of the nuclei to a small extent, so that there is a state
of $\Heap$ in which the $\alpha$-nucleus + $\ap$ are in a state that differs
only slightly from the ground state of these nuclei on their own, then it
can be expected that $\He\aH$ will have a ground state whose leptonic wave
function is very similar to that of $\Ps\H$. It would thus have a bound state.
Furthermore, removal of a positron from the system in this state would give
rise to $\Heap$ in a state with electronic wave function very similar to
$\Hy^-$ in its ground state, which we have seen earlier is a bound state.
Thus, if the presence of the leptons has only a small effect on the
nuclear ground state, both $\HeaH$ and $\Heap$ will form bound states.
It will be of interest to see whether the existence of such states can be
proved. For a recent study of $\HeaH$, see \cite{Sin03}.

\subsection{Perspectives}

Positrons are nowadays commonly used in fundamental and applied physics.
The physics of antiprotons is becoming accessible thanks to the development of sophisticated trapping
and cooling devices.

Light antinuclei ($\ad, \bar{\Pt}, \overline{{}^3\mathrm{He}}$) are currently 
produced in antiprotons factories ($\ad$ was identified in 1965 by Zichichi \etal~\cite{Mai95}). The rate is much lower than for $\ap$, but is is sufficient to
form and study a few atoms.

So, in principle, one could conceive of delicate systems with $\ad$ instead of
$\ap$ or a mixture of $\ap$ and $\ad$, such as the Borromean molecule discussed in Sec.~\ref{4u:sub:domain}.

 \markboth{\sl Stability of few-charge systems}{\sl Stability in two dimensions}
\clearpage\section{Few-charge systems in two dimensions}
\label{se:2D}
Two-dimensional (2D) charged-particle systems are often used as a model
of excitonic complexes confined in semiconductor quantum wells.
When a thin layer of a material on the $xy$ plane is sandwiched between 
two layers of another material that has a different dielectric constant, 
electrons and holes  in the middle layer are basically confined by a 
potential  $V_c(z)=0$ if  $\vert z\vert \le d/2$  and $V_c(z)=V_0$ otherwise, where $d$ is the thickness of the middle layer, with a large value of the confining potential $V_0$. Hence every wave-function  approximately  factorises into  a $z$-component which is the ground state of $V_c$, and an horizontal component where the interesting dynamics takes place.  This is why 2D-systems are thoroughly studied, in addition to their intrinsic interest. Some basic properties of bound states in 2D-quantum mechanics are reminded in Appendix~\ref{SO:sub:2D}. 

The most important property of the confined 2D system is the increased binding
energy. This is even more pronounced for large systems than for the small ones of which the dissociation threshold consists, and the binding energy increases as compared to the three-dimensional case, even if measured in units of the threshold energy.

\begin{figure}[H]
\centerline{\includegraphics[width=.6\textwidth]{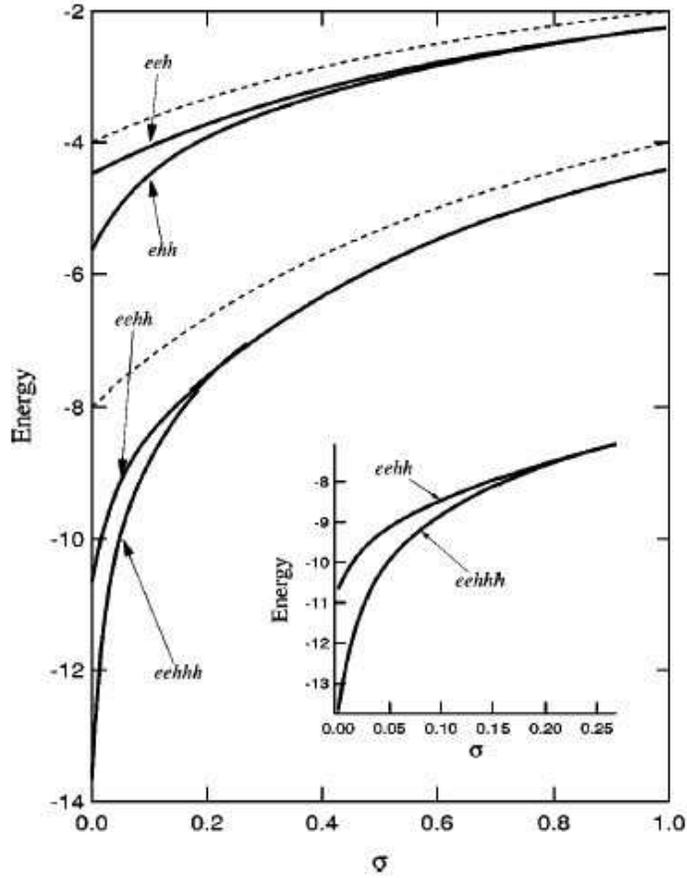}}
\caption{\label{vkf8}The total energies of 2D electron-hole systems 
as a function of the electron hole mass ratio $\sigma=m_\rme/m_\rmh$.}
\end{figure}

The binding energies of some 2D and 3D electron--hole systems are compared 
in Table~\ref{vkt3}. 
\begin{table}[H]
\caption{\label{vkt3}Energies and binding energies of the 2D and 3D excitonic
complexes in the hydrogenic $(\sigma=0)$ and positronium $(\sigma=1)$ limits.
$\sigma=m_\rme^*/m_\rmh^*$ is the  ratio of   electron and   hole effective
masses. An asterisk indicates that the state  is found to be unbound. The unit
of energy is the excitonic Rydberg.}
\begin{tabular}{lcccccccc}
\hline\hline
 & \multicolumn{4}{c}{2D} & \multicolumn{4}{c}{3D} \\
\cline{2-5}\cline{6-9}
System  &$E\, (\sigma=0)$ &$E\, (\sigma=1)$ & $B\, (\sigma=0)$ & $B\,
(\sigma=1)$
 &$E\, (\sigma=0)$ &$E\, (\sigma=1)$ & $B\, (\sigma=0)$ & $B\,
(\sigma=1)$ \\
\hline
eh      & $-4.000$     & $-2.000$   & $4.000$     & $2.000$&
$-1.000$     & $-0.
500$   & $1.000$     & $0.500$ \\
eeh     & $-4.480$     & $-2.242$   & $0.480$     & $0.242$&
$-1.055$     & $-0.
524$   & $0.055$     & $0.024$ \\
ehh     & $-5.639$     & $-2.242$   & $1.639$     & $0.242$&
$-1.204$     & $-0.
524$   & $0.204$     & $0.024$ \\
eehh    & $-10.66$     & $-4.385$   & $2.660$     & $0.385$&
$-2.349$     & $-1.
032$   & $0.348$     & $0.032$ \\
eeehh  & $  *   $     & $ *    $   & $  *  $     & $ *   $& $  *
$     & $ *
   $   & $  *  $     & $ *   $ \\
eehhh  & $-13.65$     & $ *    $   & $2.992$     & $ *   $&
$-2.687$     & $ *
   $   & $0.338$     & $ *   $ \\
\hline\hline
\end{tabular}\end{table}
The total energies of 2D electron-hole systems 
as a function of the mass ratio
$\sigma=m_\rme/m_\rmh$ are shown in Fig.~\ref{vkf8} (see Fig.~\ref{vkf0} for
the corresponding 3D case). The binding energy
of the 2D system is almost by a factor of 10 larger than that of the
corresponding 3D system.

The binding energy in general decreases by increasing $\sigma$ from
0 to 1, and this trend is especially dramatic in the case of X$_2^+$.
The 2D $(eeh)$, $(ehh)$ and $(eehh)$ (Fig. ~ref{vkf9} 
systems are bound irrespective of
the mass ratio, just as in the corresponding 3D cases.
In 2D and 3D the (eehhh) forms a bound system for small values 
of $\sigma$, while the (eeehh) system is unbound for any values 
of $\sigma$.  The critical value of $\sigma$ is 
\begin{equation}\eqalign{
\sigma_{\rm cr} < 0.27 \quad \hbox{for}\quad 2D~,\cr
\sigma_{\rm cr} < 0.23 \quad \hbox{for}\quad 2D~.}
\end{equation}.

\begin{figure}[H]
\centerline{\includegraphics[width=.6\textwidth]{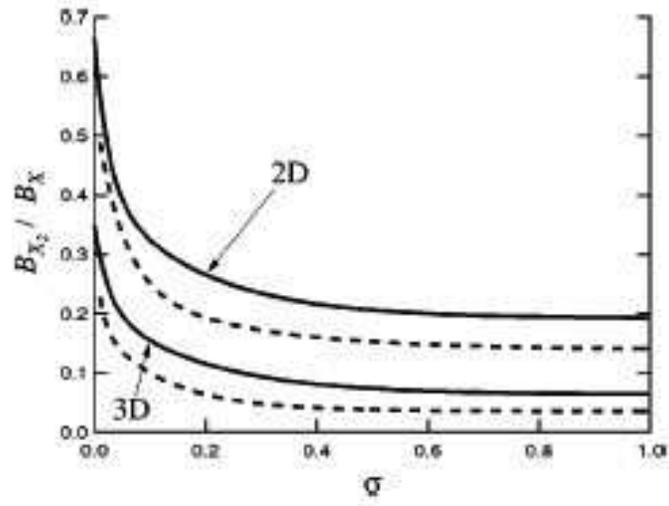}}
\caption{\label{vkf9}The binding energy of the biexciton X$_2$ ($eehh$)
compared to the binding energy of the exciton X as a function of
the mass ratio $\sigma$. The dashed curves refer to a variational
calculation described in Ref.~\protect\cite{Kle83}.}
\end{figure}

The properties of the 2D and 3D systems are found to be generally very
similar.
Even  though the binding energy and therefore the relative distances
between particles are very different in 2D and 3D, the binding energies
show very similar behaviour as a function of $\sigma$ (see e.g., Fig.~\ref{vkf9}).  
It is a striking similarity that the stability of the
($\e,\e,h,h,h$) is lost at nearly  the same value of $\sigma_{\rm cr}$ in 2D and 3D.

We should point out here that the electron-hole complexes are
highly nontrivial quantum mechanical systems in both 2D and 3D. The
relative
motion of the particles is complicated and it is impossible to
model these systems using some rigid geometrical picture, e.g., by assuming
that the ($\rme,\rme,h,h$) forms a static square in 2D. The interpretation
of the ($\rme,\rme,h,h$) molecule as a system of two exciton atoms is also
an oversimplification.
%

\markboth{\sl Stability of few-charge systems}{\sl Conclusions and Outlook}
\clearpage\section{Conclusion and outlook}
%
In this review, we have addressed the question whether a system of point-like charges, interacting through their Coulomb potential, form a collective bound state with all particles remaining at finite distance the ones from the others, or split into clusters or isolated charges.

In a few cases, the answer is obvious: a positive charge $q=+2$ can bind two electrons, since after having attached a first electron, it still exerts a Coulomb attraction on the second one. On the other hand, a tiny positive charge $q\to 0$ cannot bind two electrons.

In most case, the answer depends on the mass ratios. For three unit charges, the positronium ion $(\e^+,\e^-,\e^-)$ and other symmetric systems are stable, as well as nearly symmetric systems, while very asymmetric systems such as $(\e^-,\p,\e^+)$ or $(\p,\ap,\e^-)$ do not form a three-body bound state.

For the case of four unit charges, $(m_1^+,m_2^+,m_3^-,m_4^-)$, there is also a premium to symmetry, since all systems with two identical particles, e.g., $m_3=m_4$, are stable, and among them the twice-symmetric positronium molecule, $\Ps_2=(\e^+,\e^+,\e^-,e^-)$. However, starting from $\Ps_2$, it is observed that breaking charge-conjugation symmetry into $(M^+,M^+,m^-,m^-)$ with $M\neq m$ improves stability, whilst breaking permutation symmetry into $(M^+,m^+,M^-,m^-)$ quickly result into a loss of stability.  The question  is whether symmetry breaking benefits more to the collective state than to its threshold, or vice-versa. 

This problem of stability has been investigated by many authors since the beginning of quantum theory.
Rigorous results have been obtained, with simple or sophisticated methods of mathematical physics. This subject has also stimulated extremely accurate and innovative variational calculations. 

In nuclear physics, there are striking differences between compact, double-magic nuclei, where nucleons are tightly bound together, and halo nuclei with a tail of neutrons extending very far, though in both cases, we are dealing with stable bound states.  
One can similarly notice wide differences among bound states of charged particles. In the case of three-body bound states with unit charges, it is customary to distinguish ``molecular'' from ``atomic'' ions. The former ones, whose prototype is $(m_1^-,m_2^+,m_3^+)=(\e^-,\p,\p)=\H_2^+$, have a large excess of binding below the threshold, easily survive symmetry breaking $m_2\neq m_3$ \cite{Kor04}, and have several excited states. The latter ones, as the hydrogen ion $\H^-=(\p,\e^-,\e^-)$ have not stable excited state, and a small binding energy which quickly disappear if $m_2$ is allowed to differ from $m_3$. The change from atomic to molecular states is analysed in \cite{Kai00}, with a different language, and also identified in \cite{Kor04}.

For four-body systems, there is even a wider variety of situations. While $\Ps_2$ is weakly bound, the $\H_2$ ground-state lies more deeply below its threshold. Some states are well described as two atoms with a relative orbital motion, while others look more as a compact ion surrounded by a light particle.  After the completion of this review, an investigation  of $(m^{+Z},e^+,e^-,e^-)$ stability was investigated by Mitroy and Novikov \cite{Mit04}, with the mass and charge of first particle being varied.

Clearly several aspects deserve further investigations, in particular for systems of more than four particles, for which the available numerical investigations  could be extended and summarised by outlining the shape of the stability domain. 

Many of the results presented here can be extended to cases where the 
potential is not of Coulomb type. Let us give an example. When 
discussing the shape of the stability domain for a 3-body system 
$(m_{1}^{+},m_{2}^-,m_{3}^-)$, we have used the property that in the 
sub-domain with  the lowest threshold $(m_{1}^{+},m_{2}^-)+m_{3}^-$, 
stability survives and is even improved when the mass $m_{3}$ is 
increased while $m_{1}$ and $m_{2}$ are kept fixed. This result holds 
for any 3-body Hamiltonian with a potential that does not depend on 
the constituent masses.

The Coulomb and other power-law ($r^\beta$) potentials provide the 
simplification that the energies scale like $E\propto 
m^{-\beta/(\beta+2)}$, when a factor $m$ is applied to the masses. 
Hence the problem of stability of a $N$-body system depends only on 
$(N-1)$ mass ratios.

There is, however, a major difference between on one hand Coulomb or other 
long-range potentials and on the other hand dipole--dipole potentials 
or sharply decreasing potentials such as the Yukawa potential: the former support 
a countable infinity of bound state however small is the coupling, 
while the latter require a minimal strength to achieve binding. Note that, as
pointed out in Sec.~\ref{3u1:sub:ppbe}, once the minimal strength has been attained, a dipole--dipole potential can support a countable infinity of of bound states.

The transition from short-range to long-range potentials is interesting.
Bressanini \etal\  \cite{Ber04}, for instance, studied how the spectrum of familiar atoms and molecules
evolveo when the Coulomb potential
becomes screened, say $1/r\to \exp(-\lambda r)/r$.  Interestingly, they found that larger systems are better armed to survive screening. For some values of the range parameter $\lambda$, the $\H_2$ molecule remains bound while all smaller systems have lost their stability. This means that the Borromean behaviour denoted for the exotic configuration $(\p,\d,\ap,\ad)$ is already underneath for more familiar ions and molecules. The subject of Borromean configurations with more than four constituents, with or without screening, is one of the many developments to be envisaged in this domain.

\appendix
\markboth{\sl Stability of few-charge systems}{\sl Results on Schr{\"o}dinger operators}
\clearpage\section{Some basic results on Schr{\"o}dinger operators}
\label{se:SO}
We summarise here some results that are often referred to in this review.
\subsection{Variational principle for the ground-state}
\label{SO:sub:var-gs}
The best known formulation reads as follows. For a system with Hamiltonian, $H$, 
the expectation value of the energy of any trial wave function, $\Phi$, is an upper
bound to the ground-state energy, $E_{0}$, i.e., if $\Psi_0$
is  the  normalised ground-state wave-function, then
\begin{equation}\label{SO:eq:var1}
E_0=\langle\Psi_0\vert H\vert\Psi_0\rangle\le E[\Phi]\equiv
{\langle\Phi\vert H\vert\Phi\rangle\over \langle\Phi\vert\Phi\rangle}~.
\end{equation}
This result if often used to prove binding, as to prove this, it is sufficient to obtain
 one wave function $\Phi$ such that $E[\Phi]$ lies below the appropriate threshold. 
 It is also the basis of the 
 Rayleigh--Ritz variational method, where $\Phi$ is parametrised to approach  $\Psi_0$ as 
 closely as possible. See, e.g., Refs.~\cite{Bra83,Pau35}.

The variational principle implies the following stationary property, which is valid for all levels. 
If $\Phi=\Psi_n+\varphi$, where $\varphi$ is small, $\Phi$ will be  ``close'' to $\Psi_n$ and $E[\Phi]$ will differ from   $E_n$ by a term that is second order in the small error function, $\varphi$.

Note that if one separates out the c.m.\ motion, 
say $H=(\sum \vec{p}_{i})^2/\sum 2m_{i}+\hint$, the variational principle 
with square-integrable wave function
 applies to $\hint$, but, in practice $\langle \Phi\vert H\vert \Phi\rangle=
 \langle \Phi\vert \hint\vert \Phi\rangle$, if one deals with a trial 
 function $\Phi$ which depends only upon relative distances.
In short, there is no need to remove explicitly the centre-of-mass motion.
\subsection{Variational principle for excited states}\label{SO:sub:var-ex}
The variational principle keeps the same simple formulation for the lowest level differing from the ground state by any quantum number corresponding to
an exact symmetry. Examples are the first state with angular momentum $J$, or the first state with odd parity, etc. 

As noted, e.g., by Karl and Novikov \cite{Kar95}, one can still use the stationary property to get reasonable estimates for radial excitations. 
For instance, a two-parameter trial-function with a node would give a good idea of the radially excited 
s-wave in a central potential. However, more parameters would lead to a wave function with an almost invisible node near $r=0$ that would  approach the ground-state. In other words, there is no guarantee that the wave function would remain close to the 2s state during the minimisation.

To get an upper bound and achieve converged variational calculation, 
one used the ``minimax'' principle. The $n$-th eigenvalue (in a sector of given conserved quantum numbers) is bound by the largest eigenvalue of the restriction of $H$ to an $n$-dimensional subspace ${\cal H}_n$ of the Hilbert space ${\cal H}$, 
\begin{equation}\label{SO:eq:2}
E_n\le \max_{\Phi\in{\cal H}_n}
{\langle\Phi\vert H\vert\Phi\rangle\over \langle\Phi\vert\Phi\rangle}~.
\end{equation}
Diagonalising the restriction of $H$ to ${\cal H}_n$  leads to an 
upper limit of the $n$ first levels. The parameters defining 
${\cal H}_n$ can be tuned to approximate best a given level, and then used to describe simultaneously several levels. This ensures the orthogonality 
between the approximate wave functions, a  requirement which is 
important for estimating some transition rates.

In practice, ${\cal H}_n$ is often constructed by adding a new term in the set of basis function defining 
${\cal H}_{n-1}$, and hence ${\cal H}_{n-1}\subset {\cal H}_n$.
The Hylleraas--Undheim theorem \cite{Bra83,Pau35} states 
that the eigenvalues of $E_i^{(n)}$ obtained by diagonalising the restriction of $H$ to ${\cal H}_n$
intertwine the $E_i^{(n-1)}$ obtained from ${\cal H}_{n-1}$, 
\begin{equation}\label{SO:eq:HU}
\ldots \le E_i^{(n)}\le E_i^{(n-1)}\le E_{i+1}^{(n)}\le \dots
\end{equation}

A consequence of Eq.~(\ref{SO:eq:2}) is that if there are $n$ linearly-independent approximate eigenfunction with energy  expectation below the threshold energy $E_\text{th}$, 
then the number of bound states, $N_\text{b}$, fulfils
\begin{equation}\label{SO:eq:3} N_\text{b}\ge n~.\end{equation}
\subsection{Comparison Theorem}\label{SO:sub:comp}
If an upper bound on the number of bound states is required, a lower 
rather than an upper bound to the eigenvalues of the system is 
required.  This is, in general, much more difficult to obtain \cite{Spr69}.

The comparison theorem (see, for example, Ref.~\cite{Hil77}) is a very 
useful method for obtaining an upper bound on the number of bound 
states of a system.  This theorem states that if two Hamiltonians 
$\hat{H}_1$ and $\hat{H}_2$ are such that
 \begin{equation}
\langle \Psi | \hat{H}_1 | \Psi \rangle
\le \langle \Psi | \hat{H}_2 | \Psi \rangle~,
 \end{equation}
for any allowed square-integrable function $\Psi$, then
 \begin{equation}
E_{\mathrm{th}}^{(1)} \le E_{\mathrm{th}}^{(2)} \;,
 \end{equation}
where $E_{\mathrm{th}}^{(i)}$ is the energy at which the continuum 
threshold (if any) of $\hat{H}_i$ begins.  Also if $\hat{H}_1$ and 
$\hat{H}_2$ have at least $n$ bound states and the $m$th bound state 
of $\hat{H}_i$ has energy $E_m^{(i)}$,
 \begin{equation}
E_m^{(1)} \le E_m^{(2)} \;\;\;\; m \le n \;.
 \end{equation}
 In addition if $\hat{H}_1$ and $\hat{H}_2$ have the same continuum
threshold, i.e.,
 \begin{equation}
E_{\mathrm{th}}^{(1)} = E_{\mathrm{th}}^{(2)} \;,
 \end{equation}
and $\hat{H}_1$ has exactly $n$ bound states, $n$ is an upper
bound to the number of bound states of $\hat{H}_2$.

 In practical applications of this theorem, $\hat{H}_2$ is taken
to be the original Hamiltonian for the problem, while $\hat{H}_1$
is the Hamiltonian of a more tractable system.
 It follows from this theorem that if $E_{\mathrm{th}}^{(1)} = 
E_{\mathrm{th}}^{(2)}$, and $\hat{H}_1$ has no bound states
then $\hat{H}_2$ also has no bound states.
\subsection{Scaling}\label{SO:sub:scaling}
Consider a typical Coulomb Hamiltonian
\begin{equation}\label{SO:eq:CoulH}
H=\sum_i{\vec{p}_i^2\over 2 m_i}+\sum_{i<j}{q_iq_j\over r_{ij}}~,
\end{equation}
and multiply {\em all} masses by $m$ and {\em all} charges by $q$, and expand all distances
by a factor $\rho$, in particular $r_{ij}\to \rho r_{ij}$.  The kinetic term
gets a factor $m^{-1}\rho^{-2}$ and the potential ones a factor $q^2\rho^{-1}$. 
These factors become equal to an overall energy rescaling by a factor $\epsilon=m q^4$,
if the distances are scaled by $\rho=m^{-1}q^{-2}$.

This scaling can also be understood from the stationary property of the eigenstates. The
expectation value of $H$ has a vanishing derivative at $\rho=1$, if the exact wave function
$\Psi(r_{12},\ldots)$ is used to build a trial wave function 
$\Phi(\rho)=\Psi(\rho r_{12},\ldots)$.

As noted by  Hylleraas \cite{Hyl29}, Fock \cite{Foc30} and many others, 
most basic properties of the exact solutions remain true for variational solutions, 
with the mild restriction that the set of trial wave functions is globally invariant under
 rescaling. This is true for the scaling 
properties of energies and wave functions.

Scaling considerably reduces the parameter dependence of the 
few-body Coulomb problem. In particular, all hydrogenic systems
\begin{equation}\label{SO:eq:hydr}
\hint={\vec{p}^2\over 2m_\mathrm{r}}-{\alpha\over r}~,\end{equation}
have similar energy spectra and eigenfunctions. The properties of a Coulomb 
system containing $N$ particles depend only on $(N-1)$ mass ratios and $(N-1)$ charge ratios.

Scaling laws are easily generalised to other power-law 
potentials $r_{ij}^\beta$. See, e.g., Ref.~\cite{Qui79}.

Partial scaling holds for most other potentials. Consider for instance the spectrum of 
\begin{equation}
\label{SO:eq:Yuk }
\hint={\vec{p}^2\over 2m_\mathrm{r}}-g{\exp(-\mu r)\over r}~.
\end{equation}
This Hamiltonian is proportional to 
\begin{equation}
\label{SO:eq:Yuk-red }
\tilde{h}_\mathrm{int}=-\vec{\nabla}^2-G{\exp(-r)\over r}~,\quad 
G={2m_\mathrm{r}g\over \hbar^2 \mu}~,
\end{equation}
and thus, it is sufficient to know its properties for all but one of the 
parameters $\hbar$, $\mu$, $m_\mathrm{r}$, $g$ set to 1.
\subsection{Virial theorem}\label{SO:sub:Virial}
For a bound state in a Coulomb Hamiltonian (\ref{SO:eq:CoulH}), 
the energy $E=\langle H\rangle$, is the sum of a kinetic term $T=\langle\sum\vec{p}_i^2/(2 m_i)\rangle$ and a potential energy $U=\langle\sum q_iq_i/r_{ij}\rangle$, such that
\begin{equation}
\label{ SO:eq:Virial}
T=-{U\over 2}~.
\end{equation} 
With the same conditions as for scaling, this property extends to variational estimates, 
thus reducing by one the number of parameters to be varied 
(see, e.g., Appendix \ref{se:elem}¥ below).

For other power-law potentials, the rule is $T=\alpha U/2$ \cite{Qui79}. 
For more general potentials, the potential energy $\alpha U$ is 
replaced by the expectation value of $r\Pd V/\Pd r$.
\subsection{Hellmann--Feynman theorem}\label{SO:sub:FH}
Suppose $H$ depends on a parameter $\lambda$ and let $E(\lambda)$ be a particular eigenvalue, with normalised wave function $\Psi(\lambda)$.
Then
\begin{equation}\label{SO:eq:5}
{\partial E\over\partial\lambda}=
\langle\Psi(\lambda)\vert{\partial H\over\partial\lambda}\vert\Psi(\lambda)\rangle~.
\end{equation}
\subsection{Concavity property of the ground state}
\label{SO:su:conc}
We further assume that $H$ depends on $\lambda$ linearly.
Then, the ground-state energy is a concave function of $\lambda$. 
Indeed, the second order derivative, as given by second-order perturbation theory, 
is always negative. Another, simpler, derivation makes use of the 
variational principle \cite{Thi79}.

The result does not hold for individual radial excitations, 
but it is true for the {\em sum} of $n$ first levels.

Again, the result can also be written for variational solutions, 
provided the same set of trial functions is used over the whole range of values
of the  parameter $\lambda$.
\subsection{Symmetry breaking}\label{SO:sub:sym-break}
We often use the property that the ground-state of an Hamiltonian $H$ with 
a certain symmetry has its energy ``lowered'' if  a symmetry-breaking term is added to $H$.

Consider first the one-dimensional 
oscillator (with $\hbar=1$) $h_0=p^2+x^2$, whose ground state 
lies at $\epsilon_0=1$. It becomes $\epsilon(\lambda)=1-\lambda^2/4$ 
if $h_0\to h_0+\lambda x$, as can be seen by direct computation. 

More generally, if 
\begin{equation}
\label{SO:eq:even-odd 1} H(\lambda)=H_\mathrm{even}+\lambda H_\mathrm{odd}~,
\end{equation}
is an even Hamiltonian modified by an odd term, one can use the variational principle 
with the {\em even} ground-state $\Psi(0)$ of $H(0)$ as trial wave function, 
and derive the following relation for the ground-state energy
\begin{equation}
\label{ SO:eq:even-odd2}
E(\lambda)\le\langle\Psi(0)\vert H(\lambda)\vert\Psi(0)\rangle=
\langle \Psi(0) \vert H_\mathrm{even} \vert \Psi(0)\rangle =E(0)~.
\end{equation}
Examples are given in this review where parity is replaced by 
charge conjugation or permutation symmetry. Consider, for instance, an 
exotic Helium-like atom $(\alpha,\mu^-,\pi^-)$,  in the limit 
where $m_\alpha=\infty$. A symmetric $(\alpha,\mu^-,\mu^-)$ or 
$(\alpha,\pi^-,\pi^-)$ system is easily deduced from the 
familiar benchmark $(\alpha,\mathrm{e}^-,\mathrm{e}^-)$, 
by a scaling factor $m_\mu/m_\mathrm{e}$ or $m_\pi/m_\mathrm{e}$. 
The $(\alpha,\mu^-,\pi^-)$ Hamiltonian can be written as
\begin{equation}
\label{SO:eq:amp1 }
H(\alpha,\mu^-,\pi^-)=\overline{m^{-1}}\,\left[{\vec{p}_1^2\over2}+{\vec{p}_2^2\over2}\right]
-{2\over r_1}-{2\over r_2}+{1\over r_{12}}+\delta\left(m^{-1}\right)\left[
{\vec{p}_1^2\over2}-{\vec{p}_2^2\over2}\right]~,
\end{equation}
where $\overline{m^{-1}}=(m_\mu^{-1}+m_\pi^{-1})/2$ and $\delta(m^{-1})=(m_\mu^{-1}-m_\pi^{-1})/2$, and hence the ground state of $(\alpha,\mu^{-1},\pi^{-1})$ is close, but slightly below  a rescaled $(\alpha,\mathrm{e}^-,\mathrm{e}^-)$, i.e.,
\begin{equation}
\label{SO:eq:amp2 }
E(\alpha,\mu^-,\pi^-)\alt E(\alpha,\mathrm{e}^-,\mathrm{e}^-)\,{2 m_\mu m_\pi\over m_\mathrm{e}
(m_\mu + m_\pi)}~,
\end{equation}
the difference being of second order in $\delta(m^{-1})$. 
A similar averaging of inverse masses is adequate  for isospin-breaking 
effects in simple nuclear models or quark models with isospin-conserving forces, 
to account for the mass difference between proton and neutron or between u-quark and d-quark.

A current but not always correct statement is that  ``a symmetry-violating term will push 
the lowest (even) level down and the first excited (odd) state up''. 
This is true if the additional interaction has negligible matrix elements between any of two first levels and the higher states. 
Then  the effect of symmetry breaking can be described by a $2\times2$ matrix,
\begin{equation}
\label{SO:eq:mat22}
\begin{pmatrix}
a&\epsilon\\
\epsilon&b 
\end{pmatrix}~,\end{equation}
and  the sum of eigenvalues (trace) is conserved as the 
mixing strength $\epsilon$ is varied. This is, however, a closed system, 
a condition that  often does not hold for physical examples.
For instance, in the case of $p^2+x^2+\lambda x$, all levels are pushed down when the odd term is switched on.
\subsection{Jacobi coordinates}\label{SO:sub:Jacobi}
\begin{figure}
\centering{
\includegraphics[width=.5\textwidth]{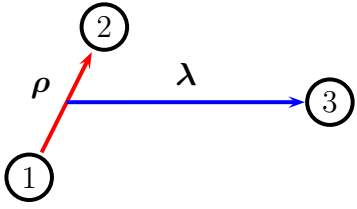} }
\caption{Jacobi variables for three particle systems. }
\label{SO:fig:Jac1} \end{figure}

For two-particles systems, the transformation
\begin{equation}\label{SO:eq:Jac1}\eqalign{%
\{\vec{r}_1,\ \vec{r}_2\}&\to\{\vec{r}=\vec{r}_2-\vec{r}_1,\ 
\vec{R}={m_1 \vec{r}_1+m_2\vec{r}_2\over m_1+m_2} \}~,\cr
\{\vec{p}_1,\ \vec{p}_2\}&\to \{\vec{p}={m_1 \vec{p}_2-m_2\vec{p}_1\over m_1+m_2},\ 
\vec{P}=\vec{p}_1+\vec{p}_2\}~,}
\end{equation}
makes it possible to separate out the motion of the centre of mass, 
\begin{equation}
\label{ SO:eq:Jac2}
{\vec{p}_1^2\over m_1}+{\vec{p}_2^2\over m_2}={\vec{P}^2\over m_{12}}+{\vec{p}^2\over \mu}~,\quad  m_{12}=m_1+m_2~,\quad \mu_{12}^{-1}=m_1^{-1}+m_2^{-1}~.
\end{equation}
For three particles, one can apply this transformation to particles 1 and 2, and 
then to particle 3 and the centre of mass of 1 and 2. This means introducing
\begin{equation}
\label{SO:eq:Jac3}
\vec{\rho}=\vec{r}_2-\vec{r}_1~,\quad 
\vec{\lambda}\propto \vec{r}_3-{m_1 \vec{r}_1+m_2\vec{r}_2\over m_1+m_2}~,\quad \vec{R}={\sum m_i\vec{r}_i\over\sum m_i}~,
\end{equation}
as illustrated in Fig.~\ref{SO:fig:Jac1},
and the conjugate momenta, satisfying the relation
\begin{equation}
\label{SO:eq:Jac4 }
\sum{\vec{p}_i^2\over m_i}={\vec{p}_\rho^2\over \mu_{12}}+{\vec{p}_\lambda^2\over\mu_\lambda}
+{\vec{P}^2\over \sum m_i}~.
\end{equation}
The coefficient of $\vec{\lambda}$ can be adjusted 
to simplify $\mu_\lambda$. For identical particles, 
the choice $\vec{\lambda}=(2\vec{r}_3-\vec{r}_1-\vec{r}_2)/\sqrt3$ 
simplifies the treatment of permutation symmetry.

\begin{figure}
\centering{\includegraphics[width=.7\textwidth]{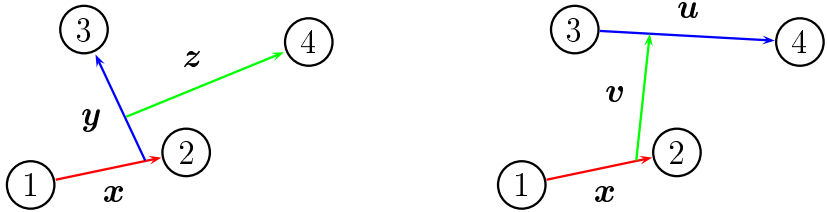} }
\caption{\label{SO:fig:Jac4}Possible choices of Jacobi variables for four particles.}
\end{figure}

For four particles, one can use
\begin{equation}
\label{SO:eq:Jac5 }
\vec{x}\propto \vec{r}_2-\vec{r}_1~,\quad
\vec{y}\propto \vec{r}_3-{m_1\vec{r}_1+m_2\vec{r}_2\over m_1+m_2}~,\quad
\vec{z}\propto \vec{r}_4-{\sum_{i=1}^3 m_i\vec{r}_i\over \sum_{i=1}^3 m_i}~,
\end{equation}
or
\begin{equation}
\label{SO:eq:Jac6}
\vec{x}\propto \vec{r}_2-\vec{r}_1~,\quad
\vec{u}\propto \vec{r}_4-\vec{r}_3~,\quad
\vec{v}\propto {m_3\vec{r}_3+m_4\vec{r}_4\over m_3+m_4}-
{m_1\vec{r}_1+m_2\vec{r}_2\over m_1+m_2}~,
\end{equation}
as illustrated in Fig.~\ref{SO:fig:Jac4}.

A more systematic way of constructing Jacobi coordinates consists first of rescaling $\vec{r}_i=\vec{x}_i/\sqrt{m_i}$, and
$\vec{p}_i=\sqrt{m_i}\vec{q}_i$, so that the kinetic energy 
reads $(\vec{q}_1^2+\cdots+\vec{q}_N^2)/2$. 
Then, any orthogonal transformation leaving invariant the total momentum 
defines a possible choice of Jacobi coordinates.

The choice of the Jacobi coordinates is often dictated by the approximation scheme 
one intend to use. For instance, if a four-body system is described as a weakly 
bound atom--atom  system [(1,2)(3,4)], the choice (\ref{SO:eq:Jac6}) seems 
convenient, with a wave function of the form
$\Psi=\varphi(\vec{x})\tilde{\varphi}(\vec{u})\Phi(\vec{v})$.

Note that for the Gaussian parametrisation described in Appendix \ref{se:VarMet},
\begin{equation}
\label{SO:eq:Jac7 }
\Psi=\sum_a c_a \exp[-\tilde{\mathsf{q}} A^{(a)} \mathsf{q}/2]~,
\quad \tilde{\mathsf{q}} A \mathsf{q}=\sum_{i,j=2}^N A_{ij}\vec{q}_i\vec{q}_j~,
\end{equation}
a change of Jacobi variables can be taken into account
 by a mere transformation of all matrices $A^{(a)}$, 
 and thus should not modify the quality of the approximation.
\subsection{2D binding}\label{SO:sub:2D}
It is well known that an attractive potential ($V\le 0$, but $V$ not always vanishing) 
in one dimension always supports a bound state with $E<0$. 
For instance, one may consider $V$ deeper than a tiny square well $\{ \tilde{V}=0$ for $\vert x-x_0\vert >a$, $\tilde{V}=-V_0<0$ if 
 $\vert x-x_0\vert <a \}$, which is explicitly shown to support at least a bound state, however tiny are the size $a$ and the depth $V_0$.

The same result holds in two dimensions. 
See, e.g., Refs.~\cite{Yan89,Gro97}. 
A more precise statement is that $V$ supports at least one
bound state for $n=1$ and $n=2$ dimensions if $\int V(x) {\rm d}^n x<0$.

A possible way of being convinced that $n=2$ is the critical 
value of the dimension for binding in any attractive potential 
relies on the simple ``delta-shell'' interaction 
$V=-g\delta(\vert \vec{x}\vert -R)$. 
 There is a bound state  if the zero-energy wave function has a node. 

One can choose the zero-energy wave function as $\phi(\vec{x})=1$ for $r=\vert\vec{x}\vert<R$. At  $r=R$, it receives an admixture of the other isotropic solution of
$\Delta \phi=0$, and becomes, for $r >R$,
\begin{equation}
\label{SO:eq:td1 }
\phi(\vec{x})=1-g (r^{2-n}-R^{2-n})~,
\end{equation}
with a logarithmic term in the limit where $n=2$. 
For $n>2$, $r^{2-n}$ has  finite variation between $r=R$ and $r=\infty$, 
and thus requires a minimal strength $g$ to direct $\phi$ toward 
the negative region. For $d\le 2$, the negative region is reached, however weak 
 the strength $g$.

\subsection{Potentials with asymptotic Coulomb attraction}\label{SO:sub:asymCoul}
It is known that any Coulomb potential $V=-g/r$ ($g>0$) applied (we are back in
3D) to a particle of mass $m$ gives an infinite number of bound states, with
energy $E_n=-mg^2/(2n^2)$, where $n=1,2,\ldots$.

Besides $r=0$ and $r=\infty$, the reduced radial wave function has $n-1$
nodes  located at $r=r_i^{(n)}$ ($i=1,\, n- 1$), and it is observed that
$r_{n-1}^{(n)}\to\infty$ as $n\to \infty$.

Now the ($n$ s) state can be considered as the ground state of a potential with 
$V=-g/r$ for $r>r_{n-1}^{(n)}$ and $V=+\infty$ at $r< r_{n-1}^{(n)}$. Hence a
potential with an attractive  Coulomb tail and a hard core of radius $R$ always
supports bound state, since one can arrange $n$ such that $r_{n-1}^{(n)}>R$. 

Therefore, a potential which is anything below $r<R$, and exactly Coulomb beyond
$R$ will support bound states with deeper binding than in the case of a hard
core of radius $R$.

One can thus understand that a potential with a Coulomb tail and an arbitrary
short-range part  decreasing very fast beyond some radius $R$, 
 will support bound states.

 \markboth{\sl Stability of few-charge systems} {\sl Methods for solving the few-body problem}
\clearpage more that\section{Variational methods}
\label{se:VarMet}
The principle of a variational method is to write down an empirical expansion of
the wave function on a set of basis functions involving free parameters and
to adjust the parameters to optimise the results.

In this section we overview the most popular forms of variational 
basis functions used in correlated few-body calculations and compare
their relative efficiency and accuracy on selected few-electron systems.

We also extend our discussion to hyperspherical harmonics, numerical grid,
Monte-Carlo methods, and compare them to conventional variational methods in
terms of convergence and accuracy.
\subsection{Parametrisation of the wave function}
With some appropriate form of the basis functions, 
appropriate (e.g., exponential, harmonic oscillator, Coulombic or Gaussian),
the matrix elements can be calculated analytically. Expanding the wave
function in such a basis leads to an a accuracy of the method that can be
improved almost arbitrarily by increasing the number of basis functions. The
convergence, however, depending on the form of the basis functions,
is often very slow.
\subsubsection{Hylleraas and Exponential basis functions}
For small systems (two- or three-electron systems) traditionally the 
best results have been obtained by variational calculations 
in Hylleraas coordinates in which the trial wave function for the example 
of a two-electron system is written in the following form
\begin{equation}\label{VarMet:eq:Hyl1}
\Psi({\bf r}_1,{\bf r}_2)={\cal A} 
\left\{ {\rm exp}(-\alpha r_1 - \beta r_2)
\sum_{i,j,k}^{i+j+k\le\Omega} a_{ijk} s^i u^j t^k \right\}~,
\end{equation}
where $u=\vert {\bf r}_1-{\bf r}_2\vert $, $t=r_2-r_1$ and $s=r_1+r_2$.
${\cal A}$ is the antisymmetriser, $a_{ijk}$ are the linear,
$\alpha$ and $\beta$ are the nonlinear variational parameters.
The convergence is controlled by $\Omega$. 

The results can be significantly improved by combining linearly 
 two or three wave functions  of the above type with different values of $\alpha$ and $\beta$
\cite{Dra02}. The different values of $\alpha$ and $\beta$  control the
asymptotic, intermediate and short range sectors:
\begin{equation}\label{VarMet:eq:Hyl2}
\Psi({\bf r}_1,{\bf r}_2)={\cal A} \left\{
\sum_{n=1}^{p} {\rm exp}(-\alpha_n r_1 - \beta_n r_2)
\sum_{i,j,k}^{i+j+k\le\Omega} a_{ijk}^{(n)} s^i u^j t^k \right\}~ .
\end{equation}

Pekeris et al.~\cite{Fra66} improved the Hylleraas functions 
by including terms which are essential for the short-range behaviour 
( $(r_1^2+r_2^2)^{1/2}\rightarrow 0)$
\begin{equation}
\Psi({\bf r}_1,{\bf r}_2)={\cal A} \lbrace {\rm exp}(-\alpha r_1 - \beta r_2)
\sum_{i,j,k,l,m} a_{ijklm} s^i u^j t^k(s^2+t^2)^{l/2}({\rm ln} s)^m \rbrace~.
\end{equation}

Thakkar and Koga \cite{Tha94} later introduced an other variant 
of the above ansatz with fractional powers.

A different exponential basis, without the explicit inclusion of 
the powers in the radial coordinates, is also very powerful \cite{Tha77}
\begin{equation}
\Psi({\bf r}_1,{\bf r}_2)={\cal A} \lbrace\sum_{i}^{i\le N} 
a_{i} {\rm exp}(-\alpha_i r_1 - \beta_i r_2-\gamma_i r_{12})\rbrace~,
\end{equation}
where $\alpha_i$, $\beta_i$ and $\gamma_i$ 
are chosen in a quasirandom way, so the calculation is similar 
to a Monte-Carlo calculation with a random distribution of the 
exponential factors.

Korobov \cite{Kor00} significantly improved Thakkar's method by 
introducing complex ${\alpha_i,\beta_i,\gamma_i}$ parameters making the 
variational ansatz more flexible. Frolov \cite{Fro01} used the
similar variational function but carefully optimised the nonlinear 
parameters instead using   random distributions. 

These forms represents a different direction in the computations. In the 
early days when the computational power was very limited, compact 
analytical representations were essential. With the advance of computers
simpler basis functions (but larger basis dimensions) became more
popular. 

The exponential/Hylleraas type functions are very powerful but the analytical
evaluation of the basis functions is complicated or even impossible for more
than two or three electrons. Goldman \cite{Gol98} suggested a quite a different approach:
his idea is to uncouple the correlations by using appropriate new 
variables. To calculate the radial part 
of the matrix elements of the Hamiltonian  one has to integrate 
\begin{equation}
I=\int_0^{\infty} r_1 \d r_1\int_0^{\infty} r_2 \d r_2
\int_{\vert r_1-r_2\vert}^{r_1+r_2} r_{12} f(r_1,r_2,r_{12})\d r_{12}~,
\end{equation}
where $f(r_1,r_2,r_{12})$ comes from the basis functions and Hamiltonian.
By introducing the perimetric variables
\begin{equation}
s=r_1+r_2~,  \quad v=r_{12}/s~, \quad w=(r_1-r_2)/r_{12}~,
\end{equation}
the integral can be rewritten as
\begin{equation}
I=\int_0^{\infty} s^5 \d s\int_0^1 v^2 \d v \int_0^1 (1-v^2w^2) f(s,v,w) \d w~,
\end{equation}
and if the $f$ function has the form
\begin{equation}
f(s,v,w)=S(s)V(v)W(w)~,
\end{equation}
then the integral is fully uncoupled into a products of one-dimensional
integrals. With the help of this trick certain forms of correlated basis
functions can be used for systems with larger number of electrons.

Significant progress has also been made in handling integrals with an exponential on the relative distances for the four-body problem. See \cite{Kin93,Zot02,Har04}, and references therein. This offers an alternative to the Gaussian parametrisation.
\subsubsection{Gaussian and correlated Gaussian basis}
The Correlated Gaussian (CG) of the form
\begin{equation}
G_{A}({\bf r})={\rm exp} \lbrace -{1\over 2} \tilde{{\bf r}} 
A {\bf r}\rbrace =
{\rm exp} \lbrace -{1\over 2} \sum_{i,j=1}^4
A_{ij}{\bf r}_i \cdot {\bf r}_j\rbrace
\end{equation}
is a very popular choice of basis functions 
in atomic and molecular physics. Here $\tilde{\bf r}$ 
stands for a one-row vector whose $i$th element is ${\bf r}_i$. 
The merit of this basis is that  the matrix elements 
are analytically available and unlike other trial functions 
(for example, Hylleraas-type  functions) one can relatively easily 
extend the basis for the case of more than three particles. The well-known 
defects of this basis are that it does not fulfil the cusp condition
and its asymptotics does not follow the exponential falloff. This latter
problem, especially for bound states, can be cured by taking 
linear combinations of adequately chosen CGs. 
The translational invariance of the wave function is ensured by requiring 
that the parameters $A$ fulfil some special conditions. 
These conditions  ensure that the 
motion of the centre-of-mass is factorised in a product form. 

\par\indent
By combining the CG with the spin function parts, the full basis function
takes the form
\begin{equation}
\Phi_{kLS}={\cal A}\lbrace \chi_{SM_S} G_A({\bf r}) \rbrace , 
\end{equation}
with an appropriate spin function $\chi_{SM_S}$, where  
``$k$'' is the index of the basis states and ${\cal A}$ 
is an antisymmetriser for the identical fermions. 

Instead of optimizing the parameters of $A$ it is more advantageous 
to rewrite the equation as
\begin{equation}
{\rm exp}\Big\lbrace -{1\over 2} \sum_{i<j} \alpha_{ij} 
({\bf r}_i-{\bf r}_j)^2-{1\over 2}\sum_{i} \beta_i r_i^2 \Big\rbrace.
\label{basis}
\end{equation}
The relationship between ${\alpha}_{ij},{\beta}_i$ and $A$ is
\begin{equation}
{\alpha}_{ij}=-A_{ij} \ \ \ \ (i\ne j),  \ \ \ \ 
{\beta}_i=\sum_{k} A_{ki},
\label{apara}
\end{equation}
where $\alpha_{ji}\, (i<j)$ is assumed to be equal to $\alpha_{ij}$.
There are two reasons to choose this form. The first is that 
in choosing $\alpha_{ij}$ in this way we deal with a correlation function
between the particles $i$ and $j$, while $A_{ij}$ has no such direct 
meaning and during the optimisation it is more difficult to limit 
the numerical interval of $A_{ij}$ to be chosen from. Secondly, one can
utilize this specific form to make the individual steps of the
parameter selection very fast. 

\subsection{Optimisation of the parameters}

To obtain very precise energy, one has to optimize the variational parameters 
$u_{ki}$ and $A_{kij}$ of the trial function.
The dimension of basis sets is typically between 100 and 1000,
and each basis state has nine nonlinear parameters.  
The optimisation of a function with a few thousands nonlinear parameters cannot be
done efficiently by using a straightforward deterministic optimisation method, since this 
could entail the complete reconstruction of the Hamiltonian matrix and 
diagonalisation every time when some of the nonlinear parameters are altered.
Moreover, the deterministic search for the optimal value of  such a 
large number of parameters is likely to get trapped in a local minimum. 

Different strategies have been proposed. Examples are given below. A comparison of results obtained for the celebrated benchmark case of the He atom is proposed in Table \ref{VarMet:tab:He1} .

\begin{table}[H]
\caption{\label{VarMet:tab:He1} Variational ground state energy of the He atom obtained by
various trial functions.}
\begin{center}\begin{tabular}{rrrl}
\hline\hline
Method   &Ref.         &Basis size&  Energy    (a.u.)                    \\
\hline
Pekeris  &\cite{Fra66} &    246   & $-2.9037243770340$           \\
Thakkar  &\cite{Tha94}&    308   & $-2.9037243770341144$         \\
Drake    &\cite{Dra94}  &   1262   & $-2.90372437703411948$        \\
Goldman  &\cite{Gol98}&   8066   & $-2.903724377034119594$       \\
Korobov  &\cite{Kor00}&   2114   & $-2.903724377034119598296$    \\
Drake    &\cite{Dra02}  &   2358   & $-2.903724377034119598305$    \\
Baye     &\cite{Hes01}   & 18900    & $-2.9037243770340$            \\
\hline\hline
\end{tabular}\end{center}
\end{table}

\subsubsection{Frolov's method}
This author mainly used exponential functions to perform a variety of three-body calculations and some studies of larger systems. His so-called ``two-stage'' strategy is presented, e.g., in Ref.~\cite{Fro98}. Only the first non-linear parameters are optimised exactly. The next terms in the variational expansion are chosen to vary in a regular manner that is described by just a few parameters.
\subsubsection{Kamimura's method}
The group of Kamimura has used Gaussian wave functions in a variety of few-body calculation. A comprehensive description of the method is given in a review article \cite{Hiy03}.  There are two main ideas.
Firstly, in a given variable $r$, the range parameters $\alpha_i$ of $\psi\propto\sum c_i\exp(-\alpha_i r^2/2)$, instead of been independent, are assigned to belong to a geometric series, say $\alpha_{i+1}/\alpha_i=\alpha_i/\alpha_{i-1}=\cdots$. Hence, only the first and last terms have to fitted.
Secondly, for more than two Jacobi variables, say $\vec{x}$ and $\vec{y}$, it is not attempted to write down the most general Gaussian $\exp[-(\alpha \vec{x}^2+2\beta \vec{x}.\vec{y}+\gamma \vec{y}^2)/2]$. Instead, if $Ú\vec{x}$ describes the relative motion inside the $(1,2)$ cluster, and $\vec{y}$ the motion of the third particle relative to this cluster, one seeks at a component
\begin{equation}
\left(\sum_i^N c_i \exp[-\alpha_i\vec{x}^2/2]\right)\left(\sum_j^{N'}d_j \exp[-\beta_j\vec{y}^2/2]\right)~,
\end{equation}
in the wave function, to be supplemented by analogous terms with clusters (1,3) and (1,2). 
The physical interpretation is to construct interferences of several virtual-decay channels.
\subsubsection{Stochastic variational method}

A procedure based on the stochastic search for the best set of nonlinear
parameters can be programmed efficiently  
and is capable of achieving highly accurate results for most 
few-body systems. The essence of the strategy can 
be summarised as follows: Let $\lbrace u_i,A_i \rbrace$ be the nonlinear 
parameters 
of the $i$th basis function out of the set of $K$ such basis functions. 
Then the procedure is
\vskip 0.3 cm
\par\noindent
(1) A succession of different sets of $(\lbrace u_i^{1},A_i^{1} \rbrace,...,
\lbrace u_i^{n_s},A_i^{n_s} \rbrace)$ are generated randomly.
\par\noindent
(2) By solving the eigenvalue problem, the corresponding energies 
$(E^1_i,...,E_i^{n_s})$ are determined.
\par\noindent
(3) The parameter set $\lbrace u_i^{m},A_i^{m} \rbrace$ 
which produces the lowest 
energy is then used to replace the existing $\lbrace u_{i},A_{i} \rbrace$ set.
\par\noindent
(4) The procedure cycles through the different parameter sets 
($\lbrace u_{i},A_{i} \rbrace,i=1,...,K)$, successively choosing different 
sets to minimize the energy until convergence is reached.
\vskip 0.3cm
\par\noindent
The essential reason motivating this strategy is the need to sample 
different sets of nonlinear parameters as fast as possible. The main
advantage is that  it is not necessary to recompute the complete
Hamiltonian nor it is necessary to solve the generalised eigenvalue problem 
from scratch each time a new parameter set is generated. By changing the 
elements of parameter set for each basis function individually, it is 
necessary to recompute only one row (column) of the Hamiltonian and 
overlap matrices each time the parameter set $\lbrace u_{i},A_{i} \rbrace$
is changed. Furthermore, the solution of the generalised eigenvalue 
problem is also expedited since the Hamiltonian matrix is already diagonal
apart from one row and one column. 

A similar strategy to the above was used when adding additional terms to 
the basis. 

The above way of finding the best  parameters is certainly very restricted.
Even this simple method gives very accurate energies, as seen in applications to various fields \cite{Var95, Var98a,Suzuki:1998bn}. More sophisticated 
techniques may give better results in a smaller basis size.
\subsection{Real space approaches}
The representation of wave functions in ``real space'' on a numerical grid (mesh)
is a very simple and powerful technique. In the simplest case, the
basis functions are simple Dirac delta functions
\begin{equation}
f_i(x_j)=\delta_{ij}
\end{equation}
in the coordinate space point $x_j$.
A more sophisticated choice pioneered by Baye et al.~\cite{Hes01} uses
infinitely differentiable Lagrange-Laguerre functions
\begin{equation}
f_i(x_j)=(-1)^i x_i^{1/2}{L_N(x_j)\over x_j-x_i} {\rm exp}(-x_j/2)
\end{equation}
where $L_N(x)$ is the Laguerre polynomial of degree $N$ and $x_i$ is
one of its zeros, i.e., $L_N(x_i)=0$. The 
$f_i(x) f_j(y) f_k(z)$ product of the one dimensional basis functions
can be used to expand the functions of relative motions in an $N$-electron
problem. This representation is very flexible and approximates the exact
wave functions extremely well.

The advantage of the mesh techniques is that  matrix elements can be trivially
calculated. The potential matrix is diagonal therefore it is particularly simple
independently of the form of the potential.
The Hamiltonian matrix is very sparse so powerful linear algebraic
iteration techniques such as Conjugate Gradient method with 
efficient  multigrid preconditioning can be used to solve the
eigenvalue problems.  
\subsection{Hyperspherical harmonics approach}
The method of hyperspherical expansion is widely used in atomic physics, especially in its variant ``adiabatic approximation''. See, e.g., \cite{Mac86,Fro86a,Dec96,Fab01,Kri00}, and  the detailed review by Lin \cite{Lin95} for details and references. In the accurate calculation by Mandelzweig et al., the method is compared  with other variational methods \cite{Kri00}.

Schematically, in the case of three particles, the two Jacobi variables $\vec{\rho}$ and $\vec{\lambda}$ are replaced by an overall distance and a set of five angles,
\begin{equation}
r=(\vec{\rho}^2+\vec{\lambda}^2)^{1/2}~,\quad
\Omega_5=\{\hat{\vec{\rho}},\hat{\vec{\lambda}},\tan^{-1}(\rho/\lambda)\}~.
\end{equation}
A partial wave-function expansion
\begin{equation}
\Psi(\vec{\rho},\vec{\lambda})=\sum_{[L]}{u_{[L]}(r)\over r^{5/2}} \mathcal{Y}_{[L]}(\Omega_5)~,
\end{equation}
in terms of the  generalised harmonics  $\mathcal{Y}$ results into a set of  coupled differential equations for the radial components $u_{[L]}$. Here, $[L]$ denotes the ``grand'' angular momentum $L$ and associated magnetic numbers compatible with rotation and permutation invariance. The convergence is studied as a function of the maximal value of $L$ which is allowed. In practice, the number of equations grows very rapidly as $L$ is increased. Several approximation schemes have been proposed, such as ``potential harmonics'' \cite{Fab01} and ``adiabatic approximation'' \cite{Mac86,Lin95}. An interesting variant is proposed in \cite{Tol95}.
\subsection{Faddeev equations}
The Faddeev equations \cite{Fad61}, first written to describe scattering in momentum space with short-range potentials, have been reformulated in position space, and applied to a variety of problems involving even confining potentials or long-range Coulomb potentials \cite{Mer76,Ric92}, after suitable modifications.

If the Hamiltonian reads
\begin{equation}
H=T+V_1+V_2+V_3~,
\end{equation}
with the kinetic energy $T$ and the pair potentials $V_1=v(r_{23})$, etc., then a decomposition
\begin{equation} \Psi=\Psi_1+\Psi_2+\Psi_3~,
\end{equation}
associated to the set of equations
\begin{equation}(E-T)\Psi_i=V_i\Psi~,\quad i=1,2,3~,
\end{equation}
implies the original Schr\"odinger equation $H\Psi=E\Psi$.
Permutation symmetry is easily implemented by imposing relations among the $\Psi_i$.

In practice, one often used a partial-wave expansion of each Faddeev component $\Psi_i(\vec{\rho},\vec{\lambda})$ to account for all internal orbital momenta $\ell_\rho$ and $\ell_\lambda$ compatible with a given overall angular momentum $J$. One obtains a set of coupled integro-differential equations in $\rho$ and $\lambda$. The convergence is studied by varying the maximal value allowed for  $\ell_\rho$ and $\ell_\lambda$.

Faddeev equations are not primarily aimed at competing to obtain the most accurate values of the binding energy. Their main merit is to produce high-quality wave-functions, accounting for the correlations, even when restricted to the lowest values of  $\ell_\rho$ and $\ell_\lambda$.
\subsection{The Variational Monte Carlo method}
The Variational Monte Carlo method (VMC) is a very powerful numerical
technique that estimates the energy and all desired properties of a
given trial wavefunction without any need to analytically compute the
matrix elements. It poses no restriction on the functional form of the
trial wavefunction, requiring only the evaluation of the wavefunction
value, its gradient and its Laplacian, and these are easily computed.
Using the VMC algorithm, essentially a stochastic numerical integration
scheme, the expectation value of the energy for any form of the trial
function can be estimated by averaging the local energy $H\Psi/\Psi$
over an ensemble of configurations distributed as $\Psi^2$, sampled
during a random walk using the Metropolis \cite{Met53} or 
Langevin \cite{Rey82} algorithms. The fluctuations of the local
energy depends on the quality of the trial functions, and the VMC
can also be used to optimize the trial functions. 

A popular and effective approach to building compact explicitly
correlated wavefunctions is to multiply a determinant wavefunction
by a correlation factor, the most commonly used being the Jastrow factor.
The inclusion of the Jastrow factor is only possible by VMC techniques.

There are various sophisticated trial wave functions have been
implemented and tested. The most simple but typical trial function is
written assuming a Pad\'e factor ${\rm exp}[(ar+br^2)/(1+cr)]$ for the
electron nucleus wave function and a Jastrow factor ${\rm
exp}[a'r/(1+c'r)]$ for the interelectronic part \cite{Ber01}: 

\begin{equation}\label{VarMet:eq:Jas}
\Psi={\cal A}\left\lbrace 
{\rm exp}\left[\sum_i{ar_i+br_i^2\over 1+cr_i}\right]
{\rm exp}\left[\sum_{i<j}{a'r_{ij}\over 1+c'r_{ij}}\right]
\Theta_{SM}\right\rbrace
\end{equation}
where $\Theta_{SM}$ is the spin function. The Pad\'e factor is simple but 
good choice for the nucleus-electron part providing flexibility with 
small number of parameters, representing the coalescence as well as
the decay of the wave function for $r\rightarrow 0$ and $r\rightarrow
\infty$, respectively. 

As an example, we compare in Table \ref{VarMet:tab:Be1} the ground state energy of the Be atom
obtained with this simple VMC ansatz with results of other approaches
including those which use correlated Gaussian basis functions. This simple form does not give very precise energy, but the energy can be improved by using more elaborate correlation factors (an example is shown in the second line of Table \ref{VarMet:tab:Be1}) or by using
diffusion Monte-Carlo method. The main advantage of the VMC is that 
while it is not necessarily the most  precise method for small systems 
its applicable to larger systems with favorable scaling properties. 

\begin{table}[H]
\caption{\label{VarMet:tab:Be1} Variational ground-state  energy of the Be atom obtained by
various trial functions.}
\begin{center}\begin{tabular}{rrrl}
\hline\hline
Method   &\ Ref.         &Basis size&  \ \ \ Energy  (a.u.)            \\
\hline
VMC 1    &\cite{Ber01}    &       &  $-14.6528 $             \\
VMC 2    &\cite{Ber01}    &         &  $-14.6651 $             \\
CG       & \cite{Kom95}                 &    1200      &         $-14.6673$              \\
SVM      &\cite{Suzuki:1998bn}                &        500  &          $-14.6673$               \\
\hline\hline
\end{tabular}\end{center}
\end{table}
\subsection{Short-range correlations}
A quantity of interest for exotic systems involving positrons and antiprotons is the lifetime due to
internal annihilation of a $(\e^+,\e^-)$ or a $(\ap,\p)$ pair. It is proportional to the probability  per unit volume of finding these constituents at zero separation, i.e., to an expectation value of the type
\begin{equation}\label{delta:def}
\delta_{ij}=\langle\Psi\vert\delta(\br_i-\br_j)\vert\Psi\rangle~,
\end{equation}
within the exact wave function $\Psi$, assumed here to be normalised.

Computing $\delta_{ij}$  accurately is much more difficult 
than estimating the binding energy, or an observable that is spread
over the whole range of the wave function \cite{Ho83,Mel99,Kri00}. It often happens
 that an approximate variational wave function is excellent in the region relevant for estimating
the energy $E=\langle H\rangle$, but, for instance, underestimates the wave function
at large distances, and then overestimates its value at large separation, 
in order to remain properly normalised. This is the case when harmonic oscillator wave
functions $\sum c_n H_n(\alpha r) \exp(-\alpha^2 r^2)$, where $H_n$ is a polynomial, are used to describe binding by non-confining forces. 
The Gaussian parametrisation
$\sum \gamma_i \exp(-\beta_i r^2)$ is better in this respect, thanks to the freedom of introducing a variety of range parameters.

Note that if a sequence of approximate normalised wave functions $\Psi_j$ leads to energies $E_j$ 
that converge toward the exact energy $E$, 
it is only guaranteed that, if $\Phi$ denotes the exact normalised wave function,
\begin{equation}
\left\vert\langle\Psi_j\vert\Phi\rangle\right\vert\to 1~,
\end{equation}
but \emph{not} that $\Psi_j\to \Phi$ (modulo a phase) at any point.

To obtain an accurate estimate of $\delta_{ij}$ without pushing too far the variational expansion for the $\Psi_j$, one can use a trick devised by Schwinger (see, e.g., \cite{Qui79}), and further developed for estimating parity-violating effects in atoms.
The basic idea is that if 
\begin{equation}
\langle\Phi\vert A\vert \Phi\rangle = 
\langle\Phi\vert B\vert \Phi\rangle~,
\end{equation}
for the exact wave function, then the sequence
\begin{equation}
B_j=\langle\Psi_j\vert B\vert \Psi_j\rangle~,
\end{equation}
of approximate values might converge much better than the plain 
$A_j=\langle\Psi_j\vert A\vert \Psi_j\rangle$ to estimate
$\langle A\rangle$.

For $N=2$ constituents, if the wave function is written as usual
$\Phi= Y_\ell^m(\Omega) u(r)/r$, with
\begin{equation}
u''(r)-{\ell(\ell+1)\over r^2}u(r)+\mu[E-V(r)]u(r)=0~,
\end{equation}
the Schwinger rule reads \cite{Qui79}
\begin{equation}\label{VarMet:Schwinger}
\delta_{12}={\mu\over 4\pi}\int_0^\infty[V'(r)-2\ell(\ell+1)/r^3]\Pd r~,
\end{equation}
and turns out very powerful. In particular with a linear potential sometimes used for describing
quark confinement, the right-hand side always gives the right answer for $\ell=0$, however poor
is the variational approximation for $u(r)$, provided it is normalised. For $\ell>0$, $\delta_{12}$ vanishes, and the rule
(\ref{VarMet:Schwinger}) links the expectation values of $V'(r)$ to that of $r^{-3}$. In the Coulomb
case, there is, indeed, a well-known identity between $\langle r^{-2}\rangle$ and
$\langle r^{-3}\rangle$.

The generalisation of the Schwinger rule to systems with $N>2$ constituents 
has been written down by Hofmann-Ostenof \etal\ \cite{Hof78} 
and by Hiller \etal\ \cite{Hil78}. For constituents with mass $m=1$, it reads
\begin{equation}
\langle\Phi\vert\Phi\rangle=
{1\over 4\pi}\left\langle\Phi\left\vert\hat{\vec{x}}_1.\nabla_1 V- {2\vec{\ell}_1^2\over x_1^3}\right\vert\Phi\right
\rangle~,
\end{equation}
where $\vec{\ell}_1=\vec{x}_1\times\vec{p}_1$ is the angular momentum carried by the first Jacobi variable
$\vec{x}_1=\vec{r}_2-\vec{r}_1$, and the differentiation $\nabla_1$ should be understood with the other Jacobi variables
$\vec{x}_2, \ldots, \vec{x}_{N-1}$ kept fixed.

Evaluating the right-hand side  to improve the mere reading of 
$\delta_{ij}$ from variational wave functions, was used to study parity-violating 
effects in atoms \cite{Hil78},
or short-range correlations in baryons \cite{Ric92}.
Consider for instance Ps$_2$. A very crude 
approximation of the wave function of Ps$_2$ gives a probability 
$\delta(\e^+\e^-)=1.84\times10^{-2}$
by direct reading, and is corrected into $2.16\times10^{-2}$ \cite{Fle95} by the generalised Schwinger rule, close to the
value $2.19\times10^{-2}$ obtained by refined calculations \cite{Koz93}.

Several variants for calculating $\delta_{ij}$ have been proposed \cite{Dra81}. Perhaps, the method of matrix-element identities works best at the beginning of a variational expansion, to speed up the convergence of $\delta_{ij}$ towards the neighbourhood of the exact value, but does not help much once one is seeking at very high accuracy.  According to \cite{Dra92}, indeed,
``for low-lying states, there is no substitute for direct high-precision calculations of the wave-function near the origin.''

 \markboth{\sl Stability of few-charge systems}{\sl The Born--Oppenheimer approximation}
\clearpage\section{The Born--Oppenheimer approximation}
\label{se:BO}
In 1927, soon after the discovery of quantum mechanics, Born and Oppenheimer
\cite{Bor27} were able to obtain an approximate solution of the Schr\"odinger
equation for a molecule by separating the electronic and nuclear motions.  This was
possible on account of the much larger masses of the nuclei, as compared to the
mass of the electron.  Born and Oppenheimer's treatment can be formally
justified in terms of a perturbation expansion of the exact non-relativistic energy
and wave function in terms of the parameter, $(m_\e/M)^{1/4}$,
where $m_\e$ is the mass of the electron and $M$ is the average mass of the nuclei.

This approximate quantum mechanical treatment of molecules, known as the
Born--Oppen\-heimer (BO) approximation, has always played a central role in the
study of molecular structure.  Descriptions of the BO approximation are given in,
for example, Bransden and Joachain \cite{Bra83} and Pauling and Wilson \cite{Pau35}.

For simplicity, let us consider how the BO approximation is applied to a diatomic
molecule, containing $N$ electrons and two nuclei, $A$ and $B$, with masses
$M_A$ and $M_B$ and charges $Z_A$ and $Z_B$, respectively.  Our treatment can be
extended to more complicated molecules.

The first step in the BO approximation is to fix the nuclei in space with a given $R$ value and
solve the Schr\"odinger equation for the electrons interacting with each other and
with the nuclei through the Coulombic force.  The equation is of the form
\begin{equation}\label{BO:eq:el-eq}
\hat{H}_\text{el} \Psi ( \vec{r};\vec{R})=E(R) \Psi (\vec{r};\vec{R})~, 
\end{equation}
where $\hat{H}_\text{el}$ is the electronic Hamiltonian.  $\Psi (\vec{r};\vec{R})$,
where $\vec{r}\equiv ( \vec{r}_1,\vec{r}_2, \dots, \vec{r}_N)$, is the electronic
wave function and $E(R)$ is the sum of the electronic energy at the given $R$ value and
the energy due to the Coulombic repulsion of the nuclei.

$\hat{H}_\text{el}$ is of the form
\begin{equation}\label{BO:eq:Hel}
\hat{H}_\text{el}=-\frac{1}{2m_e} \sum _{i=1}^N \nabla_i^2 +
\sum _{{i,j=1\atop(i<j)}}^N \frac{1}{r_{ij}}-
\sum _{i=1}^N \frac{1}{r_{iA}} - \sum_{i=1}^N \frac{1}{r_{iB}}+
\frac{Z_A Z_B}{R}~, 
\end{equation}
where $r_{ij}$ is the distance between electrons $i$ and $j$ and $r_{iA}$ and
$r_{iB}$ are the distances of electron $i$ from nuclei $A$ and $B$, respectively.

The electronic wave function, $\Psi ( \vec{r};\vec{R})$, depends parametrically on
the internuclear vector, $\vec{R}$.  Its dependence on the direction of $\vec{R}$ is
easily taken into account.  The $z$-axis of the electronic coordinates is normally
taken to be along $\vec{R}$; the probability of a given orientation is determined at a
later stage with the calculation of the nuclear wave function, $\chi (\vec{R})$.

$E(R)$ and $\Psi (\vec{r},\vec{R})$ are determined as functions of $R$.  Except
in very simple cases such as $\H_2^+$, for which $N=1$, for which equation (\ref{BO:eq:el-eq})
is separable in prolate spheroidal coordinates, only approximate solutions can be
obtained using the Rayleigh--Ritz variational method.  The quality of the approximate
solutions for a given molecule has improved very significantly over the years with
increases in computer power and the ingenuity of quantum chemists.

The nuclei are considered as two particles moving in the central potential, $V(R)$,
where
\begin{equation}\label{BO:eq:pot}  V(R)=E(R)~.  \end{equation}
Thus their centre of mass motion is separated out in the usual way.  The nuclear
wave function, $\chi (\vec{R})$, for the internal motion of the nuclei, is determined
by the nuclear Schr\"odinger wave function,
\begin{equation}
\hat{H}_n \chi (\vec{R})=E_T \chi (\vec{R}) 
\end{equation}
where
\begin{equation}
\hat{H}_n=-\frac{1}{2\mu} \nabla _{\vec{R}}^2 + V(R)~, 
\end{equation}
and
\begin{equation}
\mu = \frac{M_A M_B}{M_A + M_B}  
\end{equation}
is the reduced mass of the nuclei.  $E_T$ is the energy, in the BO approximation,
of the molecule under consideration.  The associated BO wave function,
$\Psi _T ( \vec{r},R)$ is of the form
\begin{equation}\label{BO:eq:wf}
\Psi _T ( \vec{r},\vec{R})=\chi (\vec{R}) \Psi (\vec{r}; \vec{R})~. 
\end{equation}

The most common states to be treated using the BO approximation are the states
associated with the electronic ground state, i.e.\ the solution, (or solutions, in the
event of a degeneracy), of equation (\ref{BO:eq:el-eq}) that corresponds to the lowest minimum
value of $E(R)$, as a function of $R$.  However, excited electronic states can also
be treated.  In principal, it is possible to obtain a complete set of BO solutions of
the form in equation  (\ref{BO:eq:wf}).  This would necessarily involve continuum states.

In the BO approximation the terms in the full Schr\"odinger equation for the
internal motion of the molecule that represent coupling between the nuclear and
electronic motion are neglected.  They are the term containing
\begin{equation}
\vec{\nabla} _R \chi (\vec{R}) \cdot \nabla _{\vec{R}} \Psi (\vec{r};\vec{R})
\end{equation}
and the term containing
\begin{equation}
\chi (\vec{R}) \nabla _{\vec{R}}^2 \Psi (\vec{r}; \vec{R}),
\end{equation}
which result from the operation of $\vec{\nabla}_{\vec{R}}$ on the electronic wave
function.

Also the ratio of the mass of the electrons to the masses of the nuclei is assumed
to be infinite, resulting in their neglect in the centre of mass separation.  This would
have to be corrected in an exact treatment.  Allowance for the large but not infinite
electronic-nuclear mass ratio leads to the introduction of small mass polarisation
terms in the Hamiltonian that couple the momenta of the electrons and also, if
$M_A \neq M_B$, the momenta of the electrons and the nuclei.  See, for example,
Ko{\l}os and Wolniewicz \cite{Kol63}.

A simple improvement to the BO approximation, called the adiabatic approximation,
is to set
\begin{equation}
V(R)=E(R)+A(R)~,
\end{equation}
where
\begin{equation} \label{BO:eq:eqI}
A(R)=\big( \Psi (\vec{r};\vec{R}) \mid - \frac{1}{2 \mu}
\nabla _{\vec{R}}^2 \mid
\Psi (\vec{r}; \vec{R}) \big) . 
\end{equation}
In equation (\ref{BO:eq:eqI}), the round brackets indicate integration with respect to the
electronic coordinates.  See, for example, Ko{\l}os \cite{Kol63} and Messiah \cite{Mes61}.
Further improvements to the BO approximation are considered by Pack \cite{Pac85}.

Calculations of corrections to the BO approximation as applied to $\H _2^+$
and $\H \D^+$ are referenced in Sec.~\ref{3u1:sub:H2+}.  The corrections to the BO
approximation for the ground state of diatomic two-electron molecules have been
calculated by Ko{\l}os and Wolniewicz \cite{Kol63}.  The correction to the energy
is small in comparison with the binding energy of these molecules, 4.74 eV, calculated
using the BO approximation.  This is a consequence of the large value of
$\mu/{m_\e} \geq 918.1$.

However, as pointed out in Sec.~\ref{3u1:sub:muonic}, the BO approximation does not work well
for muonic molecular ions.  In the case of $\d\t \mu$, for example, the ratio of the
reduced mass of the $\d$ and $\t$ to the mass of the electron is 10.7, which is much
smaller than the value of ${\vec{\mu}}/{m_\e}$ in the case of diatomic, two-electron
molecules.

A comparison is made of the binding energies of the five bound states of $dt \mu$
in the table in ref.\ \cite{Arm97} as calculated using the BO approximation, the
adiabatic approximation and by a very accurate variational calculation using the full
non-relativistic Hamiltonian for the internal motion of $\d\t \mu$.  The binding
energies for the most weakly bound state, the $J=1$, $v=1$ state that plays a
crucial role in muon catalysed fusion, obtained by the BO and adiabatic
approximations and the variational method are 9.7, 7.7 and 0.66 eV respectively,
showing the considerable errors that result from the two approximations.

It is of interest to note that the adiabatic representation method referred to in Sec.~\ref{3u1:sub:muonic},
 which was used by Ponomarev and his co-workers \cite{Vin82} to correct the
error in the BO approximation, involved the use of many BO basis functions of the form
in equation (\ref{BO:eq:wf}) in a close-coupling type calculation using the full Hamiltonian for
$\d\t \mu$ referred to above.  For details of applications of refinements of the BO
approximation to the $\t \mu + \D _2$ reaction in the muon catalysed fusion cycle,
see Zeman et al.\ \cite{Zem00}.

Corrections to the BO approximation in the case of $\H \aH$ have been
calculated by Armour et al.\ \cite{Arm99} using the method described by Hunter
et al.\ \cite{Hun66a}.

A very similar approximation to the BO approximation, the adiabatic-nuclei or
nuclear-impulse approximation, can be used to separate the nuclear and electronic
motion in scattering calculations involving atoms and/or molecules.  See, for
example, Chase \cite{Cha56}, Chang and Temkin \cite{Cha69}, Lane \cite{Lan80}.
For applications to positron scattering, see Armour \cite{Arm88}.

\markboth{\sc Stability of few-charge systems}{\sl Elementary calculations}
\clearpage\section{Elementary three- and four-body calculations}
\label{se:elem}
\subsection{Introduction}
\label{elem:intro}
In this Appendix, we present a few basic calculations concerning 
systems of 3 or 4 charges.  Some of them are borrowed from pioneering 
papers that illustrate the ability of quantum theory and 
variational methods to reproduce or even predict the properties of 
systems more complicated than the hydrogen atom.  The analytic 
results presented below, when properly generalised, are  basic 
ingredients of  powerful variational methods presented in  Appendix \ref{se:BO}.
\subsection{Helium and Helium-like atoms with fixed nucleus}
\label{elem:He}
This corresponds to the Hamiltonian
    \begin{equation}
       H={\vec{p}_{1}^{2}\over 2}+{\vec{p}_{2}^{2}\over 2}-
       {Z\over  r_{1}}-{Z\over  r_{2}}+{1\over  r_{12}}~,
        \label{eq:Helium-H}
    \end{equation}
where the electron mass is set to $m=1$. A rescaling is often used to 
set the attraction strength to unity and the repulsion strength to 
$1/Z$, leading to the well-known ``$1/Z$'' expansion. See, for 
instance, \cite{Bak90,Iva95}.

When the repulsion among electrons is neglected, one can solve
exactly the unperturbed Hamiltonian $H_{0}=H-1/r_{12}$ and obtain for the 
ground state an energy $E_{0}=-Z^{2}$ and a wave 
function $\Psi_{0}\propto\exp(-Zr_{1}-Zr_{2})$.

Assuming $\Psi$ to be  normalised, the repulsion gives a first order 
correction
\begin{equation}
    E_{0}\to E_{0}+\langle \Psi_{0}\vert r_{12}^{-1}\vert\Psi_{0}\rangle~.
    \label{eq:Helium-E1}
\end{equation}
The integral corresponding to the repulsive term is often calculated by means of an expansion 
of $r_{12}^{-1}$ in terms of spherical harmonics. See, for instance, \cite{Eyr63}. For the ground-state, one can also use the Gauss theorem, as, e.g., in~\cite{Pee92}. The field created at distance $r$ 
by the first electron is
\begin{equation}
    E(r)=-{1\over r^{2}}\int_{0}^r u(r)^{2}\,{\rm d} r=-{1\over r^2} + \left({1\over r^2} + 
  {2Z\over r} +  2Z^2\right) \exp(-2Zr)~,  \label{eq:Field-el1}
\end{equation}
where $u(r)=2 Z^{3/2}r\exp(-Z r)$, corresponding to a potential
\begin{equation}
    \eqalign{
    V(r)=&{}\int_{r}^\infty E(x)\,{\rm d}x=
   - {1\over r}\int_{0}^r u(x)^{2}\,{\rm d}x
    -\int_{r}^\infty{ u(x)^{2}\over x}\,{\rm d}x\cr
       =&{}{-1\over r} + \left({1\over r} +Z\right) \exp(-2Zr)~,\cr
            }
\label{eq:Field-el3}
\end{equation}
and thus a repulsion energy
\begin{equation}
   \langle r_{12}^{-1} \rangle =- \int_{0}^\infty V(r) u(r)^{2}\, {\rm 
   d} r  ={5 Z\over 8}~\cdot
    \label{eq:Field-el4}
\end{equation}
To first order, the energy  of helium is thus $-11/4=-2.75$.

A standard improvement consists of using a normalised wave function
\begin{equation}
    \Psi[\alpha]\propto\exp(-\alpha r_{1}-\alpha r_{2})~, 
    \label{eq:Field-el5}
\end{equation}
with kinetic energy $\alpha^{2}$ and potential energy $-2 Z \alpha$ 
for the attractive part, and $5\alpha/8$ for the repulsive part 
\cite{Kel27}.  The parameter $\alpha$ is interpreted as the 
effective charge seen by each electron.  One should thus minimise the 
variational energy
\begin{equation}
    \widetilde{E}(\alpha)=\alpha^{2}-2 Z \alpha+5\alpha/8
    =\left[\alpha-(Z-5/16)\right]^{2}-(Z-5/16)^{2}~,
    \label{eq:eq:Field-el6}
\end{equation}
this giving an improved minimum $\widetilde{E}=-729/256\simeq-2.85$ at
$\alpha=27/16\simeq1.69$, in the case of $Z=2$.

Note that the threshold for dissociation of charges $(Z,-1,-1)$ into an ion 
$(Z,-1)$ and an isolated electron is $E_\mathrm{th}=-Z^{2}/2$,
so that $\min_{\alpha}\widetilde{E}-E_\mathrm{th}=-(Z-5/16)^{2}+Z^{2}/2$ 
remains negative only for $Z>(10+5\sqrt{2})/16\simeq 1.07$. For 
smaller values of $Z$, demonstrating the stability of $(Z,-1,-1)$ 
requires trial wave functions more elaborated than (\ref{eq:Field-el5}). 
This is in particular the case for $\H^-$, where $Z=1$. The stability of $\H^-$ 
cannot be reached using factorised wave function $f(r_{1})f(r_{2})$, however refined 
 the individual function $f$.

Looking more closely at $\H^-$, again in the limit where the proton mass is 
infinite, the simplest wave function which achieves binding is
\cite{Bet29,Hyl29,Hyl47,Cha44,Hil77}.
\begin{equation}
    \Psi[a,b]= \exp(-a r_{1}-b r_{2}) + \{1\leftrightarrow 2\}~,
    \label{eq:Hminus1}
\end{equation}
or a factorised wave-function with $a=b$ but an explicit account of the correlation, e.g., by a factor $(1+ c r_{12})$, or  a combination of $a\neq b$ and of the $r_{12}$ term. For the choice (\ref{eq:Hminus1}),
the matrix elements of normalisation, potential energy and kinetic 
energy are easily calculated
\begin{equation}
    \eqalign{
    n(a,b)=&{}{2\pi^{2}\over a^{3}b^{3}}+{2\pi^{2}\over \bar{a}^{6}},\cr
    \langle\vec{p}_{1}^{2}\rangle=&{}{\pi^{2}(a^{2}+b^{2}\over a^{3} b^{3}}
    +{2 a b\pi^{2}\over \bar{a}^{6}},\cr
    \langle r_{1}^{-1}\rangle=&{}\pi^{2}(a+b)(a^{-3} b^{-3} + 
    \bar{a}^{-6}),\cr
    \langle r_{12}^{-1}\rangle=&{}{\pi^{2}(a^{2}+3 a b +b^{2})\over 
    a^{2}b^{2}(a+b)^{3}}+{5\pi^{2}\over 8 \bar{a}^{5}},\cr
           }
   \label{eq:matrix-Hill}
\end{equation}
where $\bar{a}=(a+b)/2$. 

In the case of Helium, one obtains an energy $E=-2.87566$, i.e., a 
modest gain with respect to the case where $a=b$ ($E=-2.8476$).  
However, the values $a\simeq 1.19$ and $b\simeq 2.18$ of the 
parameters show a non-negligible anticorrelation between the two 
electrons.

In the case of $\H^-$, binding is reached since the variational energy 
$E=-0.513303$ lies below the threshold $E_\mathrm{th}=-1/2$ by a 
fraction $x=\simeq0.027$. Anticorrelation is very strong, as
$b/a\simeq 3.67$. This means that if one electron is close to the 
proton, the second is far away, and vice-versa.

The Hughes--Eckart or mass-polarisation term $\vec{p}_{1}.\vec{p}_{2}$ 
has obviously a 
vanishing expectation value within this wave function 
(\ref{eq:Hminus1}).  Thus, if  the finite mass of the 
nucleus is taken into account, both the threshold energy and the variational energy are 
rescaled by the same factor $M/(M+m)$, where $m$ is the mass of the 
electron and $M$ that of the nucleus.  Thus, with this wave function, 
the fraction of binding is the same for all isotopes of Helium 
(neglecting finite-size effects) as for a static $Z=2$ source.  
Similarly, as noticed by Hill \cite{Hil77}, all systems 
$(M^{+}m^-m^-)$ are demonstrated to be stable (and within this 
approximation, with the same fraction $x$ and same anticorrelation 
$b/a$).

The wave function (\ref{eq:Hminus1}) or even a superposition of wave 
functions of this type will never approach very closely the exact 
solution, as it lacks one degree of freedom, namely some dependence on 
$r_{12}$, the distance among the electrons. A wave function 
that allows for this $r_{12}$ dependence is
\begin{equation}
    \Psi[a,b,c]=\varphi[a,b,c] + \cdots,\quad \varphi=\exp(-a x - b y - 
    c z),
    \label{eq:wf-perimetric}
\end{equation}
where the dots denote terms deduced by symmetry.  Here 
$x=r_{23}$, etc.  For the matrix elements dealing with scalar states, 
the integration is over $x{\rm d}x\,y{\rm d}y\,z{\rm d}z$, subject to 
triangular inequalities. A basic integral is
\begin{equation}
    \eqalign{
   &F[a,b,c]=\int \exp(-2a x -2 b y -2 c z){\rm d}x{\rm d}y{\rm d}z,\cr
   &={1\over 2}\int\limits_{0}^\infty\! \exp(-2a x){\rm d}x\!
   \int\limits_{x}^\infty\!\exp[-(b+c)\sigma]{\rm d}\sigma\! 
   \int\limits_{-x}^{+x}\!\exp[-(b-c)\delta]{\rm d}\delta,\cr
   &={1\over 4(a+b) (b+c) (c+a)},\cr
            }
    \label{eq:generic}
\end{equation}
in term of which the normalisation of a single exponential $\varphi$ is 
\begin{equation}
    n[a,b,c]=-{1\over8}{\partial F[a,b,c]\over \partial a
    \partial b \partial c},
    \label{eq:norm-perimetric} 
\end{equation}
and a typical term of the potential energy is
\begin{equation}
    v_{12}[a,b,c]={1\over 4}{\partial F[a,b,c]\over 
    \partial b \partial c}.
    \label{eq:pot-peri}
\end{equation}
For those local operators, off-diagonal matrix elements between 
$\varphi[a,b,c]$ and $\varphi[a',b',c']$ are deduced by the 
substitution $a\to \bar{a}=(a+a')/2$, etc.

For the kinetic energy, $\vec{p}_{1}\varphi=-i\varphi(b \hat{y} -c 
\hat{z})$, and thus the most general matrix element is
\begin{equation}
 \langle\varphi[a',b',c']\vert 
    \vec{p}_{1}^{2}\vert\phi[a,b,c]\rangle=(b b' +c c') 
    n(\bar{a},\bar{b},\bar{c})+ 
    (bc'+b'c)t_{1}(\bar{a},\bar{b},\bar{c}),
    \label{eq:p1s-perimetric}
\end{equation}
where $t_{1}$ is the matrix element of
\begin{equation}
    \hat{t}_{1}=\hat{y}.\hat{z}={x^{2}-y^{2}-z^{2}\over 2 y z},
    \label{eq:yz}
\end{equation}
and is given by a combination of suitable derivatives of the 
generating function $F[a,b,c]$.

For the Helium atom, the wave function (\ref{eq:wf-perimetric}) 
contains two exponentials related by $\{1\leftrightarrow 2\}$ exchange 
and thus three parameters $a$, $b$, and $c$ (two, if one use the virial 
theorem). It provides a variational energy $E=-2.89953$, to be compared 
with the best available result $E=-2.90372$ (see, e.g., 
\cite{Lin95}, and references therein).

For $\H^-$, the fraction of binding reaches $x_{b}=0.047696$, while for 
$\Ps^-$, it is $x=0.0268$.

\subsection{Four-body systems}
\label{sec.Four-body-calc}
Let us consider here a system ($1,2,3,4)$ with charges 
$\pm(+1,+1,-1,-1)$.  We shall only consider the ground state.  A wave 
function that is a generalisation of the one  used by Hylleraas and Ore 
\cite{Hyl47a} for $\Ps_{2}$ reads
\begin{equation}
    \eqalign{
  \Psi[a,b,c,d]=&{}\varphi[a,b,c,d]+\dots,\cr
  \varphi[a,b,c,d]=&{}\exp[-(a r_{13}+b r_{23}+cr_{14}+dr_{24})/2],\cr
            }
    \label{eq:Hyll-gen}
\end{equation}
where once again the dots means exponentials deduced from
symmetry considerations. In matrix elements not involving $r_{34}$, the 
integration runs over 
\begin{equation}
    {\rm d}\tau=r_{13}r_{14}r_{23}r_{24} {\rm d}\tau',   \quad
    {\rm d}\tau'={\rm d}r_{12}{\rm d}r_{13}{\rm d}r_{14}
                               {\rm d}r_{23}{\rm d}r_{24},
    \label{eq:integration-4body}
\end{equation}
as the system is fully identified by the two independent triangles
(1,2,3) and (1,2,4). A basic integral is 
\begin{equation}
  F[a,b,c,d,u]= \int{{\rm d}\tau'\over r_{12}}
       \exp(-a r_{13}-b r_{23}-cr_{14}-dr_{24}-ur_{12}). 
    \label{eq:basicI-4}
\end{equation}
Again, the triangle inequalities are conveniently accounted for by 
using the variables $(\sigma_{3},\delta_{3})=r_{13}\pm r_{23}$, and similarly 
for particle 4. One  obtains
\begin{equation}
    F[a,b,c,d,u]={16 \log[(b+c+u)(a+d+u)/((a+c+u)(b+d+u))]\over
         (a-b)(a+b)(c-d)(c+d)}\cdot
    \label{eq:F4}
\end{equation}
The normalisation integral, and the matrix elements of the potential 
can be written as derivatives of $F$. For instance the normalisation is 
\begin{equation}
 n(a,b,c,d)=\int \vert\varphi^{2}\vert{\rm d}\tau=
 -\left.{\partial F\over \partial u \partial a \partial b \partial c \partial 
 d }\right\vert_{u=0}.
    \label{eq:norma4}
\end{equation}
The internuclear repulsion  is
\begin{equation}
   v_{12}[a,b,c,d]=\langle r_{12}^{-1}\rangle ={\partial^4 
   F_{12}[a,b,c,d,0]\over \partial a \partial b \partial c \partial d}~. 
    \label{eq:v12}
\end{equation}
 The electronic repulsion is obviously
\begin{equation}
    v_{34}[a,b,c,d]=v_{12}[a,c,b,d].
    \label{eq:v34}
\end{equation}
The kinetic energy of particle 3 can be calculated starting from
\begin{equation}
    \vec{p}_{3}\varphi[a,b,c,d]=-i\varphi[a,b,c,d](a 
    \hat{r}_{13}+b\hat{r}_{23}),
    \label{eq:p3-phi}
\end{equation}
so that, exactly as in the 3-body case
\begin{equation}
   \langle\varphi\vert\vec{p}_{3}^{2}\vert\varphi\rangle=
   (a^{2}+b^{2})\langle\varphi\vert\varphi\rangle-2 ab
   \langle\varphi\vert(r_{12}^{2}-r_{13}^{2}-r_{23}^{2})/(2 
   r_{13}r_{23})\vert\varphi\rangle
    \label{eq:p3sq}
\end{equation}
 which can be expressed in terms of derivatives of $F[a,b,c,d,u]$.  
 The expectation values of  $\vec{p}_{i}^{2}$ for other particles are deduced by 
 suitable permutations of $a$, $b$, $c$ and $d$.

Off-diagonal matrix elements between 
$\varphi[a,b,c,d]$ and $\varphi[a',b',c',d']$ are given by the simple 
rule $a\to \bar{a}=(a+a')/2$ for the local operators. For the kinetic 
energy terms in Eq.~(\ref{eq:p3sq}), one replaces $a^{2}+b^{2}$ by 
$aa'+bb'$ and $2ab$ by $ab'+a'b$, while the operators adopt the 
average arguments $\bar{a}$, $\bar{b}$, etc.

One eventually obtains  explicit expressions for the expectation value of the normalisation, kinetic energy and  potential energy. Equation (\ref{4u:eq:Hylleraas-ev}) follows straightforwardly.

\clearpage\begin{acknowledgments}
We would like to thank  our collaborators on the topics covered by this review, 
W.~Byers Brown, S.~Fleck,  A.~Krikeb,  A.~Martin and Tai T.~Wu, 
for their encouragement and useful advice.

E.A.G.A. thanks EPSRC (UK) for support for this research through grants GR/L29170 and GR/R26672.

K.V.~is supported by OTKA grants (Hungary) T029003 an T037991 and he is sponsored by 
the U.S.\ Department of Energy under  contract DE-AC05-00OR22725 with the 
Oak Ridge National Laboratory, managed by UT-Battelle, LLC.

J.-M.~R. benefitted from the hospitality of IPNL,  Universit\'e de Lyon, where part of this work was done.
\end{acknowledgments}
\markboth{\sl Stability of few-charge systems}{\sl Bibliography}
\clearpage
\bibliography{stabrev}
\bibliographystyle{unsrt}
\end{document}